\documentclass[fleqn,usenatbib]{mnras}

\usepackage{newtxtext,newtxmath}

\usepackage[T1]{fontenc}
\usepackage{ae,aecompl}

\usepackage{graphicx}	
\usepackage{amsmath}	
\usepackage{subfig}
\usepackage{pifont}
\usepackage[table]{xcolor} 

\def\mpcoh{\,h^{-1}{\rm Mpc}}

\providecommand{\e}[1]{\ensuremath{\times 10^{#1}}}
\def\msunoh{\,h^{-1}{\rm M_\odot}}
\def\citejap#1{\citeauthor{#1}\ \citeyear{#1}}
\definecolor{japviolet}{rgb}{0.53, 0., 0.69}

\title[The future of stars and the intergalactic medium]{Evolving beyond $\mathbf{\emph{z}=0}$: insights about the future of stars and the intergalactic medium}

\author[BK. Oh et al.]{%
Boon Kiat Oh,$^{1, 2}$\thanks{E-mail: bkoh@roe.ac.uk}
John A. Peacock,$^{1}$
Sadegh Khochfar,$^{1}$
and Britton D. Smith$^{1}$
\\
$^{1}$Institute for Astronomy, University of Edinburgh, Royal Observatory, Edinburgh EH9 3HJ, United Kingdom\\
$^{2}$Center for Theoretical Physics, Department of Physics and Astronomy, Seoul National University, Seoul 08826, Korea\\
}

\date{Accepted XXX. Received YYY; in original form ZZZ}

\pubyear{2019}

\hypersetup{draft}
\begin{document}
\label{firstpage}
\pagerange{\pageref{firstpage}--\pageref{lastpage}}
\maketitle

\begin{abstract}
	We present results from seven cosmological simulations that have been extended beyond the present era as far as redshift $z=-0.995$ or $t\approx96\,{\rm Gyr}$, using the {\tt Enzo} simulation code. We adopt the calibrated star formation and feedback prescriptions from our previous work on reproducing the Milky Way with {\tt Enzo} with modifications to the simulation code, chemistry and cooling library. We then consider the future behaviour of the halo mass function (HMF), the equation of state (EOS) of the IGM, and the cosmic star formation history (SFH). Consistent with previous work, we find a freeze-out in the HMF at $z\approx-0.6$. The evolution of the EOS of the IGM presents an interesting case study of the cosmological coincidence problem, where there is a sharp decline in the IGM temperature immediately after $z=0$. For the SFH, the simulations produce a peak and a subsequent decline into the future. However, we do find a turnaround in the SFH after $z\approx-0.98$ in some simulations, probably due to the limitations of the criteria used for star formation. By integrating the SFH in time up to $z=-0.92$, the simulation with the best spatial resolution predicts an asymptotic total stellar mass that is very close to that obtained from extrapolating the fit of the observed SFR. Lastly, we investigate the future evolution of the partition of baryons within a Milky Way-sized galaxy, using both a zoom and a box simulation. Despite vastly different resolutions, these simulations predict individual haloes containing an equal fraction of baryons in stars and gas at the time of freeze-out ($t\approx30\,{\rm Gyr}$). 
\end{abstract}




\section{Introduction}

When haloes of dark matter form through gravitational collapse, baryons can fall into these dominant dark potential wells, becoming pressure supported until they are able to cool and condense to form stars \citep[e.g.][]{1978MNRAS.183..341W}. Once stars form, these rapidly provide feedback by injecting energy into the interstellar medium (ISM). When massive stars reach the end of their main-sequence lifetimes, they explode as supernovae, enriching the ISM by injecting a large amount of energy ($10^{51}$\,erg per supernova: \citejap{1972ARA&A..10..129W}) and metals. The metals will provide an additional source of radiative cooling, especially for cold gas (e.g. \citejap{2008MNRAS.385.1443S}). 

Feedback processes have been studied and implemented extensively in various simulations \citep{2003MNRAS.339..289S, 2005MNRAS.363....2K, 2006MNRAS.373.1265O, 2010MNRAS.402.1536S, 2014MNRAS.444.1518V, 2015MNRAS.446..521S, 2018MNRAS.473.4077P}. We conducted a similar study in \citet{BK20} by exploring the subgrid parameter space within the {\tt Enzo} code. After calibrating the subgrid physics for individual haloes of specific masses, we now wish to apply this prescription to the halo population within simulations of cosmological volumes, with the aim of studying the long-term evolution of baryons. Since we expect Active Galactic Nuclei (AGN) feedback to be subdominant at mass scales below Milky-Way sized haloes \citep{2006MNRAS.370..645B, 2010ApJ...717..379B, 2014IAUS..303..354S}, we do not include this feedback in our simulations. 

The gas that will fuel star formation comes from the interstellar, circumgalactic and intergalactic medium. It originates by infall from the intergalactic medium (IGM) in the outskirts of the halo, beyond the virial radius. The gas then enters the intersection between the galaxy and the IGM, the circumgalactic medium (CGM). The CGM contains gas that originates from the metal-poor IGM inflows, metal-rich supernova and feedback outflows, and recycled gas from various sources such as stripping from infalling satellite galaxies \citep{2019ApJ...873..129P}, making it a unique region. Finally, gas reaches the innermost regions of the galaxy, contributing to the interstellar medium (ISM). According to this categorisation, the IGM contains the bulk of cosmic matter \citep{2016ARA&A..54..313M}. Therefore, the IGM is vital across astrophysical scales, ranging from tests of models of structure formation \citep{2005PhRvD..71f3534V, 2005PhRvD..71j3515S}, to anisotropies in the CMB \citep{1986ApJ...306L..51O, 2000ApJ...529...12H}, to cosmological parameter inference \citep{2007MNRAS.374..159P, 2011MNRAS.415.3929W}. For these reasons, our focus in this paper will be on the properties of the IGM and their evolution over cosmological history.

This evolution is related indirectly to the cosmological history of star formation. That history is measured most effectively by stellar emission from the far-UV (FUV) to the far-infrared (FIR)  \citep{2014ARA&A..52..415M}. Short-lived massive stars except those in the oldest galaxies dominate the UV emission, allowing a direct determination of the instantaneous star formation rate density (SFRD). This measurement assumes a stellar initial mass function and dust content. At the opposite extreme of wavelength, the FIR emission of dusty starburst galaxies also acts as an effective tracer of young stars and the SFRD because interstellar dust preferentially absorbs UV light and re-emits in the thermal IR. \citet{2014ARA&A..52..415M} fitted a double power-law fit to these observations. This fit describes a cosmic star formation rate density that rises gradually, peaking at $z\approx2$ and then declines towards the present.

This UV emission from star formation helps maintain the ionisation of the IGM, together with emission from active galactic nuclei. The exact contribution from each of these sources remains under debate, but at $z>3$ the decreasing population of bright quasars leads to a correspondingly reduced contribution to the UV background, suggesting that stars must provide the majority of the ionizing flux at early times \citep{1999ApJ...514..648M, 2000ApJ...535..530G, 2003ApJ...586..693W, 2005MNRAS.356..596M, 2010Natur.468...49R}. Thus star formation is an important driver for the strength of the UV background. The transfer of energy from this diffuse background flux is most apparent in the IGM. There is a tight power law relation between the temperature and density of the gas in the IGM at high $z$ \citep{1997MNRAS.292...27H}. This relation results from the balance of the background UV photoheating and the adiabatic cooling due to the expansion of the universe \citep{2016MNRAS.456...47M} amongst other processes.

Extrapolation of the analytic fit to the cosmic SFRD \citep{2014ARA&A..52..415M} into the future predicts a continuation of the decline seen between $z\approx2$ and the present. Potential causes include the slowing of the growth of large scale structure due to the accelerating expansion of the universe and efficient stellar and AGN feedback \citep{2018MNRAS.477.3744S}. These authors demonstrated that the decline in the star formation rate could be avoided by switching off the AGN feedback in their simulations, suggesting that the future of star formation is heavily dependent on feedback processes. They ended their simulations at an age of $20\,{\rm Gyr}$, before the `freeze out' era ($z\approx-0.6$ or $t\approx30\,{\rm Gyr}$), following which haloes undergo isolated evolution. 

This raises the question of the long-term fate of the IGM gas: when provided with a infinite further amount of time, can it potentially cool and form stars, even though the cooling timescales in the IGM are too long to have any significant impact on star formation at $z=0$? The answer is not obvious, because gas cooling has to compete with the accelerating expansion of the universe, which reduces the inflow of gas. These processes affect the reservoir of gas available for long-term star formation. We therefore aim to extend predictions of star formation and the evolution of the IGM beyond the next $20\,{\rm Gyr}$, in order to understand the asymptotic fate of the baryonic components of the universe. A pioneering study of this topic was made by \cite{2004NewA....9..573N}, and we aim to expand on their work by using a diversity of methods and higher resolution.

In this paper, we start with a cosmological box simulation that is comparable to that of \citet{2004NewA....9..573N}. We aim to compare that work and its conclusions with the predictions of modern galaxy formation codes, where the treatment of feedback is rather different and where it is possible to perform more demanding convergence studies at higher resolution. We use the calibrated star formation and feedback prescriptions discussed extensively in \citet{BK20}. Building on this simulation, we vary the mass and maximum spatial resolution to test for convergence. We also apply the feedback prescription associated with Setup 1 from \citet{BK20} as a test of the sensitivity of the results to different star formation and feedback prescriptions. Lastly, we continue the zoom simulation in \citet{BK20} into the future, quantifying the impact of vastly different resolutions. Comparing these seven simulations, we look at the evolution of the dark matter haloes, gas properties and star formation.

This paper is structured as follows. Section \ref{sec:sim_setup} describes the cosmological parameters used in the generation of the initial conditions, the code, and setup for evolving them into the future.  This will be the first application of {\tt Enzo}, {\tt Grackle} and {\tt ROCKSTAR} to galaxy formation simulations of the future, i.e., negative redshifts. Since they were not designed for such a purpose, we explore the necessary changes to the codes to carry out the simulations. Section \ref{sec:naga-results} will first present the iteration of results from these changes to the simulation code. We also verify the capability of the {\tt ROCKSTAR} halo finder to accurately identify and trace haloes into the future. This will be followed up by the comparison of our results to \citet{2004NewA....9..573N} and establishing the convergence of these results. Lastly, we present and discuss the results from simulations of various resolutions and, star formation and feedback prescriptions. The halo mass functions, phase distribution of temperature and density of gas, equation of state of the IGM and star formation history of these simulations will be compared. Section \ref{sec:naga-summary} provides a summary and discussion of the results obtained.

\section{Simulations and post processing}\label{sec:sim_setup}

In this section, we provide an overview of our simulation setups. As mentioned, the codes used were not designed for evolving galaxy formation into the future, and we include a critical assessment of the ability of {\tt Enzo} and {\tt Grackle} to fulfil this requirement. It is not surprising to find that certain components of the code require modifications, and the necessary changes will be discussed in subsequent sections.

Cosmological parameters were taken from WMAP-9 \citep{2013ApJS..208...20B}, $\Omega_m=0.285$, $\Omega_\Lambda=0.715$, $\Omega_b=0.0461$, $h = 0.695$ and $\sigma_8=0.828$ with their usual definitions are assumed across all simulations. We generate the initial conditions of our simulations using MUlti-Scale Initial Conditions for cosmological simulations ({\tt MUSIC}: \citejap{2011MNRAS.415.2101H}). Other than in terms of varying resolution, all simulations are initialised and set up identically. We evolve the simulation using the AMR code, {\tt Enzo}, using the hydrodynamic solver that originated from {\tt ZEUS} \citep{1992ApJS...80..753S} and  an N-body adaptive particle-mesh gravity solver \citep{1985ApJS...57..241E}. The cooling and chemistry processes are handled by the equilibrium cooling mode of the {\tt Grackle} library \citep{2017MNRAS.466.2217S}. This makes use of the tabulated cooling rates derived from the photoionisation code, {\tt CLOUDY} \citep{2013RMxAA..49..137F}. Lastly, we apply and evolve the UV background radiation given by \citet{2012ApJ...746..125H} up to and beyond $z=0$. We will discuss the future behaviour of the UV background in Section \ref{sec:uvb_mod}. 

For star formation and thermal feedback, we adopt the model by \citet{1992ApJ...399L.113C} and \citet{2011ApJ...731....6S}'s modified version of the \citet{2006ApJ...650..560C} thermal supernova feedback. The parameter space for this framework was explored extensively in \citet{BK20}, who considered how to match the star-formation history of the Milky Way. Here we adopt their `Setup 1' and `Setup 2', which differ mainly in the conversion from gas to stars. The former uses a timestep dependent conversion efficiency, Jeans instability check and a minimum mass of $10^5\,M_\odot$ for the star particles, while the latter applies a timestep independent conversion and removes both the instability check and minimum mass. 

When evolving a simulation with Setup 1, we apply the corresponding set of feedback parameter values of (2.5\e{-4}, 1\_3, 0.2), consistent with the definition of ($\epsilon$, $r$\_$s$, $f_s$) given in \citet{BK20}. On the other hand, when we evolve the simulation with Setup 2, the feedback parameter values are set to (3.0\e{-5}, 1\_1, 0.9). For clarity, $\epsilon$ refers to the feedback efficiency implemented in the simulations. It is a user-defined factor relating the amount of feedback energy ($E_{\rm feedback}$) injected to the rest mass energy of the star forming gas ($m_{\rm form}\times c^2$) (refer to Equation 6 in \citejap{BK20}). $r\_s$ defines the volume in which this feedback energy is injected into, e.g., $r\_s=1\_1$ refers to the 6 cells adjacent to the one containing the star particle. Lastly, $f_s$ is the star formation efficiency factor which specifies the conversion efficiency of the gas mass in a cell into stellar mass (refer to sections 2.1 and 2.2 in \citejap{BK20} for details). Setup 2 will be the primary star formation and feedback prescription used in this paper. We run a total of seven simulations to investigate the convergence of the properties of baryons as they evolve into the future and they are summarised in Table \ref{tab:naga-setup}. 

\begin{table*}
	\caption{List of simulations discussed in this paper with their corresponding reference name. This table includes the number of particles, cosmological box size, the maximum number of AMR level, the maximum spatial and mass resolution, and the star formation setup and its corresponding feedback prescription in each simulation. Refer to Section \ref{sec:sim_setup} for more information about the naming convention and specifics of each simulation.}
	\label{tab:naga-setup}
	\centering
	\begin{tabular}{|p{.08\textwidth}|p{.08\textwidth}|p{.08\textwidth}|p{.05\textwidth}|p{.12\textwidth}|p{.12\textwidth}|p{.1\textwidth}|p{.15\textwidth}|}
		\hline
		\multicolumn{8}{|c|}{Simulation setup list} \\
		\hline
		Reference name & Particle number & Box size [$h^{-1}\,{\rm cMpc}$] & AMR level & Spatial resolution [ckpc] & Mass resolution [$M_\odot$] & Star formation setup & Feedback prescription\\
		\hline
		{\sl NL} & $128^3$ & 50 & 4 & 35.13 & 6.79\e{9} & 2 & (3.0\e{-5}, 1\_1, 0.9)\\
		\hline
		{\sl NL-1} & $128^3$ & 50 & 3 & 70.26 & 6.79\e{9} & 2 & (3.0\e{-5}, 1\_1, 0.9)\\
		\hline
		{\sl NL+1} & $128^3$ & 50 & 5 & 17.56 & 6.79\e{9} & 2 & (3.0\e{-5}, 1\_1, 0.9)\\
		\hline
		{\sl NLm-1} & $64^3$ & 50 & 5 & 35.13 & 5.43\e{10} & 2 & (3.0\e{-5}, 1\_1, 0.9)\\
		\hline
		{\sl NLm+1} & $256^3$ & 50 & 3 & 35.13 & 8.48\e{8} & 2 & (3.0\e{-5}, 1\_1, 0.9)\\
		\hline
		{\sl NLfb} & $128^3$ & 50 & 4 & 35.13 & 6.79\e{9} & 1 & (2.5\e{-4}, 1\_3, 0.2)\\
		\hline
		{\sl zoom} & $256^3$ & 100 & 8 & 2.196 & 1.72\e{5} & 2 & (3.0\e{-5}, 1\_1, 0.9)\\
		\hline
	\end{tabular}
\end{table*} 

The names of the simulations indicate their resolutions and feedback prescriptions. For example, {\sl NL} denotes a baseline simulation  with a resolution comparable to that of \citet{2004NewA....9..573N}, employing feedback according to Setup 2.  Simulations {\sl NL$\pm x$} and {\sl NLm$\pm x$} modify {\sl NL}, with $x$ levels added or removed from the AMR and root grid resolution respectively. For {\sl NLm$\pm x$}, both the spatial and mass resolution of the initial conditions are changed, while the maximum AMR is adjusted to keep the maximum spatial resolution constant. Only the spatial resolution is changed for {\sl NL$\pm x$}. To these we add two further simulations: {\sl NLfb} uses the feedback prescription of Setup 1; {\sl zoom} is a continuation of the zoom simulations described in \citet{BK20} beyond $z=0$. These resolution choices are discussed in Section \ref{sec:resolution}, and the results of the simulations are discussed in Section \ref{sec:naga-results}.

\subsection{Resolution}\label{sec:resolution}

\citet{2004NewA....9..573N} presented the future evolution of the IGM with a version of the parallel tree SPH code {\tt GADGET} \citep{2001NewA....6...79S}. Their simulation consisted of $64^3$ particles each for dark matter and gas within a $50 \mpcoh$ cosmological box, translating into a mass resolution of $3.4\e{10}\msunoh$ and $5.3\e{9}\msunoh$ for the dark matter and gas respectively. In contrast, we are using the particle-mesh code, {\tt Enzo}, to evolve a simulation into the future for the first time. To allow a fair comparison of results, we implement a comparable mass resolution in our simulation. 

For a direct comparison to $64^3$ dark matter and gas particles each in a {\tt GADGET} simulation, we require a $128^3$ root grid with four AMR levels in {\tt Enzo}. This increased number of particles in the grid case was prompted by deviations in the low-mass end of the halo mass functions between {\tt GADGET} and {\tt Enzo}  \citep{2005ApJS..160....1O}. This increment also resulted in a higher spatial resolution in the root grid, which allows the formation of low mass haloes, preventing the loss of small-scale power. The number of AMR levels was decided according to the agreement of the late time power spectrum determined by \citet{2005ApJS..160....1O}. This setup translates to a maximum spatial resolution of $24.42\,h^{-1}\,{\rm ckpc}$ and a dark matter mass resolution of $4.72\e{9}\msunoh$. Generally, these numbers are indicative of a low-resolution simulation. Hence, we look to quantify and establish convergence by increasing and decreasing the number of AMR and root grid resolution. As mentioned, a summary of these runs is presented in Table \ref{tab:naga-setup}.

\subsection{Final redshift}

To compare with \citet{2004NewA....9..573N}, we have to evolve our simulations for an equivalent or longer period of time. The cosmological parameters assumed in their simulation ($\Omega_m=0.3$, $\Omega_\Lambda=0.7$, $\Omega_b=0.04$, $h = 0.7$ and $\sigma_8=0.9$) are rather similar to ours, but in detail we need to know the age of the Universe at a given redshift, $t(z)$. For a flat model, this is
\begin{equation}
\label{eq:z_vs_t}
H_0t = \frac{2}{3\sqrt{1-\Omega_{m}}}\sinh^{-1}\sqrt{\frac{1-\Omega_{m}}{\Omega_{m}(1+z)^3}}, 
\end{equation} where $\mathrm{H_0}$ and $\Omega_{m}$ is the Hubble parameter and matter density parameter at $z=0$ \citep{1993ppc..book.....P}. 
We choose a final redshift of $-0.995$ in our simulations, corresponding to a scale factor $a = (1+z)^{-1} = 200$, or
$t \approx 7\,t_\mathrm{H}$. This encompasses the end point chosen by \citet{2004NewA....9..573N}, $a=166$.

\subsection{Modifications to Enzo and Grackle}

When first integrating towards this target, the results displayed some peculiarities, particularly in the distribution of the temperature and density of the gas. As a result, we re-evaluated the ability of various components in {\tt Enzo} and {\tt Grackle} to function in this non-standard regime.  Certain methods or values in the machinery proved adequate for evolution up to $z=0$, but with small errors that became important is the conditions experienced in the future. These include the evolution of the UV background, values in the {\tt CLOUDY} table and fail-safe features in {\tt Grackle}.

\subsubsection{UV background evolution} \label{sec:uvb_mod}

One of the most popular UV background models is the Haardt and Madau model obtained with {\tt CUBA}. This is a radiative transfer code that quantifies the propagation of Lyman-continuum photons through a partially ionised inhomogeneous IGM \citep{1996ApJ...461...20H}. This UV background model has undergone several iterations as a result of the improvements in the quantity and quality of observations, and a better understanding of the relevant physics over the years. 

\citet{2004NewA....9..573N} implemented a uniform UV background with a modified \citet{1996ApJ...461...20H} spectrum, having complete reionization at $z\approx6$ \citep{1999ApJ...511..521D, 2001AJ....122.2850B}. Beyond $z=0$, the authors linearly extrapolated the UV background, consistent with the extrapolated decline of cosmic star formation. This methodology ensures that the UV background approaches zero quickly. Any interpolation or extrapolation of the UV background is done linearly in $z$ in {\tt Enzo} by default. We choose to use an updated \citet{2012ApJ...746..125H} UV background model for our simulations. If we continue to use a linear extrapolation in $z$ for this model, it reverses the photoheating rates for neutral atomic hydrogen (HI), neutral helium (HeI) and singly ionised helium (HeII) from a positive to a negative value at $z=-0.195$, $z=-0.201$ and $z=-0.232$ respectively. This transition means that instead of heating the IGM, the UV background will be cooling the IGM at the mentioned $z$ for the various species. Since it was ambiguous what `linear' meant in \citet{2004NewA....9..573N}, to avoid the unphysical cooling, we modify the extrapolation scheme of the photoheating rates from linear to logarithmic in $1+z$ space in {\tt Enzo}. In other words, we use
\begin{equation}
    \log\,UV = A + B\,\log(1+z),
\end{equation} where A and B are dimensionless constants and $UV$ refers to photoheating rates for HI, HeI and HeII with the assumption that these heating rates reach zero at $z=-0.99999999$ (an arbitrary choice, which has no effect on the results in the redshift regime of our computations, $z<-0.995$).

Any extrapolation of the UV background is unsatisfactory because it assumes that the global star formation rate will decrease into future. Ideally, we want to implement a UV background that is self-consistent with the ongoing star formation rate at the specified redshift. However, this is not practical as it requires an iterative process between the SFRD and implemented UV background in order for them to match in the simulations. In any case, we only resolve haloes with mass above $\sim10^{11}\,M_\odot$, for which the star formation will not be highly sensitive to the UV background \citep{2018MNRAS.480.1740D, 2020MNRAS.494.2200K}.

\subsubsection{CLOUDY table} \label{sec:cloudy_mod}

Our simulations employ the equilibrium cooling mode from {\tt Grackle} \citep{2017MNRAS.466.2217S}, which uses tabulated heating and cooling rates as a function of density, temperature and redshift, derived from the {\tt CLOUDY} photoionisation code \citep{2013RMxAA..49..137F}.  When evolving a simulation into the future, the available {\tt CLOUDY} table fails to account for two factors: heating and cooling rates into the future and significantly lower densities due to the expansion of the universe. Therefore, there is a need to revise and modify the {\tt CLOUDY} table.

To obtain the cooling and heating rates in the future, we can choose to extrapolate from the last two data points of heating and cooling rates in redshifts ($z=0.04912$ and $z=0.0$). However, this extrapolation assumes that it encompasses all behaviour beyond these data points, leading to an increased likelihood of unphysical values as we extrapolate further into the future. Therefore, by assuming that the heating rates reach zero at some arbitrary point time in the distant future ($z=-0.99999999$), we allow the cooling and heating rates to be instead interpolated between two defined points ($z=0$ and $z=-0.99999999$). As discussed before, the intensity of the UV background is expected to decrease to zero in the far future because of the extrapolated decline in the global star formation rate.

\begin{figure}
	\centering
	\includegraphics[width=\columnwidth]{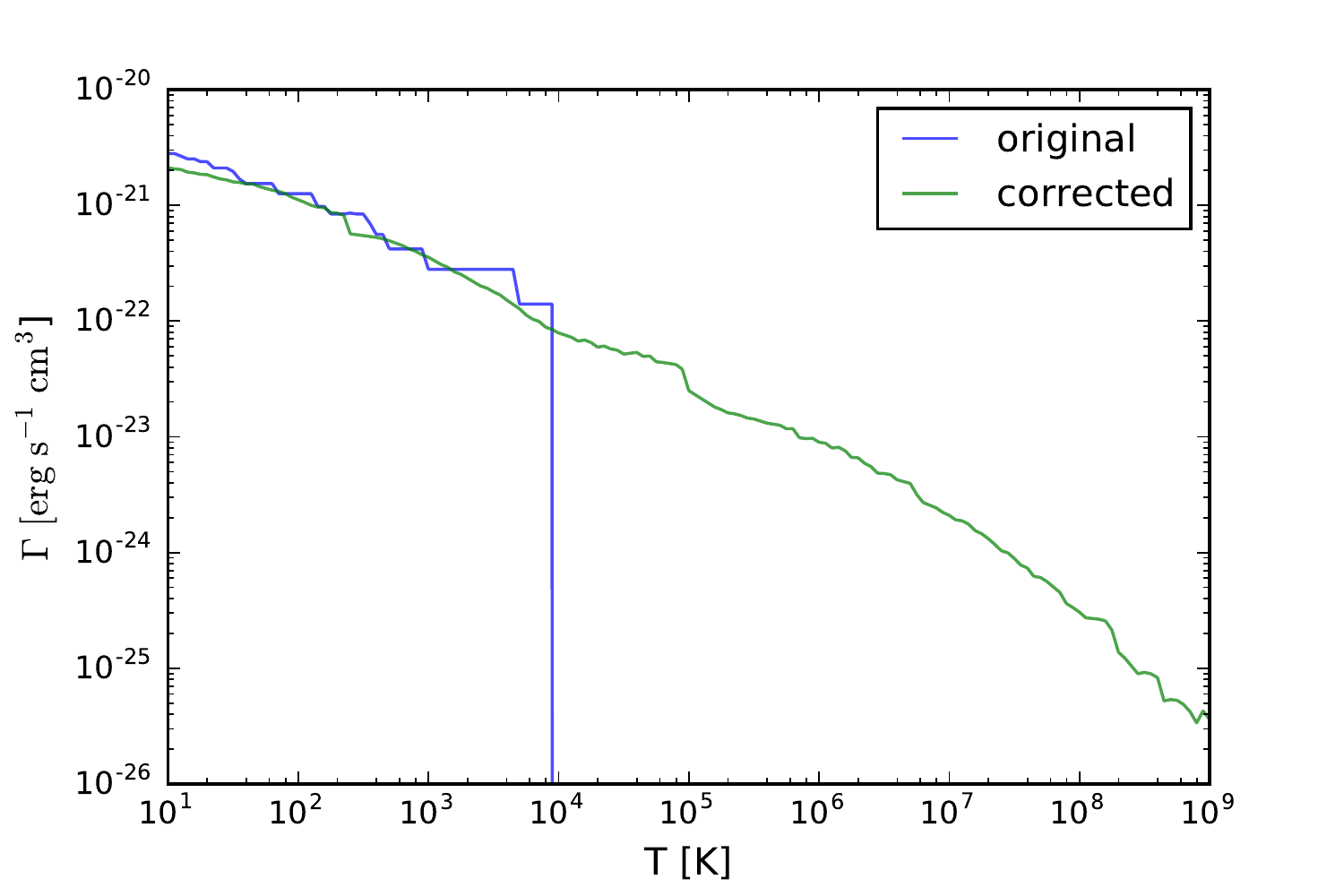}
	\caption{Graph of  the heating rate coefficient $\Gamma$ with respect to $T$ at a fixed density and $z$. The blue and green lines represent the heating rate with and without corrections respectively. The temperature range is much more extended with the corrections. In addition, the flat portions of the heating rate curve are removed, allowing for a more realistic interpolation. Refer to the discussion of the {\tt CLOUDY} table under Section \ref{sec:cloudy_mod} for details about the corrections.}
	\label{fig:grackle_fix}
\end{figure}

A further unrelated issue is illustrated by Figure \ref{fig:grackle_fix}, where we plot the original heating rates ($\Gamma$) with respect to temperature ($T$) at a fixed density and redshift in the {\tt CLOUDY} table (blue line). Although the heating rates are stored as $\Gamma$, they are calculated from $\Gamma \times n_{\rm H}^2$ where $n_{\rm H}$ is the number density of ${\rm H}$. In the far future when $n_{\rm H}$ can be much smaller than the values encountered at $z>0$, this quantity suffers from floating point underflow. Using the lowest temperature deemed to not be significantly affected by round-off error, $T_\alpha$, at a given density, $\rho_0$, we carry out a second order interpolation in logarithmic space to correct the heating rates for $T \geq T_\alpha$, 
\begin{equation} \label{eq:log_linear_interp}
\begin{split}
\Gamma(\rho_0, z, T \geq T_\alpha) &= \mathrm{L0} \times \Gamma(\rho_1, z, T \geq T_\alpha) + \mathrm{L1} \times \Gamma(\rho_2, z, T \geq T_\alpha) \\&+ \mathrm{L2} \times \Gamma(\rho_3, z, T \geq T_\alpha),
\end{split}
\end{equation} where L0, L1 and L2 are Lagrange basis functions given by
\begin{align}
    \rm L0 = (\rho_0 - \rho_2)(\rho_0-\rho_3) / (\rho_1 - \rho_2)(\rho_1-\rho_3),\\
    \rm L1 = (\rho_0 - \rho_1)(\rho_0-\rho_3) / (\rho_2 - \rho_1)(\rho_2-\rho_3),\\
    \rm L2 = (\rho_0 - \rho_1)(\rho_0-\rho_2) / (\rho_3 - \rho_1)(\rho_3-\rho_2),
\end{align}
with subscripts 1, 2 and 3 referring to the first, second and third sequentially higher density. This process corrects regions affected by underflow in a manner that preserves the density dependence of the rates as shown by the green line in Figure \ref{fig:grackle_fix}.

We carry out this correction iteratively, between $-10 < \log_{10}(n_H / {\rm cm^{-3}}) < 4$, starting with heating rates at a fixed redshift having the least amount of missing data points, i.e., from high to low density. Once we resolve the problem within $\rho_0$; we employ a slightly different process for the next density, $\rho_{-1}$. We make use of the gradient
\begin{equation}
m = 0.5 \times \frac{\Gamma(\rho_1, z, T \geq T_\beta) - \Gamma(\rho_0, z, T \geq T_\beta)}{\rho_1 - \rho_0},
\end{equation} 
where $T_\beta$ is similar to $T_\alpha$, but for the density, $\rho_{-1}$ and 0.5 is a factor to dampen the solution in order to better match the apparent density dependence, resulting in
\begin{equation}
\Gamma(\rho_{-1}, z, T \geq T_\beta) = m \times \left(\rho_{-1} - \rho_0 \right) + \Gamma(\rho_0, z, T \geq T_\beta).
\end{equation} As mentioned previously, the temperature associated with the flat portion varies between density values, we therefore use $T_\beta$ for the $\rho_{-1}$ slice to differentiate it from $\rho_0$. The solution first rectifies the heating rates of the density with fewest missing data points, before moving to progressively harder cases. This entire process then repeats for different redshifts. This methodology is adopted to address the lack of values in {\tt CLOUDY} for exceptionally low density and temperature, which are present in the far future of our universe. 

\subsubsection{Grackle} \label{sec:grackle_mod}

There are measures in {\tt Grackle} to prevent arithmetic underflow and round-off error, as in other codes. In particular, round-off error occurs when the net change in internal energy is too small to be captured by the floating point precision. In the default setup, when the absolute change in internal energy is less than $10^{-20}\,{\rm erg\ s^{-1}\ cm^{-3}}$, {\tt Grackle} replaces this small change in internal energy with a small heating value, regardless of whether or not it was cooling or heating in the first instance. In typical simulations that complete at $z=0$, this artificial heating is insignificant, but this small numerical patch can eventually dominate the results when considering extremely low density gas with long cooling times in the far future.

We experimented with reducing the threshold value from $10^{-20}\,{\rm erg\ s^{-1}\ cm^{-3}}$ to $10^{-40}\,{\rm erg\ s^{-1}\ cm^{-3}}$ and setting the heating/cooling rate to zero instead of inserting the small heating value. While this implementation resolves the issues associated with artificial heating, it introduces other numerical artefacts. In our follow-up attempt, we allowed the gas to cool or heat accordingly, even if the absolute value is below the threshold. In short, we simply removed the threshold and any corrections introduced because we did not find any consequences from the round-off error.

In addition to the issues discussed above, {\tt Grackle} also inserts a small heating value when gas in the cell is less than $1\,{\rm K}$ and still cooling. In the far future, $1\,{\rm K}$ is much larger than the temperature floor set in the simulations according to the CMB temperature,
\begin{equation}
\label{eq:cmb_floor}
T_{\rm CMB} = T_{\rm CMB, 0} (1+z), 
\end{equation} where $T_{\rm CMB, 0}$ is the CMB temperature at $z=0$ ($\approx 2.725\,{\rm K}$). Therefore, to prevent the introduction of the artificial heating term, we switch off the cooling and force the cold gas to remain at the CMB temperature according to Equation \ref{eq:cmb_floor}. While insignificant at $z \geq 0$, in the far future, this unphysical heating term can cause significant heating to the gas in the IGM as other sources of heating and cooling become negligible. Collectively, the modifications discussed thus far affect the evolution of the IGM most significantly.

\subsection{Halo finding and analysis}\label{sec:postproc}

We identify haloes in the simulations with Robust Overdensity Calculation using the k-Space Topologically Adaptive Refinement code ({\tt ROCKSTAR}: \citejap{2013ApJ...762..109B}). It makes use of the six-dimensional information of the dark matter particles to identify and locate a halo. However, its capability to find haloes in simulations into the future has not previously been explored. In the future, the freeze out of the growth of large scale structures leads to a shut-down of merging processes, so that each halo evolves independently in isolation. This scenario means that the proper size of the halo remains constant, translating to a shrinking comoving size that poses a challenge to halo finding.

In order to test the capability of {\tt ROCKSTAR}, we created an idealised halo catalogue by placing isolated haloes on a uniform grid. They are sampled from the \cite{2002MNRAS.329...61S} halo mass function (HMF). Each halo is endowed with a truncated Navarro-White-Frenk (NFW) density profile \citep{1996ApJ...462..563N}, sampled randomly by particles, down to a mass limit of two particles with a mass resolution of $4.96\times10^9\msunoh$ at three different redshifts ($z=0, -0.5, -0.9$). This number is much lower than the minimum number of particles required by {\tt ROCKSTAR} to identify gravitationally bound haloes, so we thus expect a cut-off in the number of haloes recovered at low masses. Each redshift contains the same distribution of haloes. However, we shrink the comoving virial radii of the haloes to replicate the effect of the expanding universe explained in Section \ref{sec:postproc}. Refer to Appendix \ref{sec:rockstar_test} for the plots quantifying the capability of {\tt ROCKSTAR} to locate and identify haloes at and beyond $z=0$. We then input the positions and velocities of these dark matter particles into {\tt ROCKSTAR}, repeating the halo finding five times to quantify the variation between each run, as {\tt ROCKSTAR} has some explicitly non-deterministic features.

We also repeated the halo finding using the single and multiple processors modes of the code. For the latter, the simulation box is split in half and two processors are assigned to each computational domain to locate and identify haloes. This setup involves a total of seven processors with six of them finding haloes and one master processor. We consistently obtain an identical total number of haloes across all five runs with a single processor. However, we found two different numbers within the five repeats at each redshift with multiple processors. One of these numbers is consistent with the single processor setup while the other is always smaller. We repeated the exercise again at a different times but obtain results that are consistent across the different setups. This anomaly led us to believe it is a hardware related issue. Even though we cannot replicate the issue consistently, we use a single processor for halo finding in our simulations to prevent any potential error. We then carry out post-processing of the results from the simulations and halo finding with the {\tt yt} analysis toolkit \citep{2011ApJS..192....9T}. 

\section{Results and discussion} \label{sec:naga-results}

In the following sections, we will present and discuss the evolution of various properties of the gas and dark matter in the simulations. These include iterations of the changes described in Section \ref{sec:sim_setup} and comparisons to previous work. We will first focus on the {\sl NL} simulation before extending the study to six other simulations of varying spatial and mass resolution and feedback prescription. The specifications of these simulations can be found in Table \ref{tab:naga-setup}. Lastly, we will round up this section by comparing the results from the cosmological box simulations with an extension into the future of the zoom simulation described in \citet{BK20}.

\subsection{Evolution of the distribution of the gas temperature and overdensity}\label{sec:sim_evo}

\begin{figure*}
	\centering
	\subfloat[$z=0, t=t_0$ \label{fig:d1tH}]{%
		\includegraphics[width=.46\linewidth]{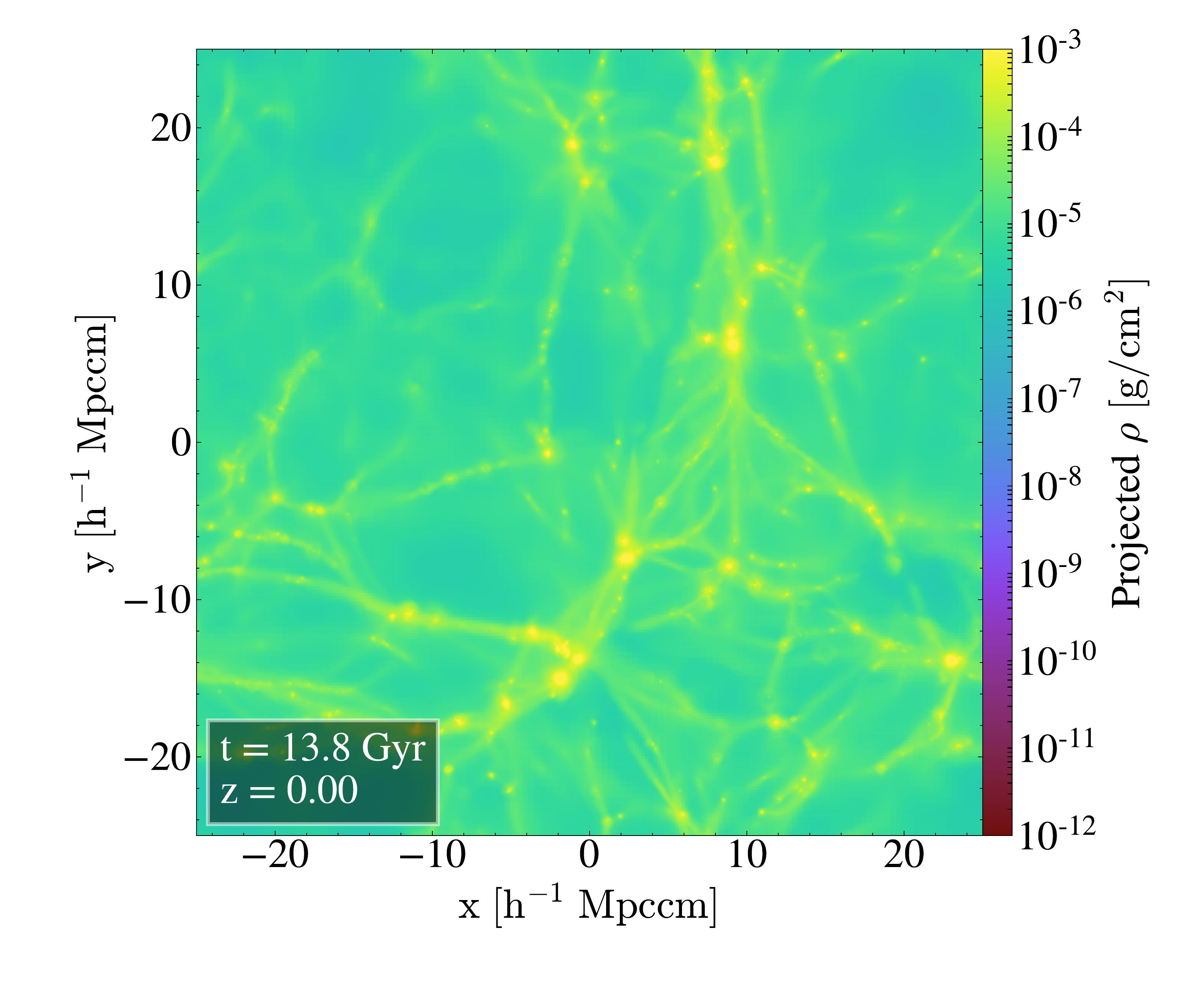}
	}
	\subfloat[$z=-0.59, t \approx t_0+t_\mathrm{H}$ \label{fig:d2tH}]{%
		\includegraphics[width=.46\linewidth]{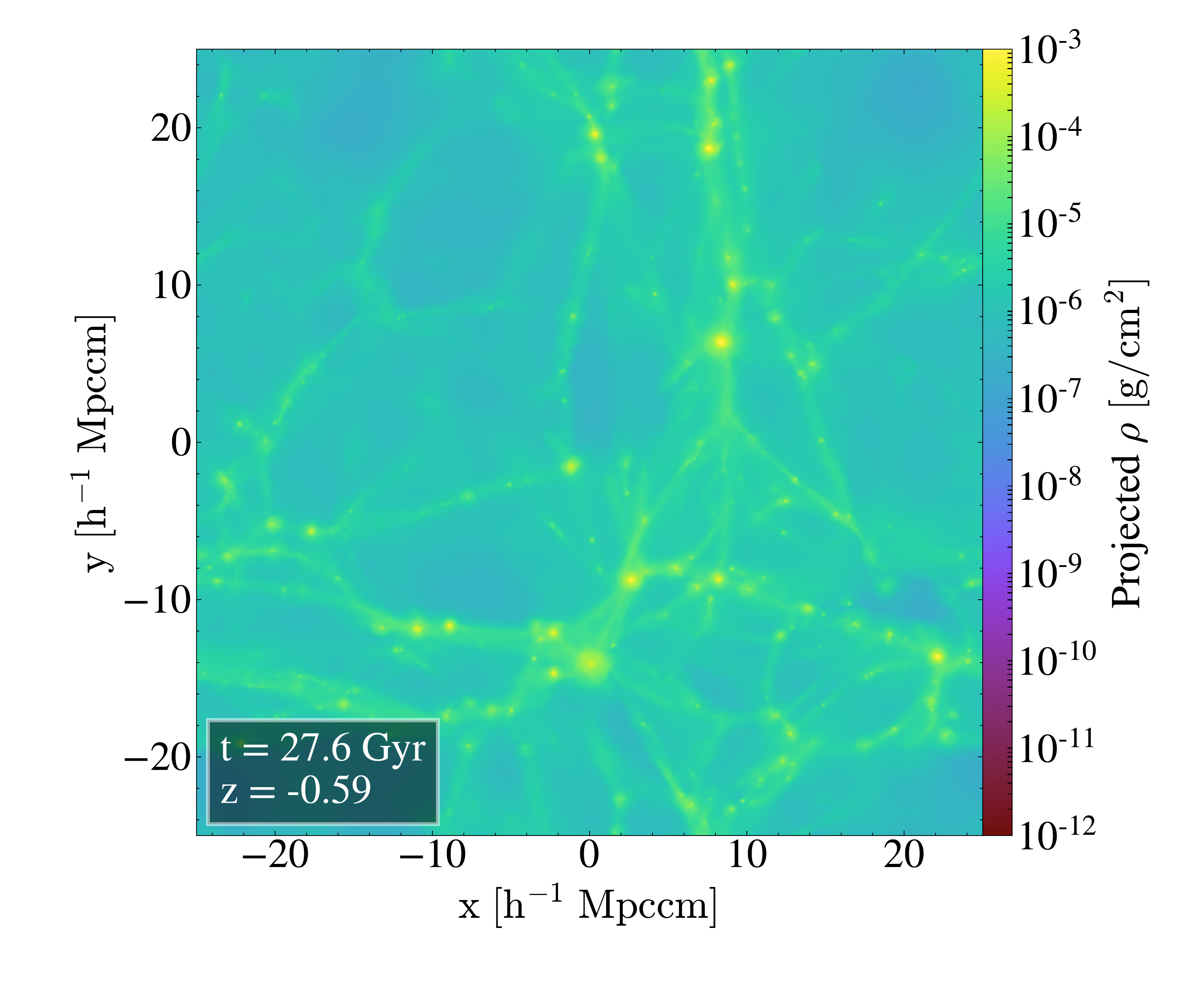}
	}
	\hfill
	\subfloat[$z=-0.82, t \approx t_0+2t_\mathrm{H}$ \label{fig:d3tH}]{%
		\includegraphics[width=.46\linewidth]{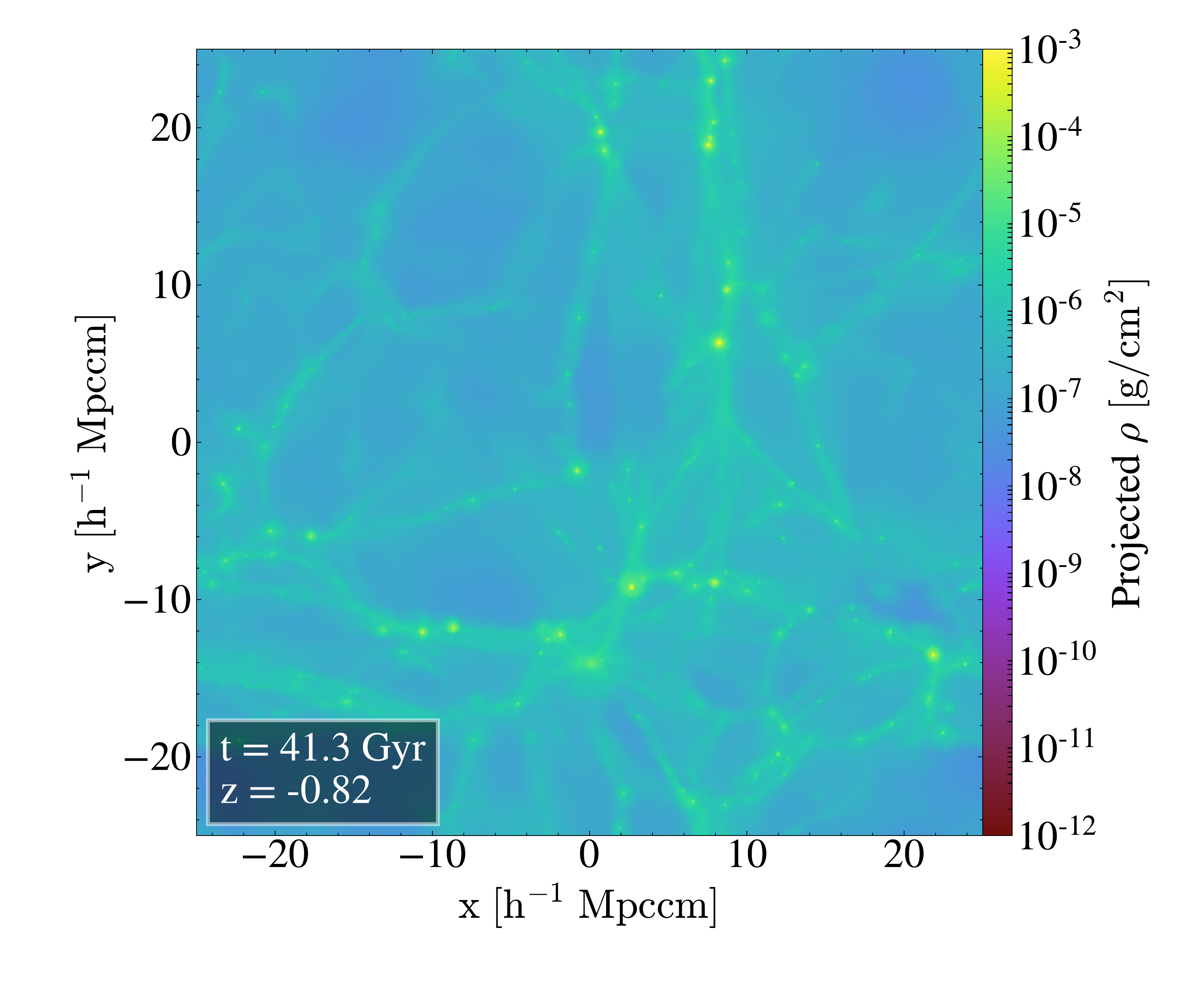}
	}
	\subfloat[$z=-0.92, t \approx t_0+3t_\mathrm{H}$ \label{fig:d5tH}]{%
		\includegraphics[width=.46\linewidth]{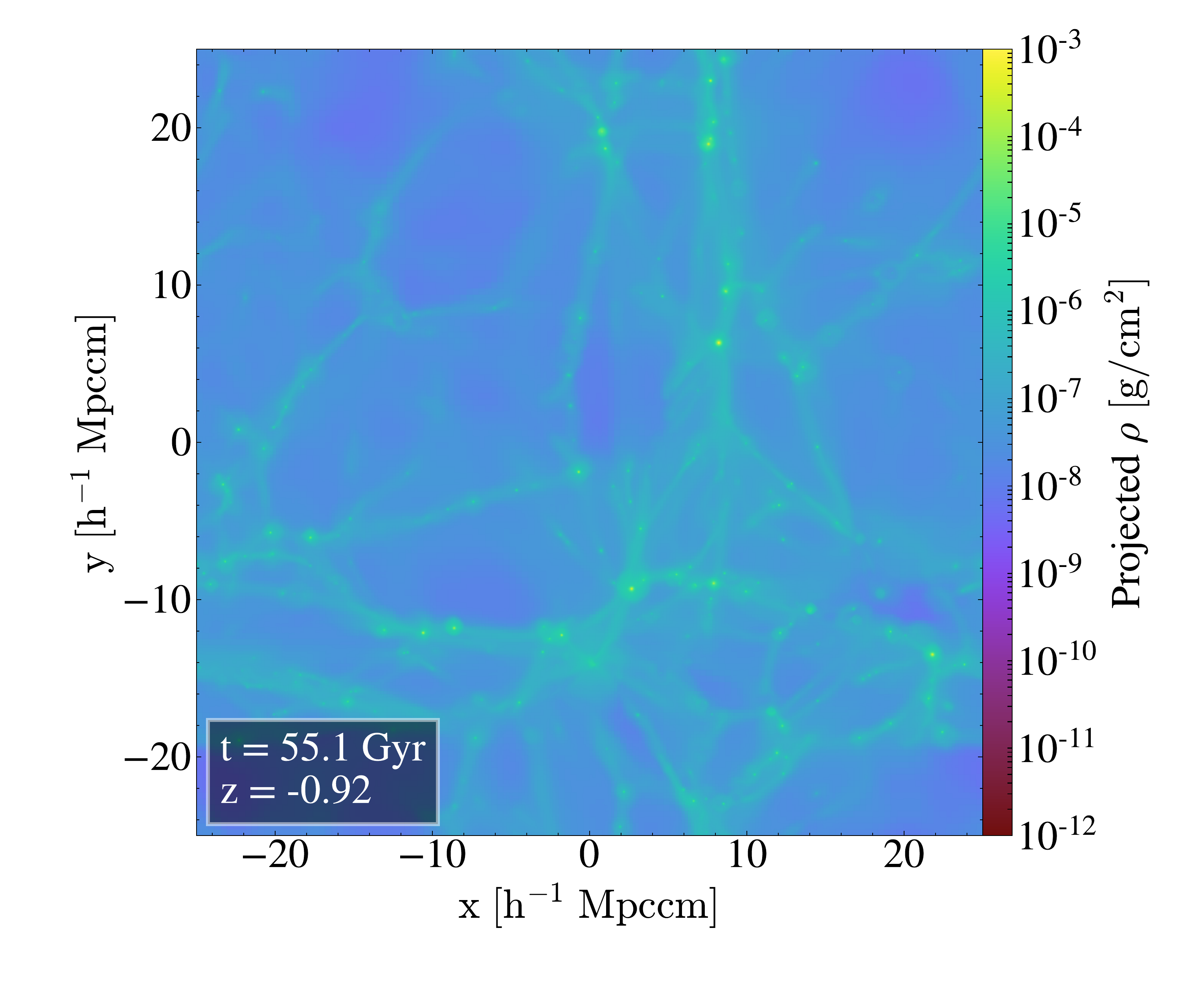}
	}
	\hfill
	\subfloat[$z=-0.98, t \approx t_0+5t_\mathrm{H}$ \label{fig:d7tH}]{%
		\includegraphics[width=.46\linewidth]{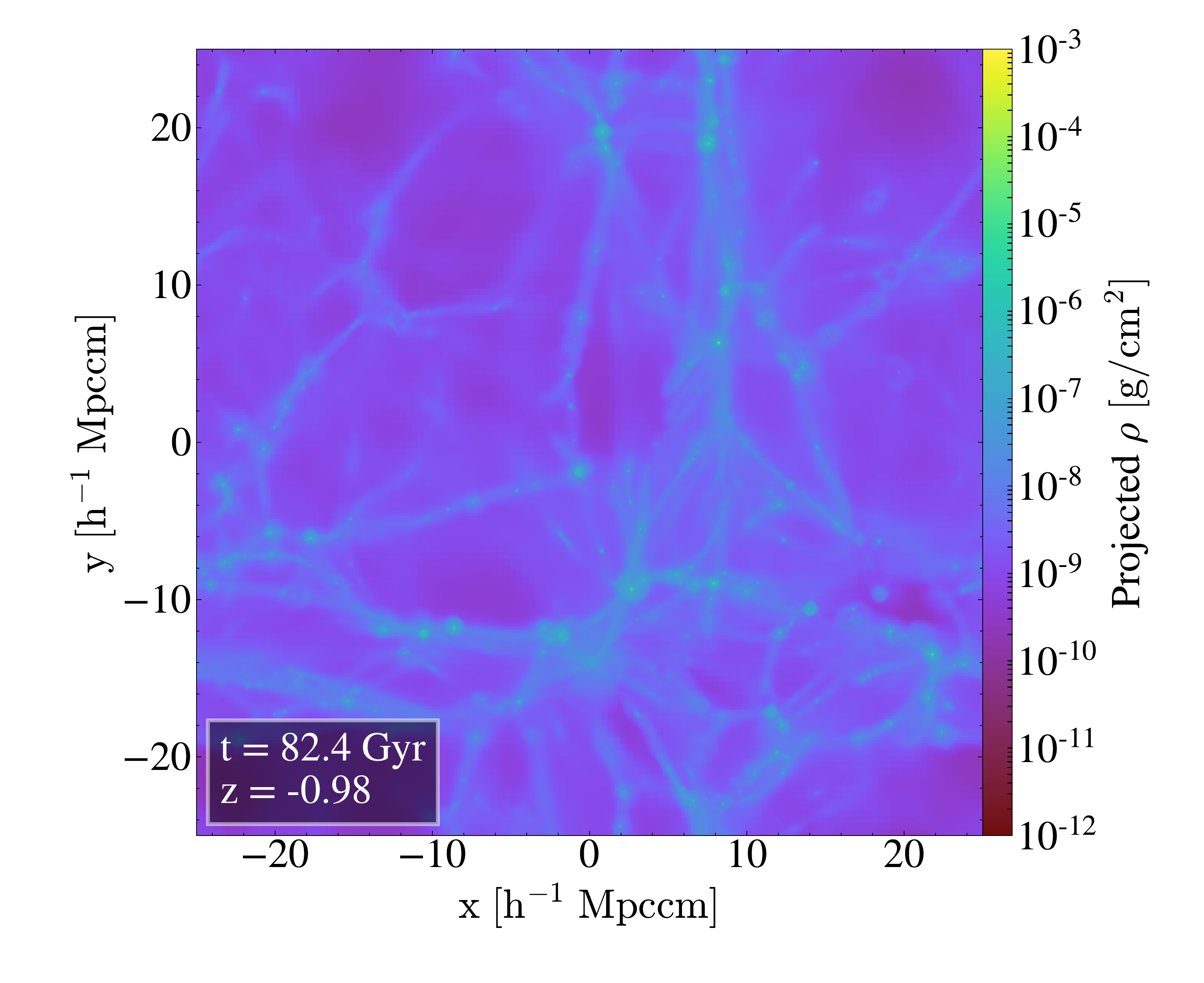}
	}
	\subfloat[$z=-0.99, t \approx t_0+6t_\mathrm{H}$ \label{fig:d8tH}]{%
		\includegraphics[width=.46\linewidth]{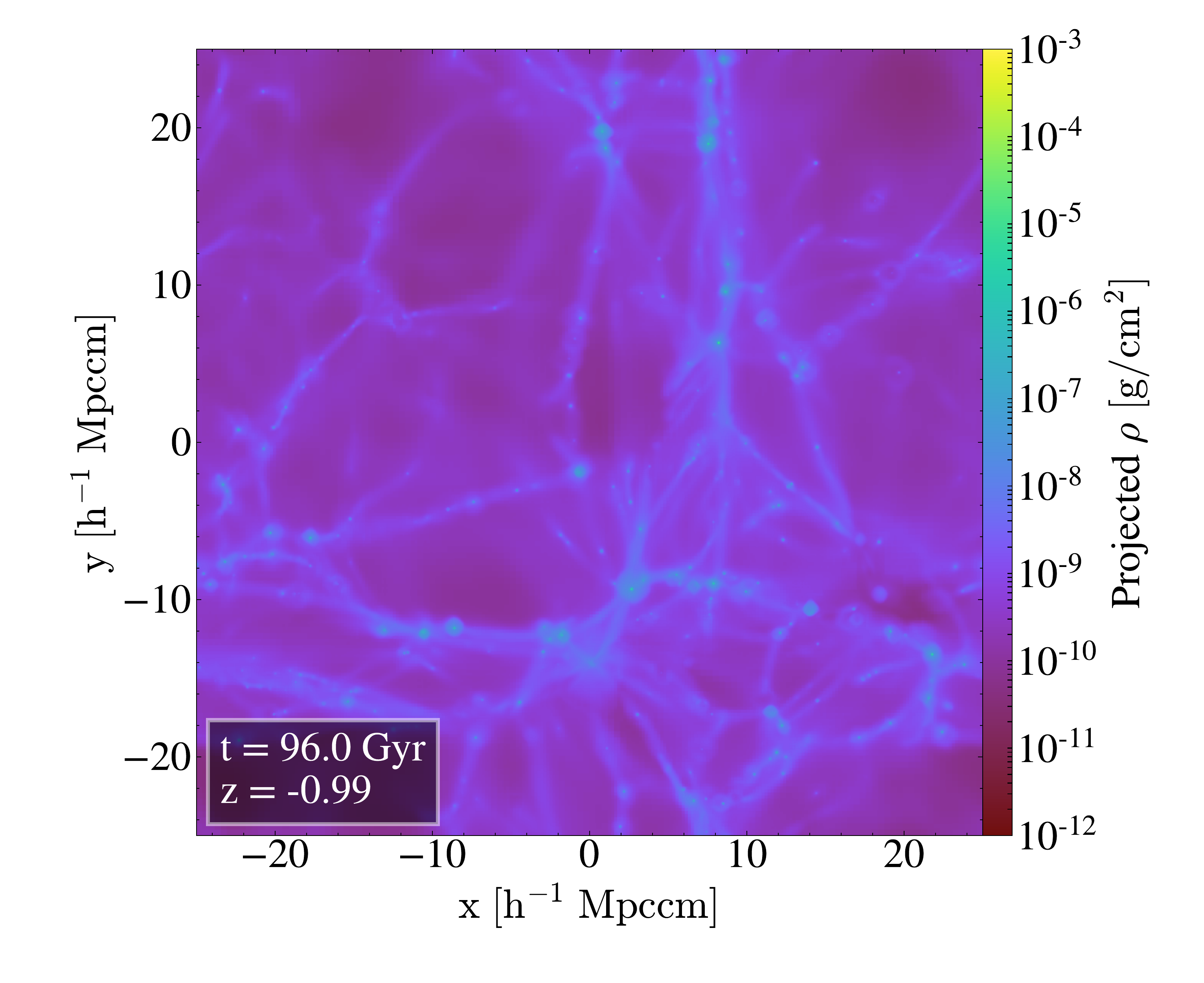}
	}
	\caption{Density projection plots of a slice with a comoving thickness of $10\mpcoh$ and a comoving width of $50\mpcoh$ at the $z$ and $t$ indicated in the captions. As the simulation evolves into the future, large scale structure growth freezes out as the universe becomes $\Lambda$-dominated. This phenomenon explains the high level of similarity of the plots, especially at late times. 
	}
	\label{fig:pproj}
\end{figure*}

We first present a plot of the gas density, projected in a slice with a comoving thickness of $10\mpcoh$ and a comoving width of $50\mpcoh$ (Figure \ref{fig:pproj}). Starting from panel (b), we notice a lack of large scale structure evolution, consistent with the prediction of freeze out within $2\,t_\mathrm{H}$ past $z=0$ \citep{2004NewA....9..573N, 2018MNRAS.477.3744S}. From this point of time, haloes evolve in isolation.

\begin{figure*}
	\centering
	\subfloat[$z=0, t=t_0$ \label{fig:T1tH}]{%
		\includegraphics[width=.46\linewidth]{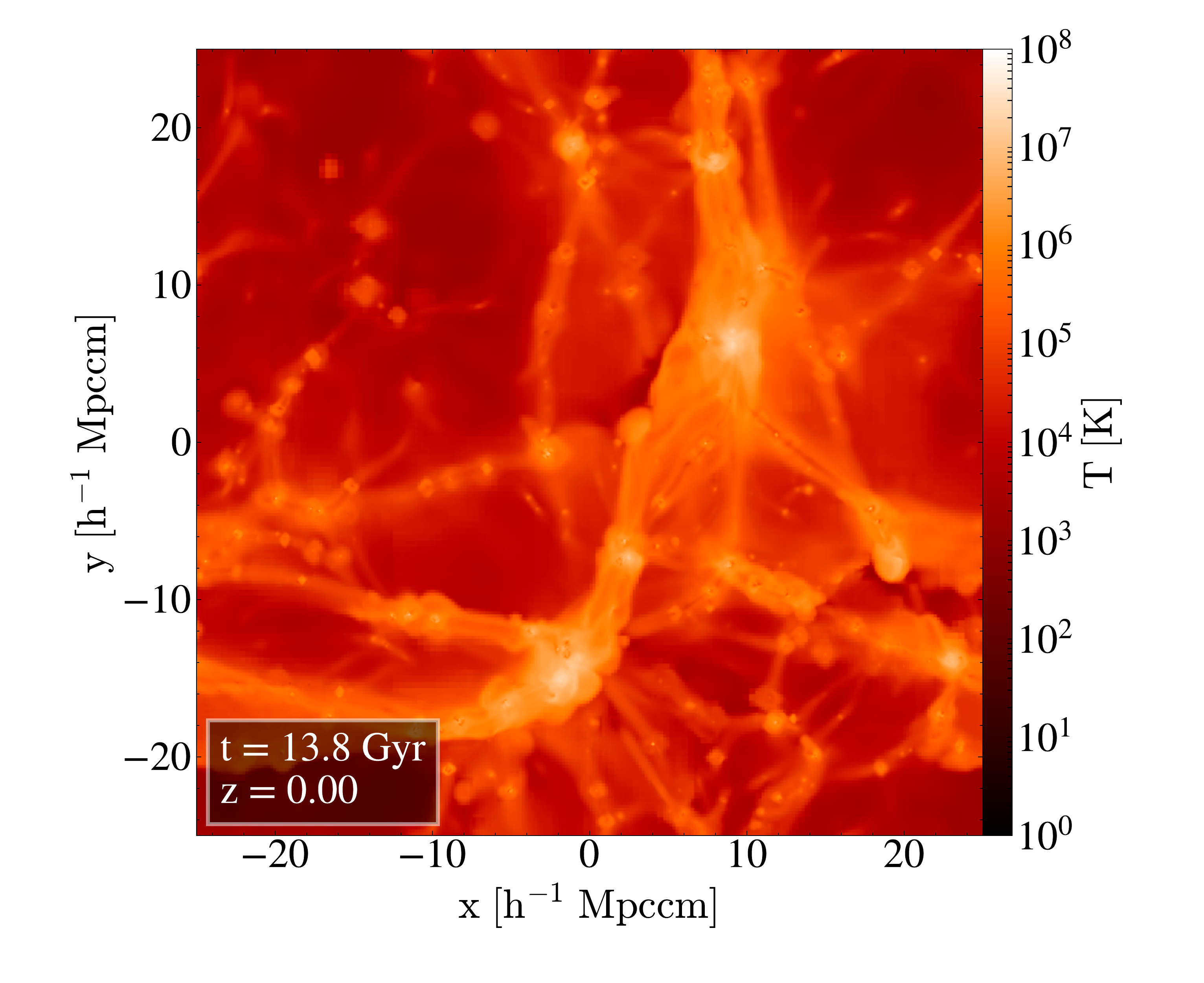}
	}
	\subfloat[$z=-0.59, t \approx t_0+t_\mathrm{H}$ \label{fig:T2tH}]{%
		\includegraphics[width=.46\linewidth]{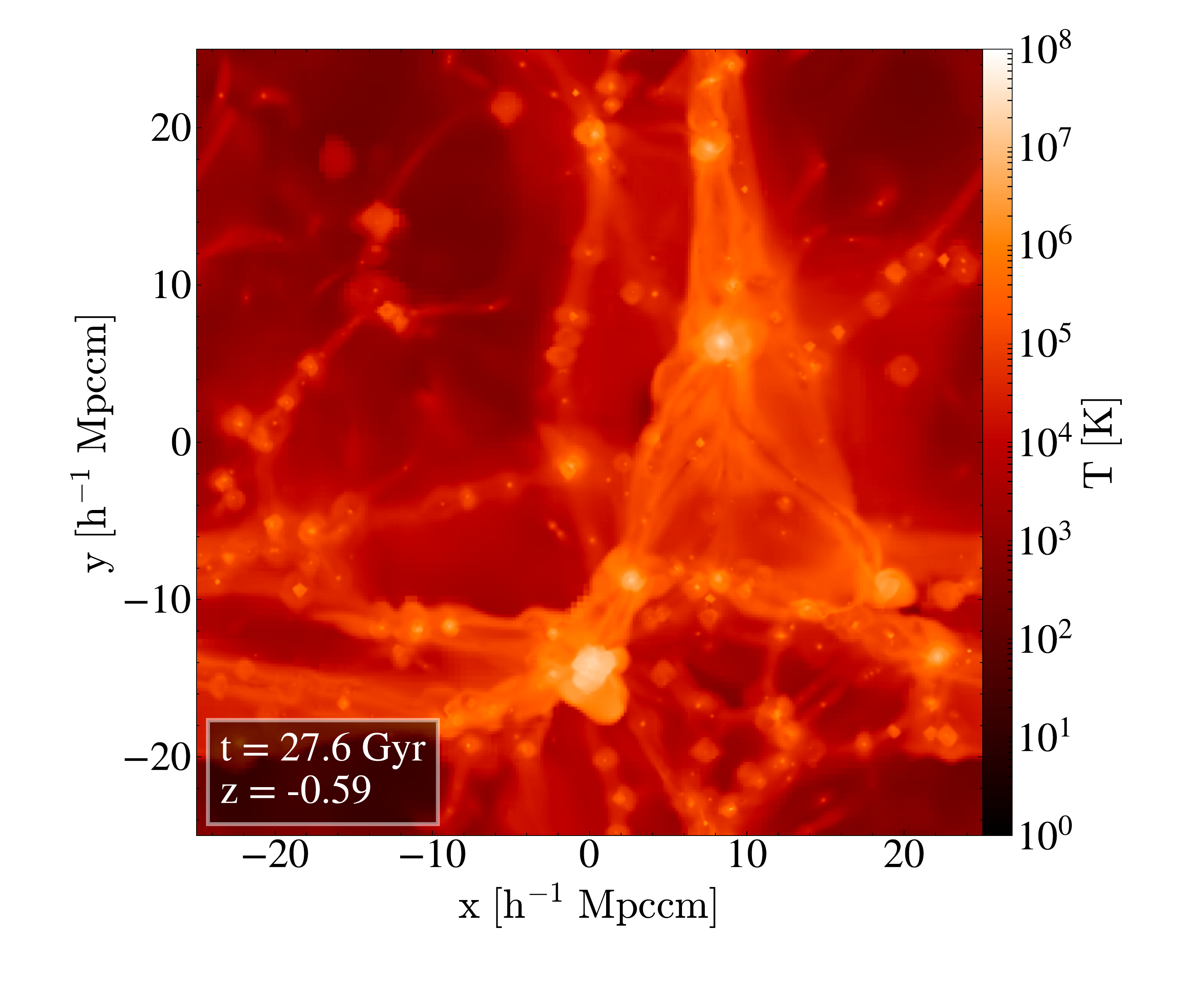}
	}
	\hfill
	\subfloat[$z=-0.82, t \approx t_0+2t_\mathrm{H}$ \label{fig:T3tH}]{%
		\includegraphics[width=.46\linewidth]{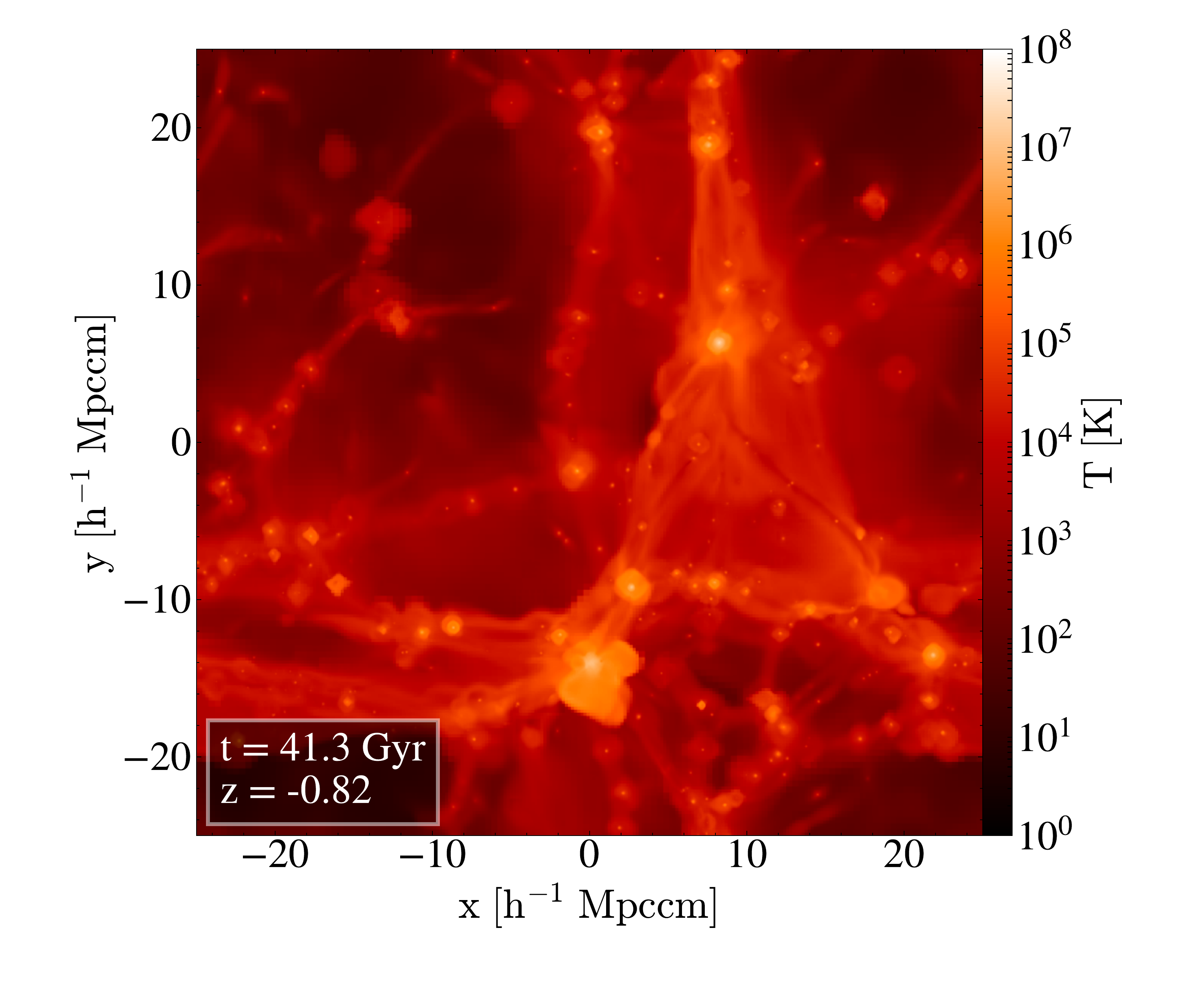}
	}
	\subfloat[$z=-0.92, t \approx t_0+3t_\mathrm{H}$ \label{fig:T4tH}]{%
		\includegraphics[width=.46\linewidth]{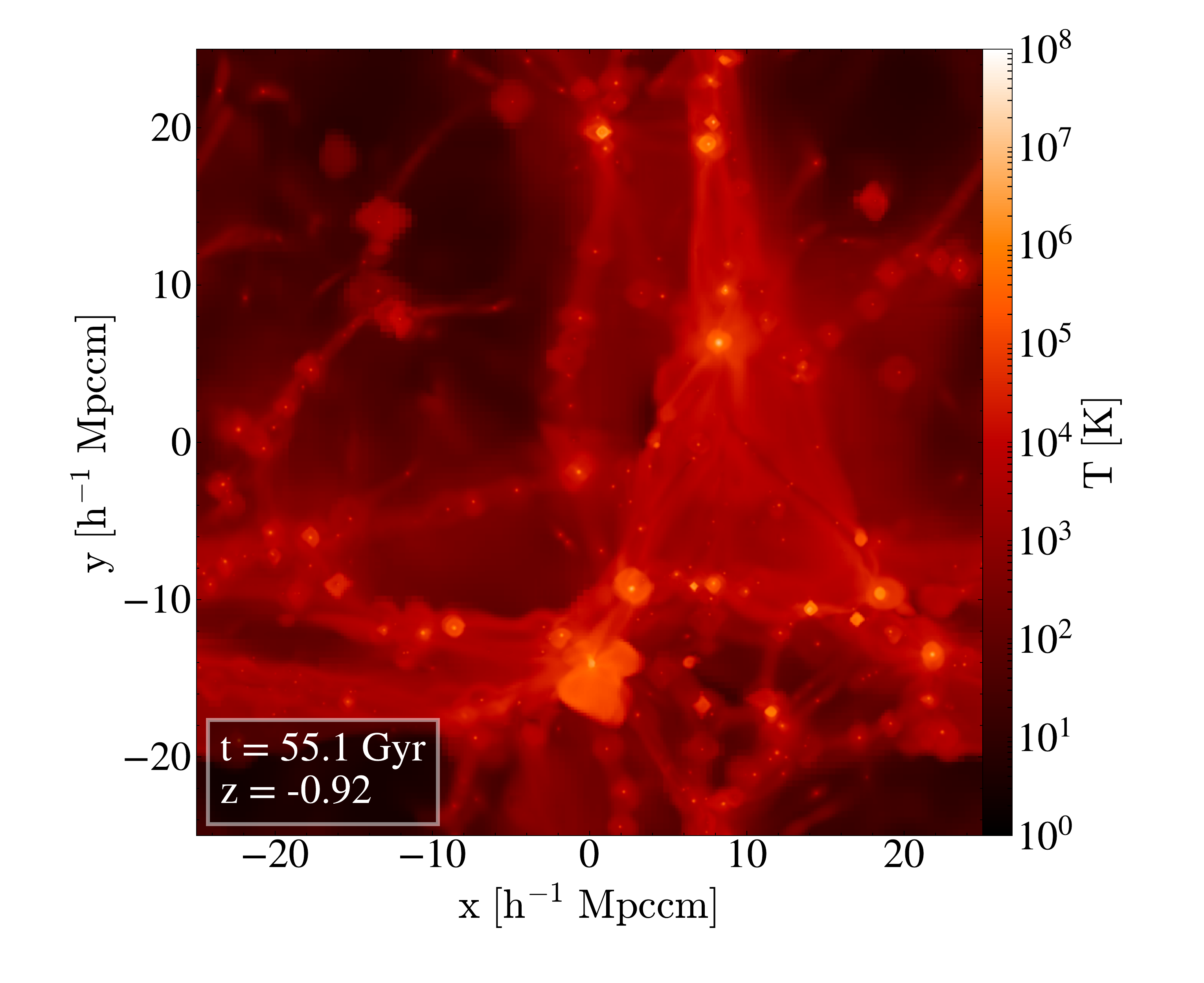}
	}
	\hfill
	\subfloat[$z=-0.98, t \approx t_0+5t_\mathrm{H}$ \label{fig:T6tH}]{%
		\includegraphics[width=.46\linewidth]{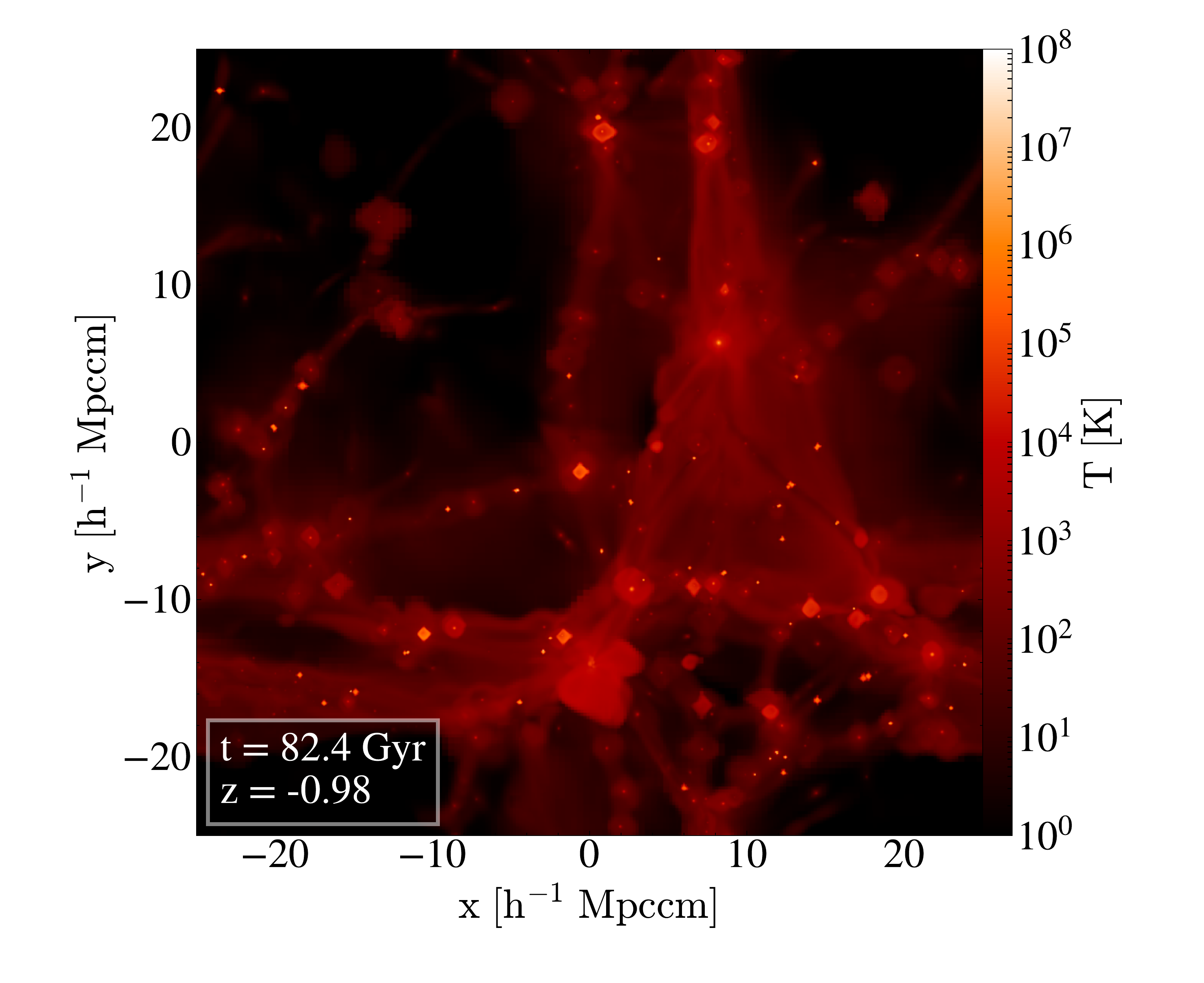}
	}
	\subfloat[$z=-0.99, t \approx t_0+6t_\mathrm{H}$ \label{fig:T7tH}]{%
		\includegraphics[width=.46\linewidth]{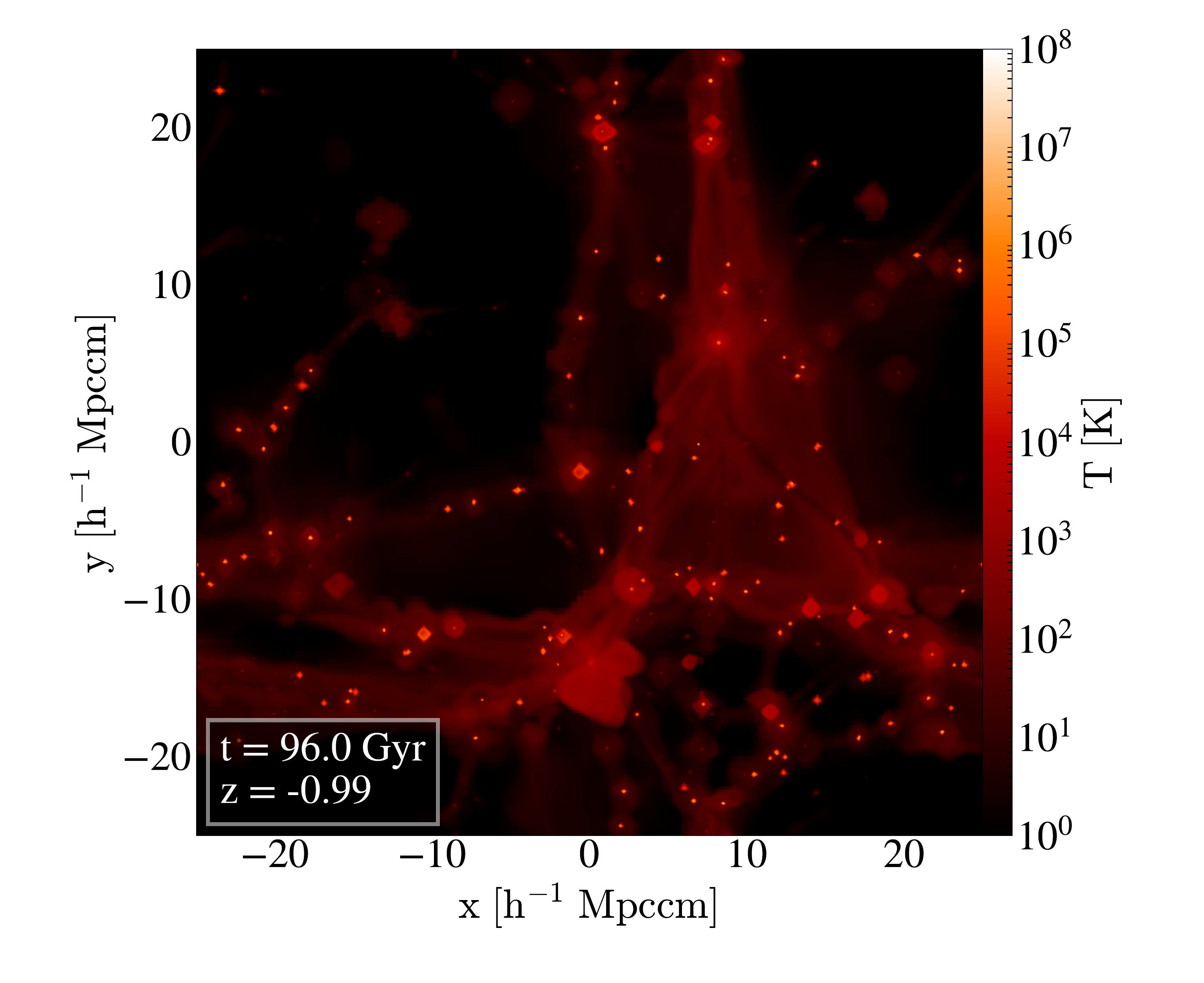}
	}
	\caption{Mass-weighted temperature projection plot of an identical slice as Figure \ref{fig:pproj}. As the simulation evolves into the future, the gas in the filaments cools and reaches an equilibrium with the background temperature. On the other hand, the haloes become hotter and their comoving sizes shrink to small dots in the plots.}
	\label{fig:tproj}
\end{figure*}

We then look at the mass-weighted temperature projection plot of an identical slice in Figure \ref{fig:tproj}. The filaments cool from $T\approx10^7\,\mathrm{K}$ at $z=0$ to $T\approx10^4\,\mathrm{K}$ at $z=-0.92$ before reaching the same temperature as most of the gas at $z=-0.99$. This temperature drop is most likely due to the adiabatic cooling from the expansion of the universe. The virialised dark matter haloes with $T > 10^4\,\mathrm{K}$ become increasingly isolated within the cold IGM. They are represented by small, brightly coloured dots, particularly in panels (e) and (f). These dots are a physical representation of their shrinking comoving size as the universe evolves into the future.

\begin{figure*}
	\centering
	\subfloat[$z=0, t=t_0$ \label{fig:pp1tH}]{%
		\includegraphics[width=.42\linewidth]{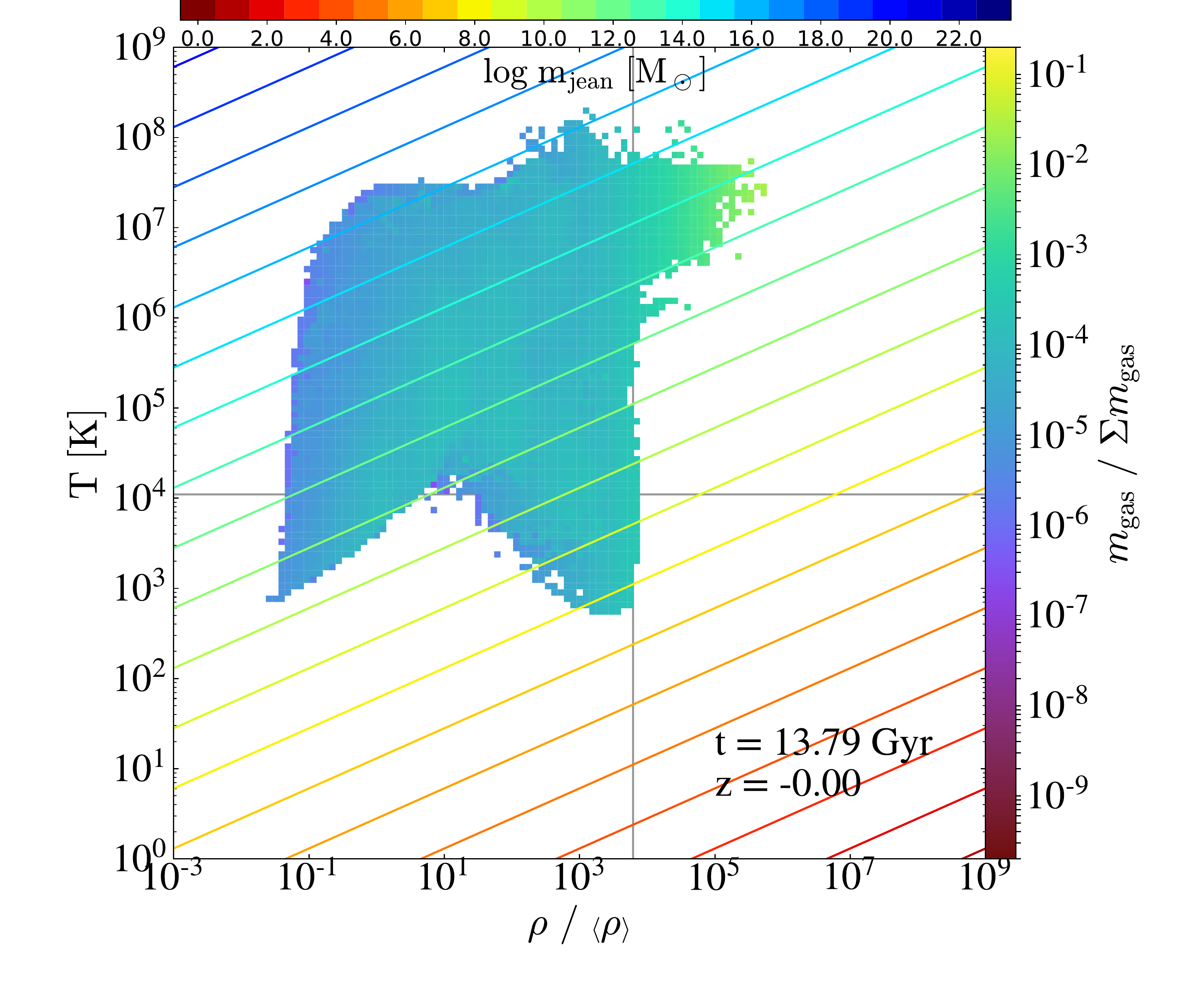}
	}
	\subfloat[$z=-0.59, t \approx t_0+t_\mathrm{H}$ \label{fig:pp2tH}]{%
		\includegraphics[width=.42\linewidth]{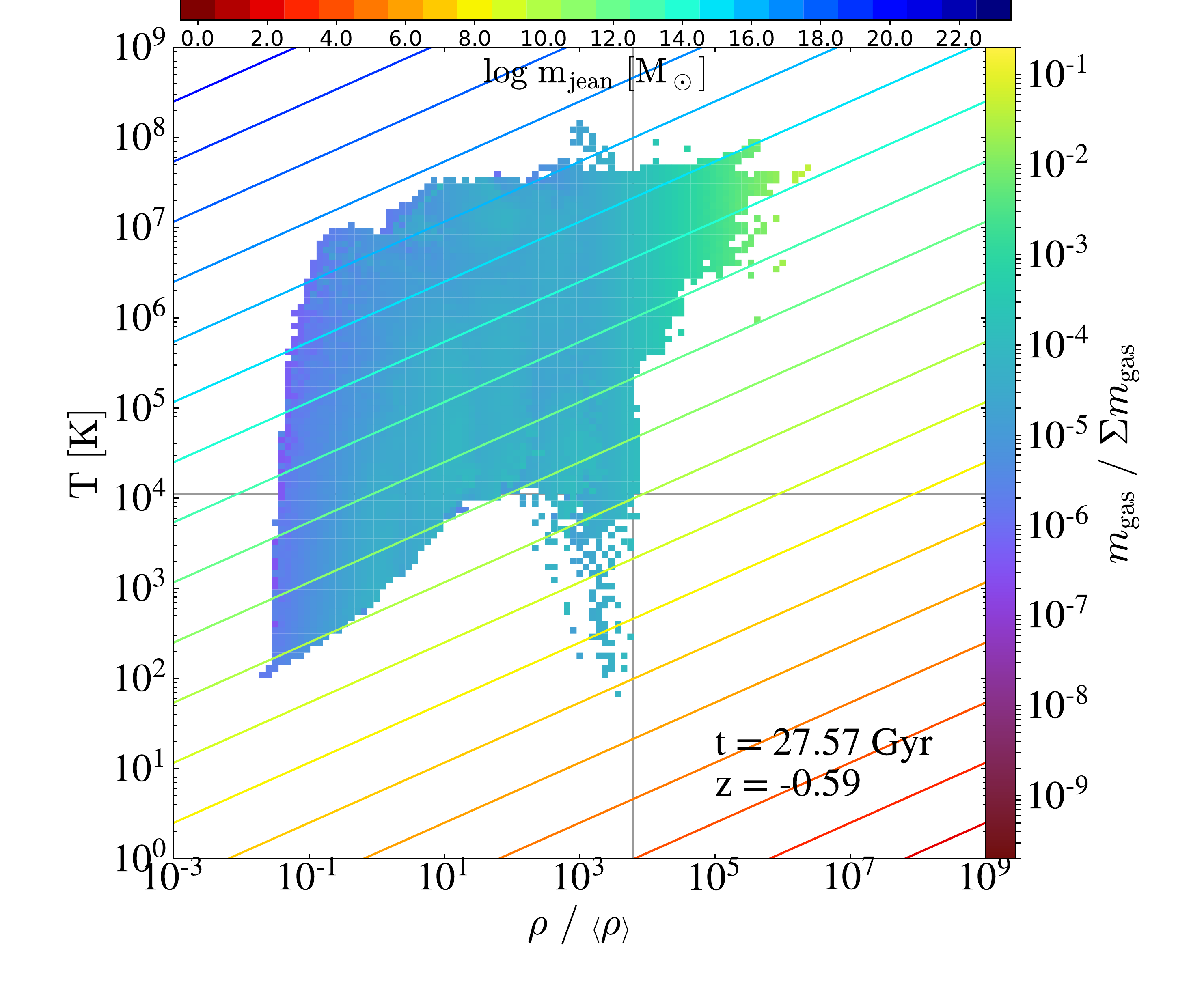}
	}
	\hfill
	\subfloat[$z=-0.82, t \approx t_0+2t_\mathrm{H}$ \label{fig:pp3tH}]{%
		\includegraphics[width=.42\linewidth]{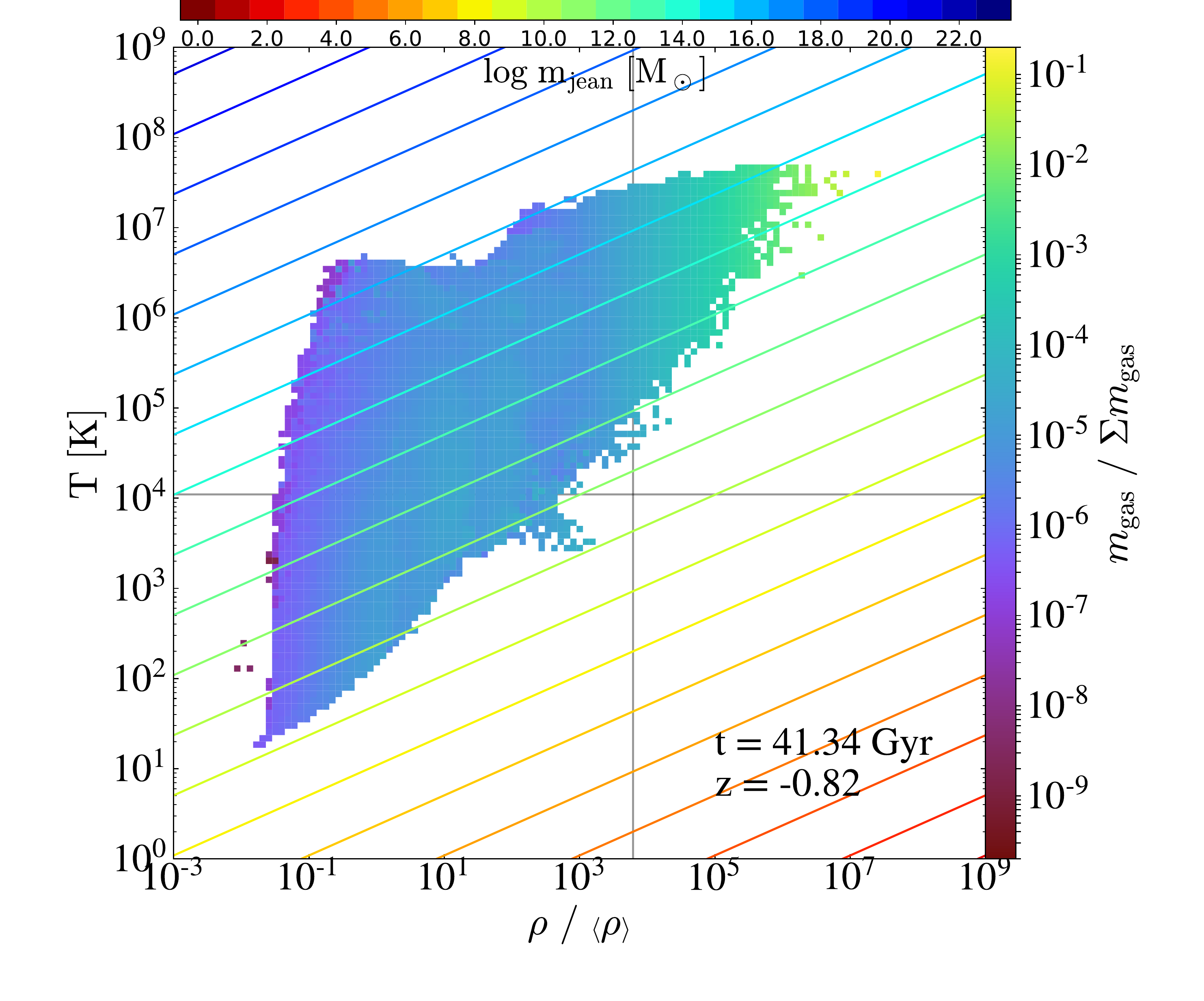}
	}
	\subfloat[$z=-0.92, t \approx t_0+3t_\mathrm{H}$ \label{fig:pp4tH}]{%
		\includegraphics[width=.42\linewidth]{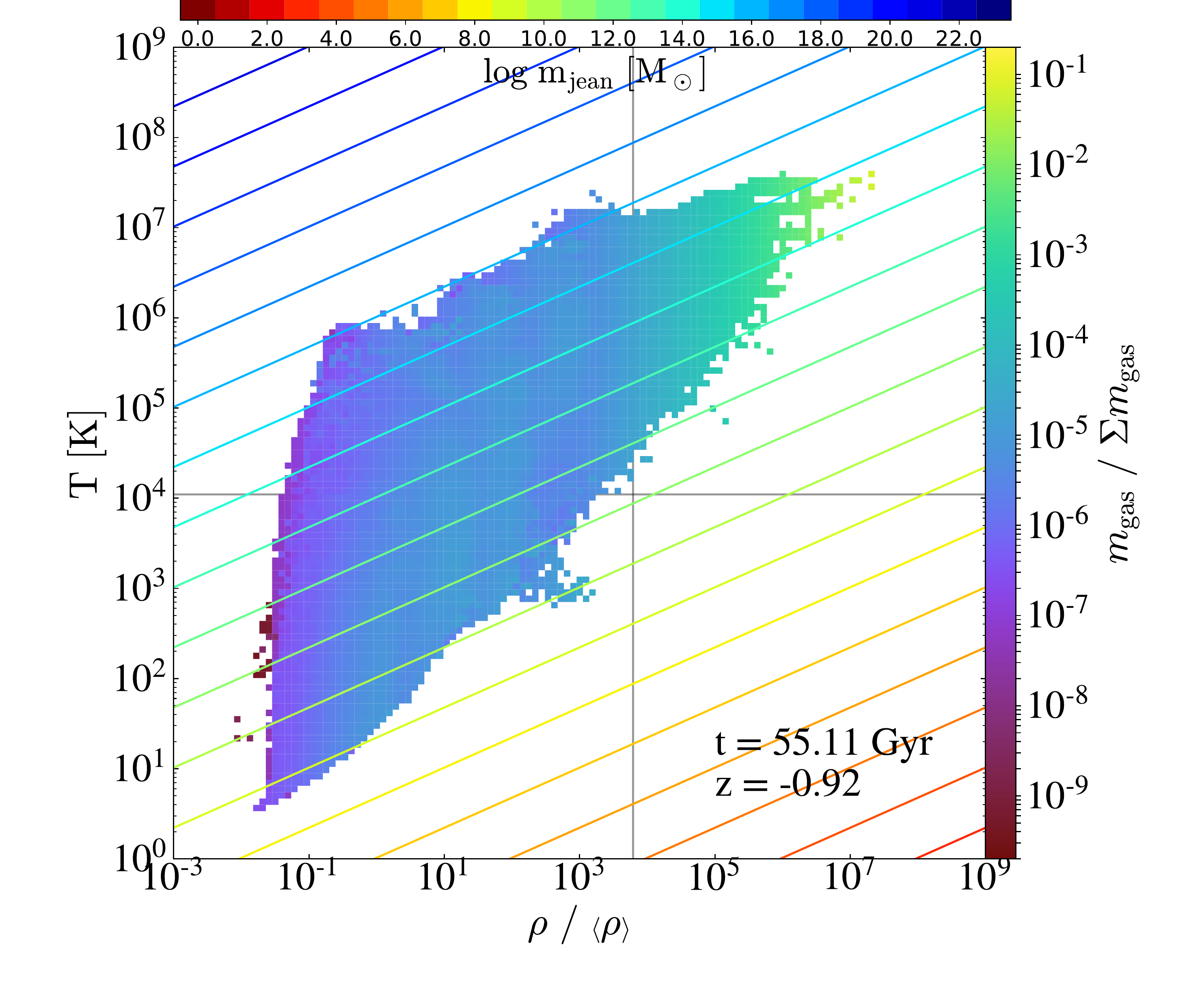}
	}
	\hfill
	\subfloat[$z=-0.98, t \approx t_0+5t_\mathrm{H}$ \label{fig:pp6tH}]{%
		\includegraphics[width=.42\linewidth]{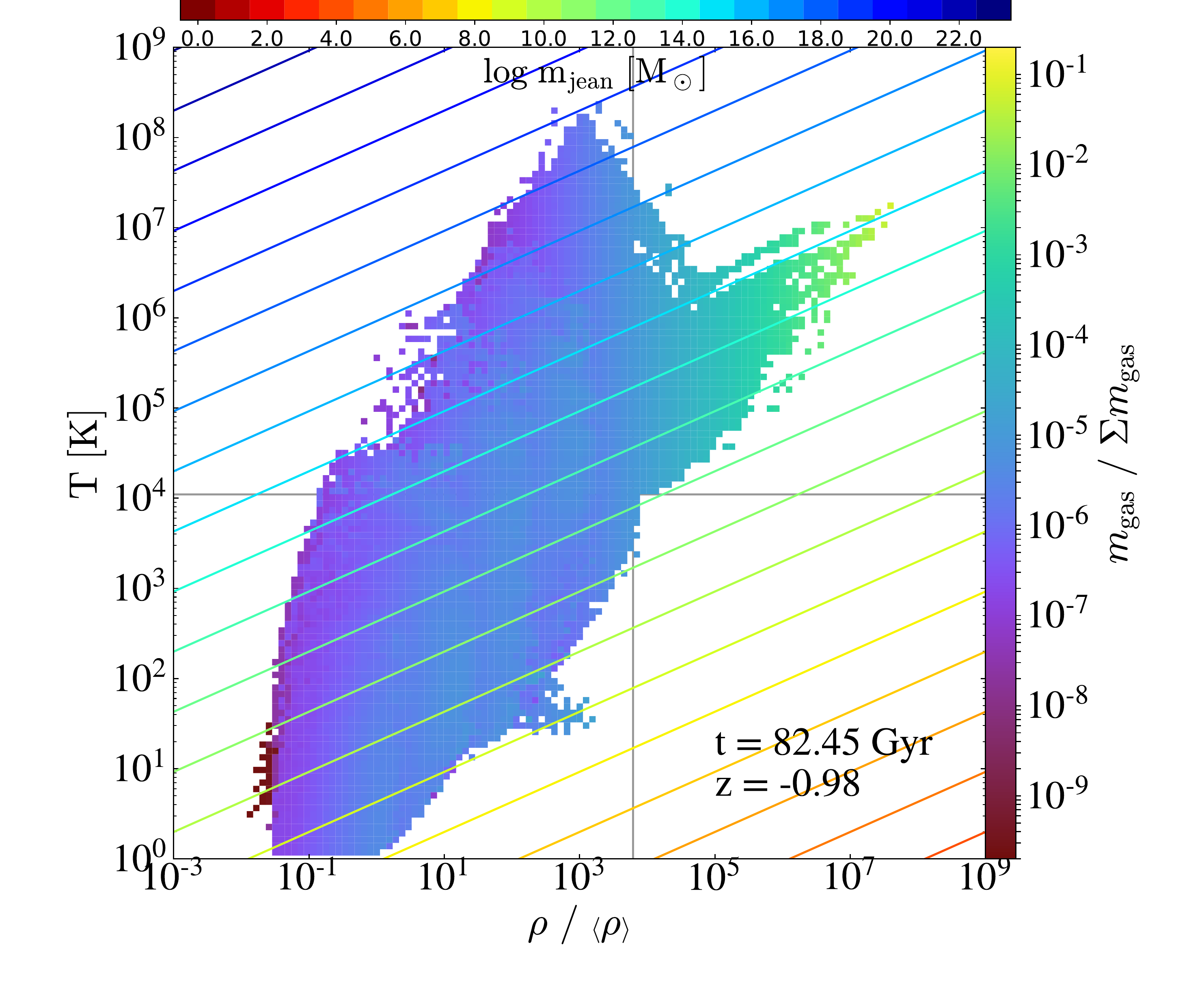}
	}
	\subfloat[$z=-0.99, t \approx t_0+6t_\mathrm{H}$ \label{fig:pp7tH}]{%
		\includegraphics[width=.42\linewidth]{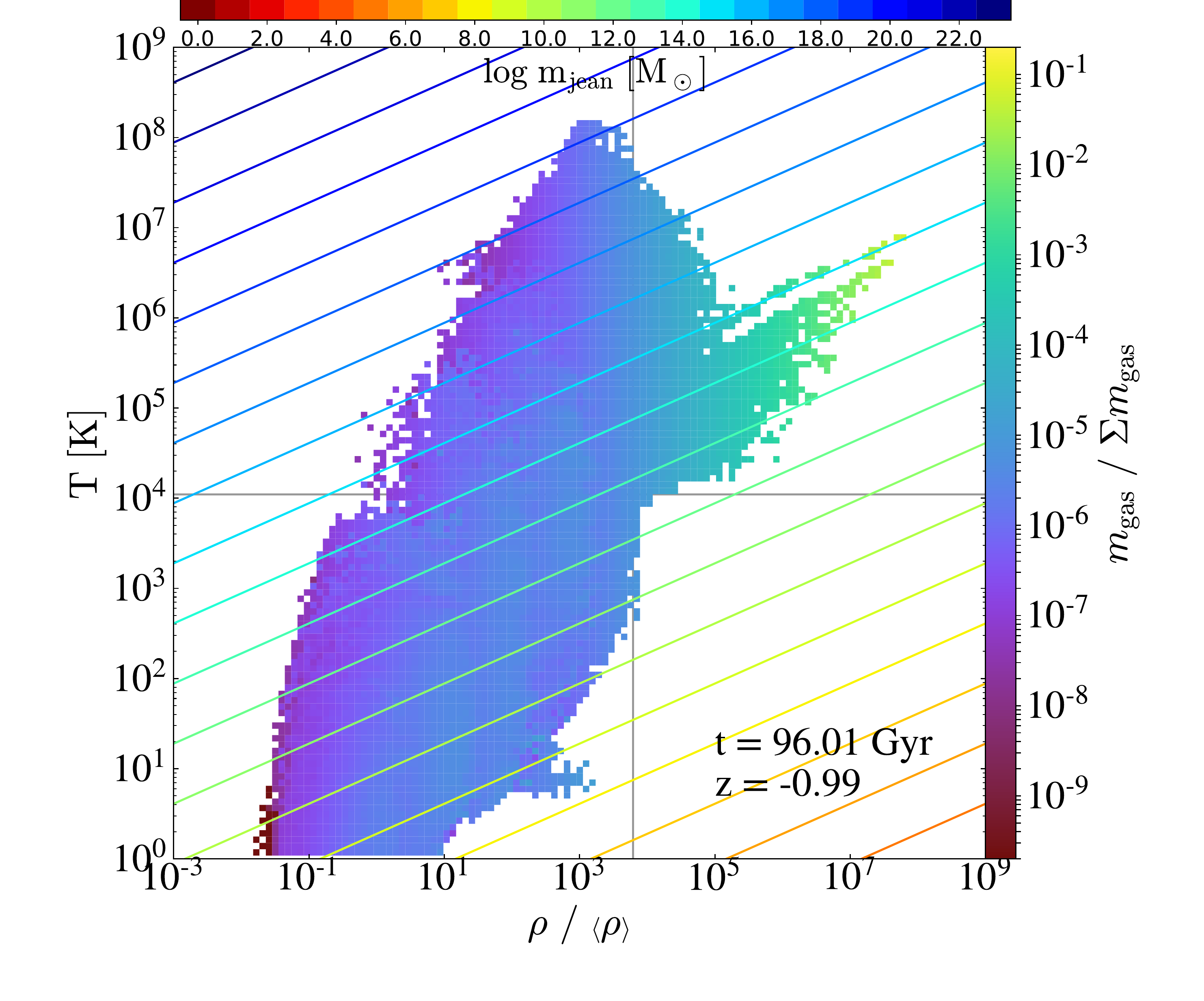}
	}
	\caption{Gas mass in bins of overdensity and temperature of the gas at the $z$ and $t$ indicated in the captions. We have included diagonal lines of constant Jeans mass of the gas according to its temperature and baryon overdensity. The colour of these lines corresponds to a value indicated by the colour bar at the top of the figure. The horizontal and vertical lines are the specified threshold baryon overdensity and temperature required for star formation in the simulations. In other words, gas in the bottom right region can potentially form stars if all other criteria are fulfilled. Refer to Section \ref{sec:sim_evo} for a detailed description.}
	\label{fig:sim_phaseplot}
\end{figure*}

Lastly, we look at the combined evolution of these gas properties with the phase distribution in Figure \ref{fig:sim_phaseplot}. Again, the panels concern the same time snapshots as in Figures \ref{fig:pproj} and \ref{fig:tproj}. We can divide the phase distribution into four quadrants using lines of threshold density and temperature for star formation in the simulation. The bottom left quadrant contains gas of low overdensity and low temperature, which constitutes the IGM. We will investigate its evolution in greater detail in Section \ref{sec:igm_evo}. Gas that has cooled radiatively inside dark matter haloes occupies the bottom right quadrant. However, there is an absence of gas in this region because it has been converted to stars according to the star formation criteria specified in the simulation. Lastly, we combine the upper left and right quadrants and classify gas in these regions as `hot gas'. It consists of the warm-hot intergalactic medium \citep{1999ApJ...514....1C} and hot dense gas in massive haloes. This dissection of the phase space is consistent with the definitions used in \citet{2001ApJ...552..473D}.

As the universe evolves into the future, the phase distribution elongates diagonally. The extremely long bremsstrahlung-dominated cooling time of the hot gas with $\mathrm{T = 10^{7.5}\,K}$ and overdensity of 200 at $z=0$ is the cause of this elongation.  Note that gas at a constant overdensity corresponds to a continuously declining $\rho$ because $\left<\rho\right>$ decreases into the future. The feedback from galaxies thus affects gas of decreasing density as the simulation evolves: for a given amount of thermal energy, the resulting temperature will be higher as the density declines, resulting in a second temperature peak ($\mathrm{T\approx10^8\,K}$) in gas of intermediate overdensity around $10^3$. At $z=-0.98$, adiabatic cooling due to the expansion of the universe begins to cause some gas to hit the temperature floor given by Equation \ref{eq:cmb_floor}. Gas of higher density is able to reach the CMB temperature, affecting the equation of state of the IGM, which we will discuss in Section \ref{sec:igm_evo}. 

Furthermore, in the future, the Jeans mass of the gas starts to become very large, making gravitational collapse difficult. As the halo mass function freezes out, the gravitational potential wells only evolves internally, restricting the inflow of gas. Therefore, the gas remains hot and unable to collapse to high densities.

\subsubsection*{Comparison to \citet{2004NewA....9..573N}} \label{sec:naga-compare}

\citet{2004NewA....9..573N} performed an analysis similar to the one in Section \ref{sec:sim_evo} with a SPH simulation, which we have attempted to match in resolution. This similarity means that differences between the simulation results must lie in the methodology of the simulation code and implementation of the baryonic processes. Taking these factors into account, we will compare the evolution of the phase distribution in our Figure \ref{fig:sim_phaseplot} to Figure 3 of \citet{2004NewA....9..573N}.

There is a similar elongation in phase space occupied by the gas into the future in both simulations. We have discussed the reasons for this evolution in Section \ref{sec:sim_evo}. We also observe a similar peak in gas mass at high overdensities ($>10^5$) and temperature ($\mathrm{>10^6\,K}$) in both figures. But despite this general agreement, there are specific differences present in the figures, which can be attributed to the different star formation and feedback prescriptions.  An island of gas in haloes with $T\approx10^4\,{\rm K}$ and $\log_{10} \rho/\left<\rho\right> > 6$ in Figure 3 of \citet{2004NewA....9..573N} is absent in Figure \ref{fig:sim_phaseplot}. This disparity reflects the difference in star formation criteria. In our simulation, gas is converted into stars when $\log_{10} \rho/\left<\rho\right> > 4$, indicated by the vertical line in each panel of Figure \ref{fig:sim_phaseplot}. This threshold is lower than the comoving baryon overdensity of $7.7\e{5}$ at $z=0$ used by \citet{2004NewA....9..573N}. Therefore, all the gas supposedly inhabiting this island in phase space is turned into stars in our simulation. 

The details of feedback prescription also differs between the simulations despite both using thermal feedback. The amount of feedback energy injected per solar mass of stars in our simulation is approximately an order of magnitude higher than $\mathrm{4\e{48}\,{\rm erg}\,M_\odot^{-1}}$ in \citet{2004NewA....9..573N}. Gas in the haloes of our simulation is thus heated to a higher temperature and is likely to propagate further into the IGM. The overdensity corresponding to gas in haloes in our simulation is then lower because the feedback is stronger and pushes the gas away from the centre of the haloes. 

Although there are differences in the specific details of the evolution of the phase distribution, the general trend agrees between both simulations. \citet{2004NewA....9..573N} claimed that the results obtained are dominated by gravitationally-induced shocks and insensitive to the exact UV background, star formation and feedback model. The comparison we have made thus far supports this conclusion to a certain extent, depending on the scale of interest. We now expand these results by analysing other aspects that were not explored by \citet{2004NewA....9..573N}. These include the halo mass function (HMF), equation of state of the IGM, star formation rate density (SFRD) and in particular the resolution convergence of these properties.       

\subsection{The future of the halo mass function} \label{sec:hmf_evo}

\begin{figure}
	\centering
	\includegraphics[width=\linewidth]{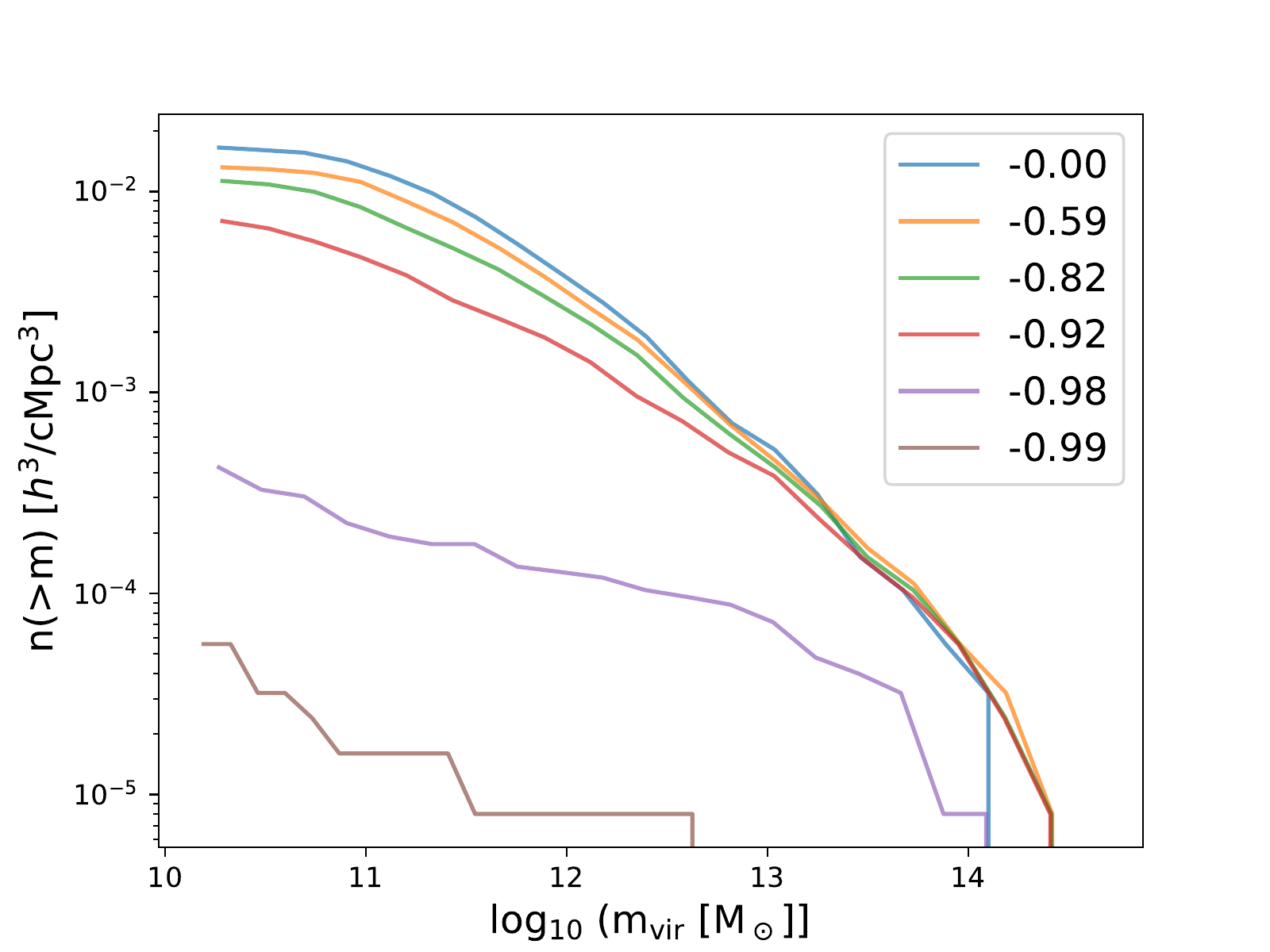}
	\caption{Evolution of the HMF with $z$ into the future. The lines are coloured according to their respective $z$ shown in the legend. The time interval between the lines is approximately one Hubble time, $t_H\approx13.7\,{\rm Gyr}$. We observe a freeze-out in the evolution at $z\approx-0.6$ when the high mass end remains constant. We discuss how the spatial resolution due to the expansion of the universe affects the later evolution of the HMF in Section \ref{sec:hmf_evo}.}
	\label{fig:hmf_evo}
\end{figure}

We have investigated the ability of {\tt ROCKSTAR} to locate haloes in an ideal environment where they are isolated. In this section, we wish to extend this study to a realistic cosmological simulation. We illustrate this evolution out to $z=-0.99$ in Figure \ref{fig:hmf_evo}. From $z=0$ to $z=-0.92$, we believe that {\tt ROCKSTAR} is locating and identifying haloes reliably over the entire halo mass range. The previously discussed period of freeze out occurs within this period at $z\approx-0.6$. As a consequence, the HMF in the figure displays a lack of significant evolution from $z=-0.59$ to $z=-0.92$ between $8\e{13}\,M_\odot \leq m_{\rm vir} \leq 5\e{14}\,M_\odot$, giving us confidence in the results within this range. But we find an increasing deviation at the low mass end of the HMF due to the worsening of proper force resolution with time, leading to low mass haloes becoming comparable in radius to the grid cell size.

At $z=-0.98$, there is a drastic drop in the number of haloes across the whole mass range. At the same time, the most massive halo found in the simulation is only $\sim2\e{14}\,M_\odot$. This trend continues to $z=-0.99$ where the effect is so significant that it causes orders of magnitude difference over the full mass range. We view this as a propagation of the poor force resolution from the low to high mass haloes. The most massive halo at each redshift is found consistently, in close proximity to its previously known location -- but its assigned mass declines with time, starting at $z=-0.92$. This decrease in mass is attributed to the loss of particles at the boundary of the halo due to the coupling of the accelerating expansion rate and the deteriorating force resolution. In a related change, the virial radius of the most massive halo is approximately an order of magnitude smaller at $z=-0.99$ ($2.964\,h^{-1}\,{\rm ckpc}$) than at $z=-0.98$ ($21.826\,h^{-1}\,{\rm ckpc}$).

We have thus shown that caution should be taken when examining the state of the simulation in the distant future. Despite the overall reasonable appearance of the large-scale structure, the simulation loses its ability to resolve haloes properly. The mock halo catalogue does not capture this effect because the haloes were generated in an ideal fashion, without force resolution effects. 

\subsection{The future of the intergalactic medium}\label{sec:igm_evo}

\begin{figure*}
	\centering
	\subfloat[$z=0, t=t_0$ \label{fig:igm1tH}]{%
		\includegraphics[width=.44\linewidth]{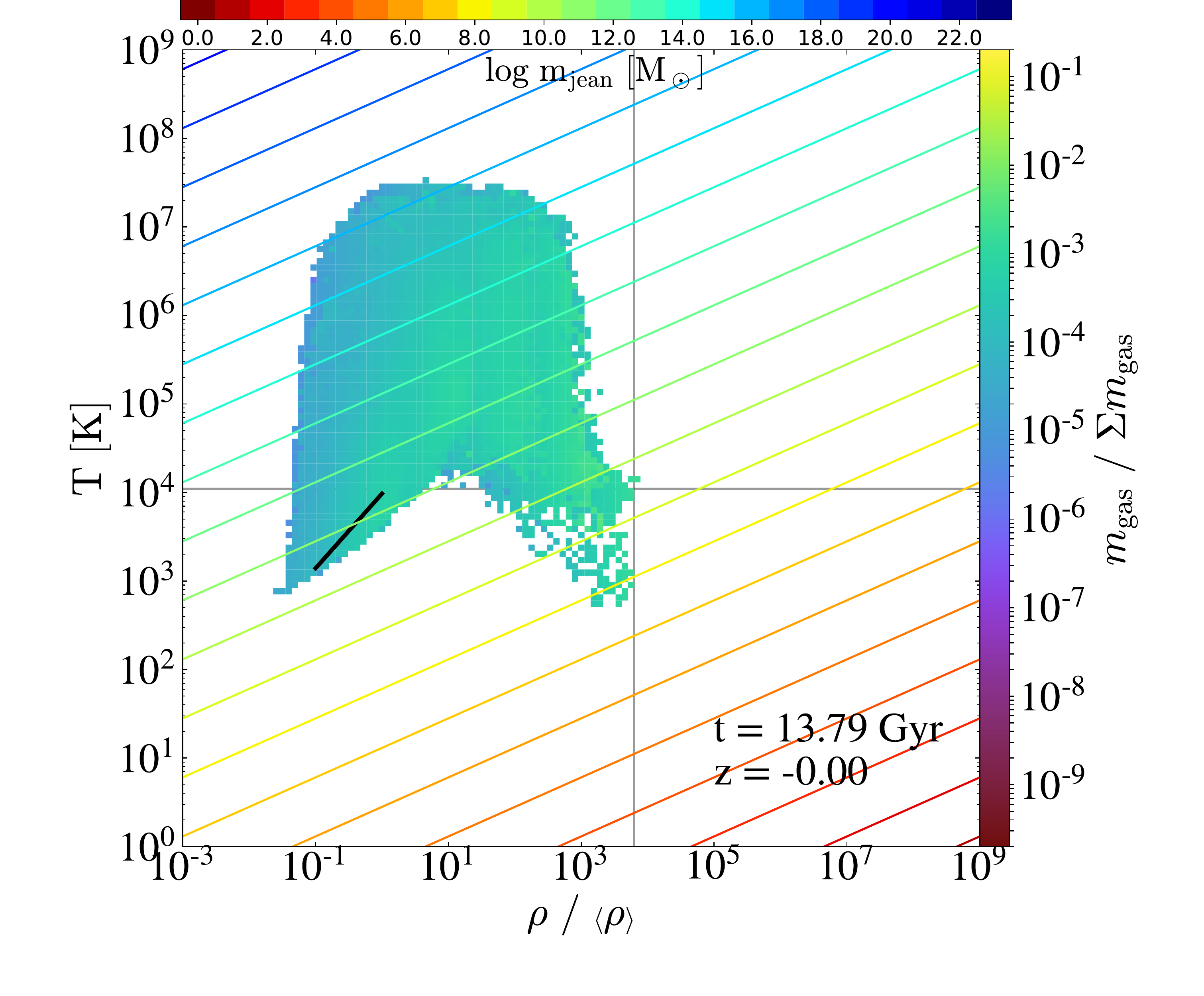}
	}
	\subfloat[$z=-0.59, t \approx t_0+t_\mathrm{H}$ \label{fig:igm2tH}]{%
		\includegraphics[width=.44\linewidth]{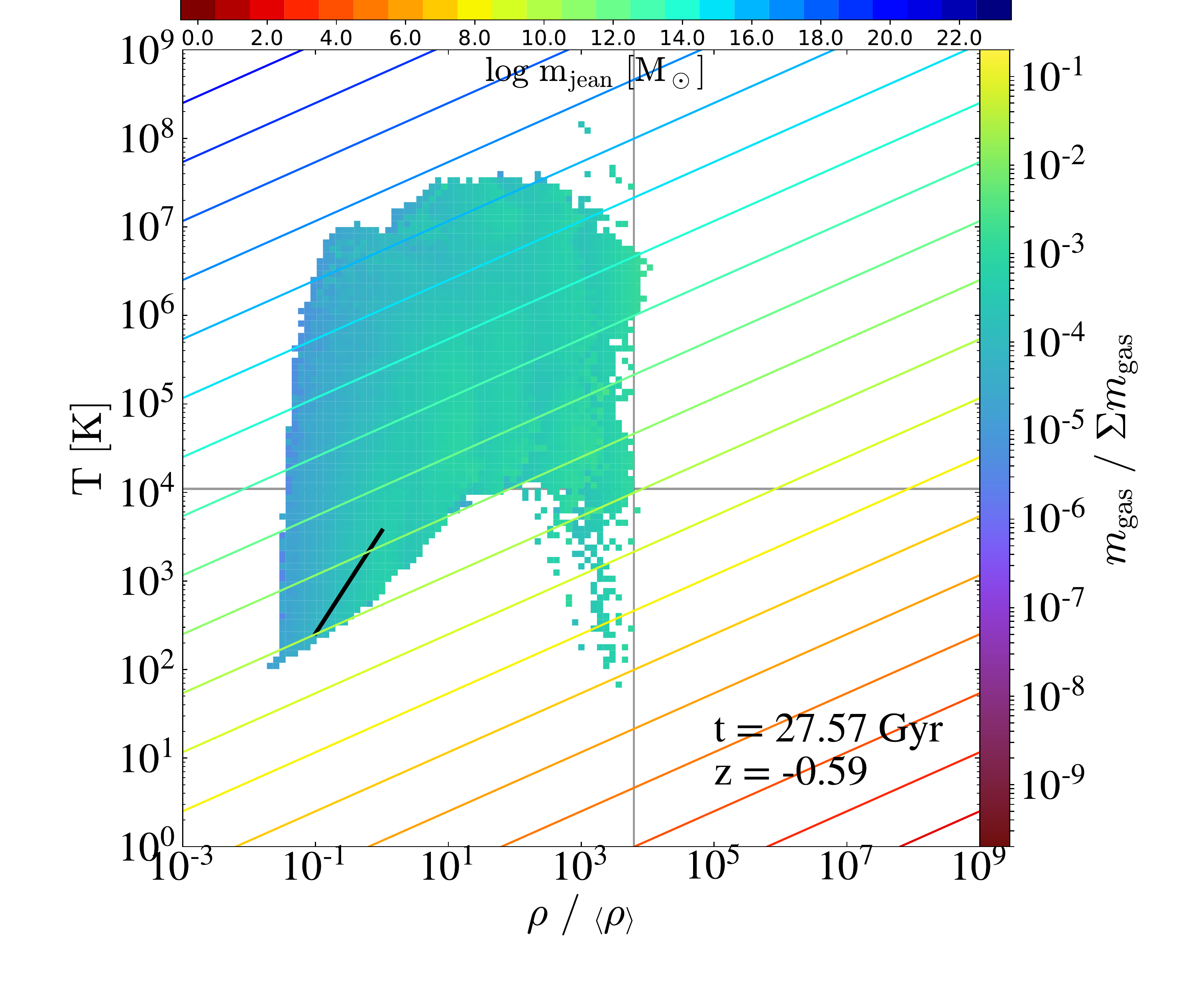}
	}
	\hfill
	\subfloat[$z=-0.82, t \approx t_0+2t_\mathrm{H}$ \label{fig:igm3tH}]{%
		\includegraphics[width=.44\linewidth]{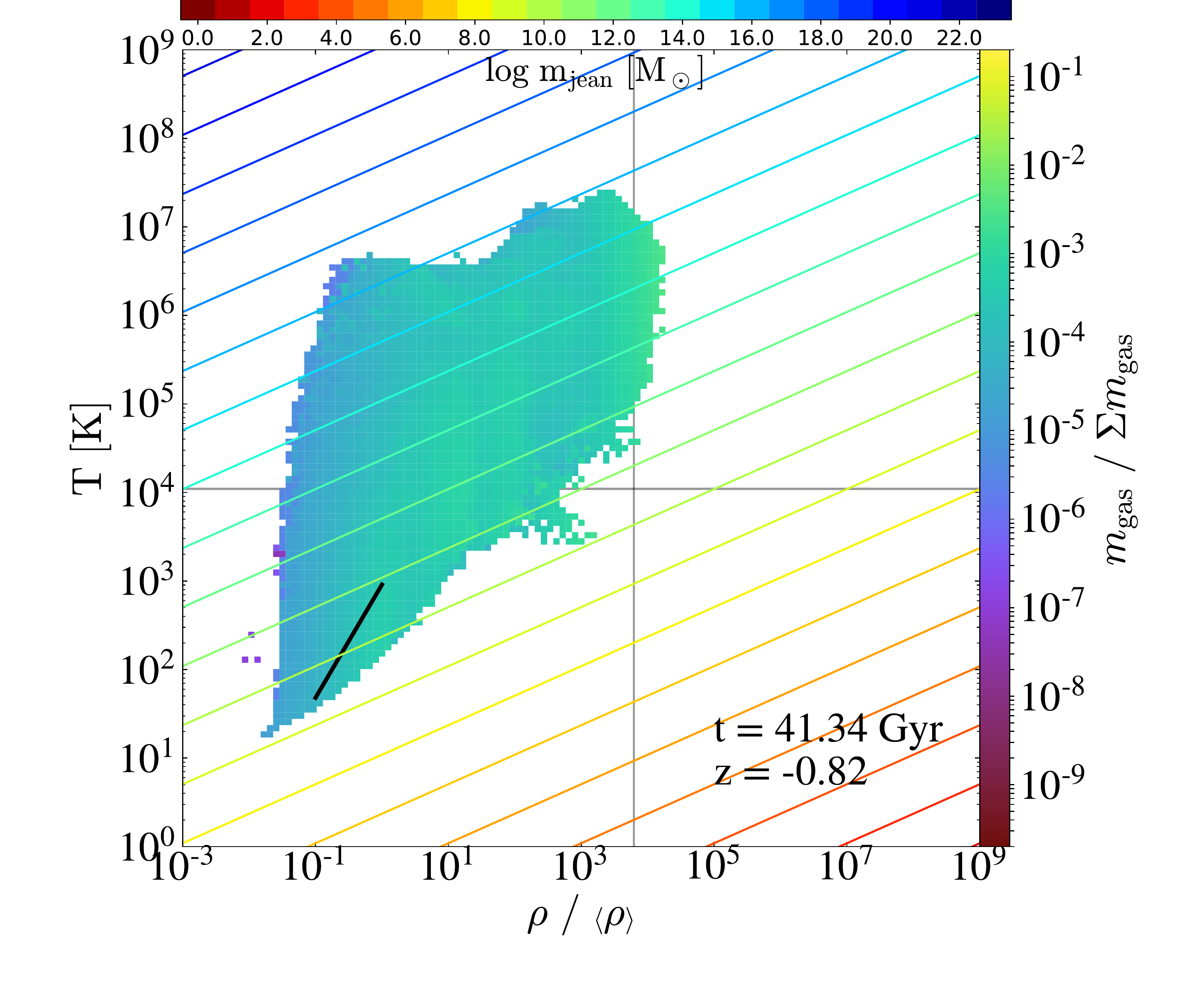}
	}
	\subfloat[$z=-0.92, t \approx t_0+3t_\mathrm{H}$ \label{fig:igm4tH}]{%
		\includegraphics[width=.44\linewidth]{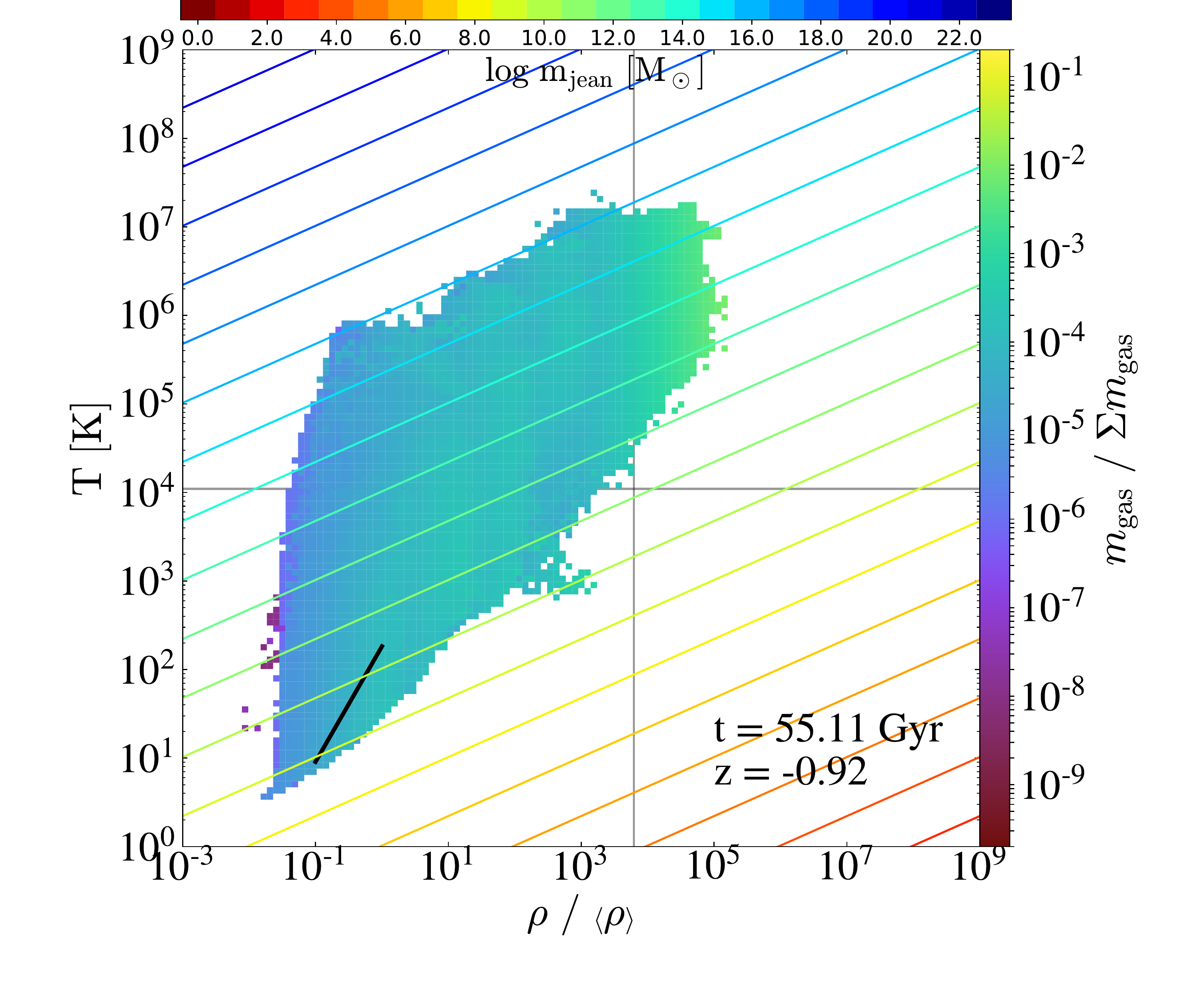}
	}
	\hfill
	\subfloat[$z=-0.98, t \approx t_0+5t_\mathrm{H}$ \label{fig:igm6tH}]{%
		\includegraphics[width=.44\linewidth]{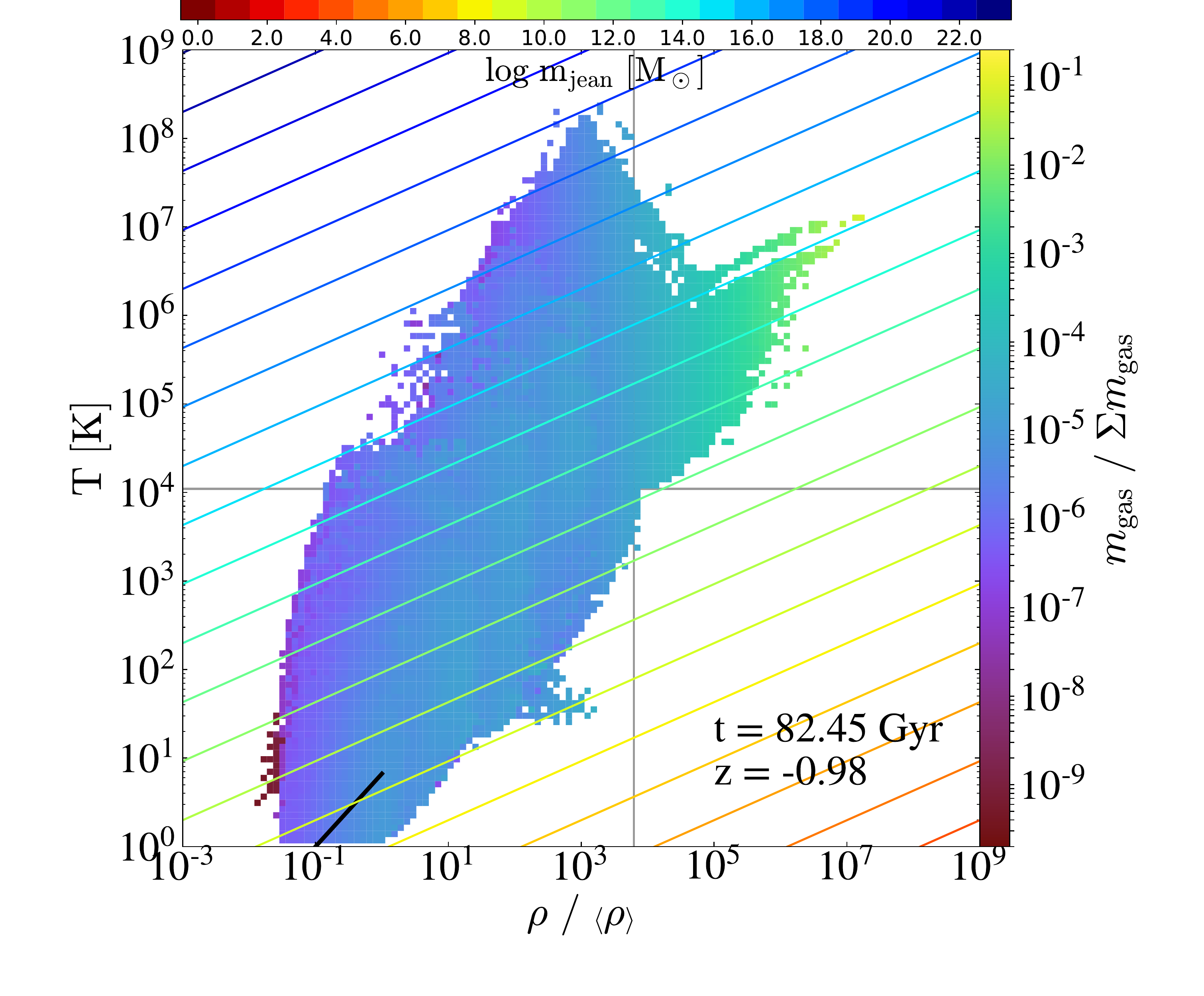}
	}
	\subfloat[$z=-0.99, t \approx t_0+6t_\mathrm{H}$ \label{fig:igm7tH}]{%
		\includegraphics[width=.44\linewidth]{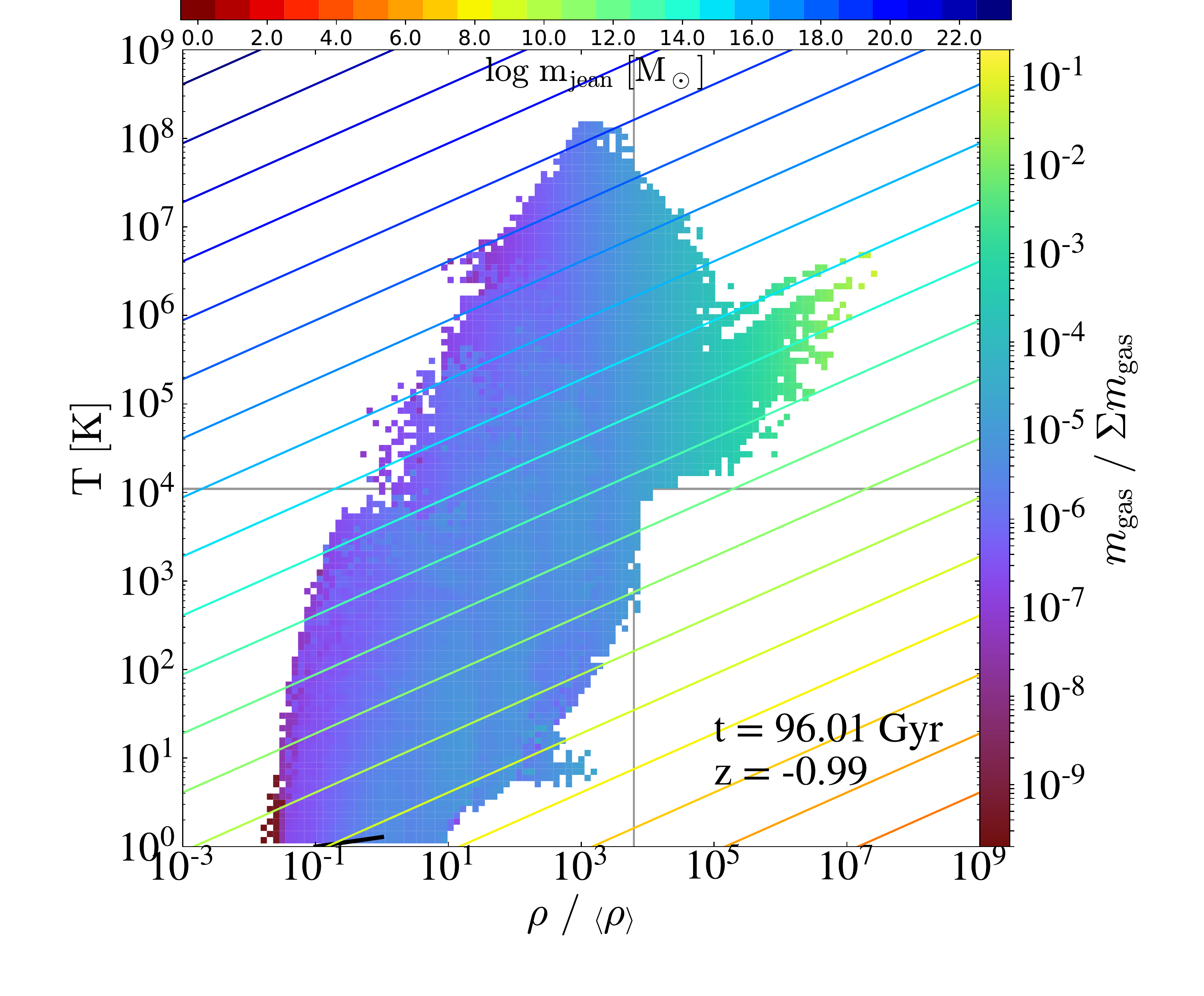}
	}
	\caption{Similar to Figure \ref{fig:sim_phaseplot} but containing only gas in the IGM. We add a black diagonal line to represent the best-fit equation of state of the IGM given by Equation \ref{eq:igm_eos}. However, the existence of gas of high overdensity suggests contamination of the phase plots. Refer to Section \ref{sec:igm_evo} for discussion on the evolution of the IGM.}
	\label{fig:igm_phaseplot}
\end{figure*}

In this section, we turn our attention to the evolution of the IGM itself. We focus mainly on the phase distribution of gas, defining IGM material as having an overdensity less than $10^3$ \citep{2001ApJ...552..473D}. \citet{1997MNRAS.292...27H} found that low density gas (overdensity $< 5$) in the IGM could be characterised by
\begin{equation}
\label{eq:igm_eos}
T = T_0(1+\delta)^{\gamma-1}, 
\end{equation} where $T_0$ is the temperature at cosmic mean density, $\delta = \rho/\left<\rho\right>-1$ is the gas overdensity and $\gamma$ is the sensitivity of the gas temperature to its overdensity. This fitted power law is shown as a diagonal black line at low overdensities in Figure \ref{fig:igm_phaseplot}, which is a modification of Figure \ref{fig:sim_phaseplot} in which we
have removed all gas within $r_{\rm vir}$ of any halo, to yield a plot governing the IGM only.

Before we look at how the IGM evolves into the future, it is interesting to note that the IGM phase distribution at late times contains gas with overdensity above $6\e{3}$. Such gas must reside within haloes and should have been excluded from the plot. This is additional evidence that {\tt Enzo} is failing to resolve low mass haloes. Since we fit the power law at much lower overdensity, these issues do not affect the evolution of the equation of state of this component of the IGM. We fit Equation \ref{eq:igm_eos} using two bins of grid cells with a width of $5\%$ centred around gas overdensities of $10^{-1}$ and $10^{0}$, consistent with previous work \citep{2015MNRAS.450.4081P, 2018ApJ...859..125S}. We then calculate the volume-weighted median temperature around these values and construct the best-fit power law.

We first compare our fit with previous results at $z\approx2.5$ from \citet{2018ApJ...859..125S}. The authors presented results from both a Nyx simulation \citep{2013ApJ...765...39A, 2015MNRAS.446.3697L} and the Illustris simulation \citep{2014MNRAS.444.1518V}. The parameters obtained from fitting Equation \ref{eq:igm_eos} are $\log_{\rm 10}T_0 = 4.01$ and $\gamma = 1.57$ at $z=2.4$ for Nyx and $\log_{\rm 10}T_0 = 4.12$ and $\gamma = 1.6$ at $z=2.44$ for Illustris. These agree well with the values from our simulation, which are $\log_{\rm 10}T_0 = 4.21$ and $\gamma = 1.58$ at $z=2.55$. They are also consistent with observational constraints from \citet{2000MNRAS.318..817S}: $\log_{\rm 10}T_0 \approx 4.20$ and $\gamma \approx 1.2$ at $z=2.5$. It should however be noted that deviations in these values can arise due to the differences in the assumed UV background \citep{2017ApJ...837..106O}. 

\begin{figure}
	\centering
	\includegraphics[width=\columnwidth]{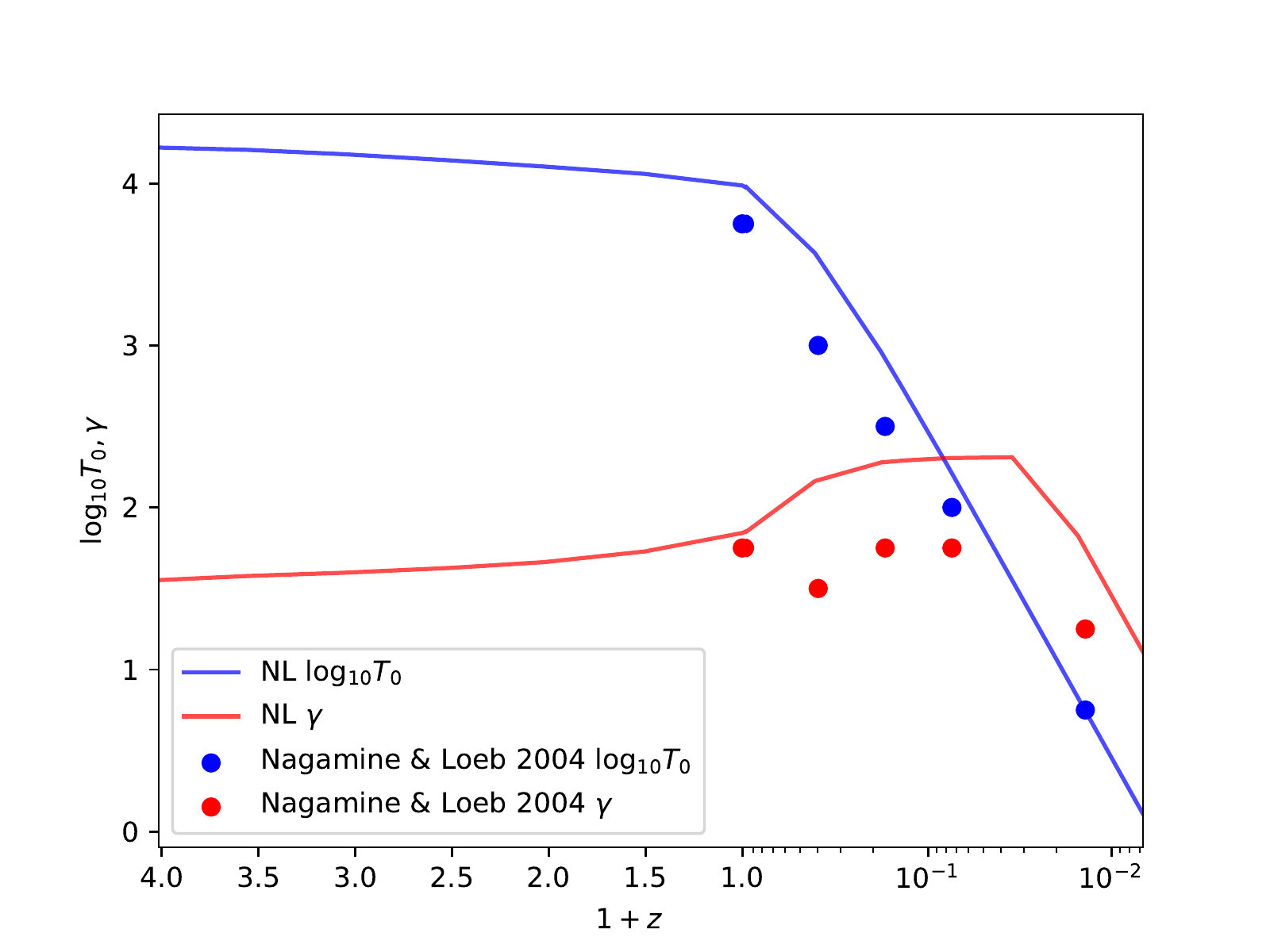}
	\caption{Evolution of the properties of the IGM, linearly for $z\geq0$ and logarithmically for $z<0$. The lines and dots are results from our and \citet{2004NewA....9..573N}'s simulation respectively. Blue and red colour refers to $\log_{\rm 10}T_0$ and $\gamma$ from Equation \ref{eq:igm_eos} respectively. The sharp decline of $\log_{\rm 10}T_0$ beyond $z=0$ suggests that we are at a special point in the evolution of the universe. Refer to Section \ref{sec:igm_evo} for discussion about this in relation to the `coincidence problem'.}
	\label{fig:igm_prop_evo}
\end{figure}

The next step is to investigate how the values of $T_0$ and $\gamma$ evolve, which is presented in Figure \ref{fig:igm_prop_evo}; we include for comparison dots that show a visual fit to Figure 3 of \citet{2004NewA....9..573N}. The median temperatures and the resulting $\log_{\rm 10}T_0$ and $\gamma$ from our simulation are insensitive to whether we use a volume-weighted or mass-weighted median. There is a consistent drop in $\log_{\rm 10}T_0$ beyond $z=0$ across both simulations. As discussed above, the expansion of the universe drives an increasing amount of gas towards the CMB temperature. This convergence with the CMB temperature is higher further into the future. Figure \ref{fig:igm_phaseplot} shows that the overdensity associated with this gas increases with time, eventually encompassing the values used to derive $\gamma$ by $1+z\approx0.01$. At this point, $\gamma$ approaches a value of unity because the median temperatures of the gas within the overdensity bins are similar to each other, close to the CMB value. As a result, it is harder to provide a meaningful interpretation for the values of $T_0$ and $\gamma$ so far into the future.

Before $\gamma$ falls abruptly to a value near unity, Figure \ref{fig:igm_prop_evo} shows that this parameter increases slowly between $4.0 \leq 1+z \leq 0.02$. \citet{2016MNRAS.456...47M} pointed out that the balance of the photoheating from the UV background and the cooling due to the cosmological expansion creates a tight relation between the temperature and density of the IGM at early times. Other processes such as Compton cooling by the CMB also shaped this relation, albeit in a minor role. What is expected to happen to this relation in the future? Following our assumptions, the extrapolated heating from the UV background decreases to zero while there is an increased adiabatic cooling rate from the expansion of the universe. Together with a continual increase in the shocked fraction of gas \citep{2016MNRAS.456...47M}, these factors create deviations in the previously tight relation, and the temperature distribution of the gas at given density gains a higher variance in the future. Since Equation \ref{eq:igm_eos} is supposed to describe a tight relation at high redshifts, this variance increases the error associated with such a fit to the IGM into the future. Combined with what we discussed in the previous paragraph, it appears that this previously tight relation at high redshifts will not be as suitable to describe the IGM into the future.

Lastly, we note that the decline of $T_0$ begins at $z \approx 0$, suggesting that the IGM is presently in a delicate balance with only just sufficient photons to remain ionised. Radiation from stars and AGN that contribute to the UV background will soon no longer be sufficient to keep the universe ionised. If this is the case, it adds another case study to the cosmological coincidence problem, in which the present is a unique point in the evolution of the universe. This is related to the history of star formation to this point: if the SFRD had not peaked at $z\approx2$ and subsequently declined, the
decrease in $T_0$ would be postponed.

\subsection{Resolution convergence}\label{sec:res-conv}

\subsubsection{Halo mass function}\label{sec:hmf-conv}

In Sections \ref{sec:hmf_evo} and \ref{sec:igm_evo}, we discussed the evolution of the HMF and the IGM into the future using simulation {\sl NL}, described in Section \ref{sec:sim_setup}, which has comparable resolution to the simulation of \citet{2004NewA....9..573N}. We have shown that our results are in reasonable agreement. However, it is not clear whether the results are converged in terms of resolution. Therefore, we introduce six other simulations that are summarised in Table \ref{tab:naga-setup}. We will be using {\sl NL} as the baseline, {\sl NL $\pm$ 1} for spatial resolution comparison, {\sl NLm $\pm$ 1} for mass resolution comparison, and {\sl NLfb} for a feedback sensitivity study. Each simulation contains only one parameter that is different from {\sl NL}. 

\begin{figure*}
	\centering
	\subfloat[$z=0, t=t_0$ \label{fig:conv-hmf1tH}]{%
		\includegraphics[width=.43\linewidth]{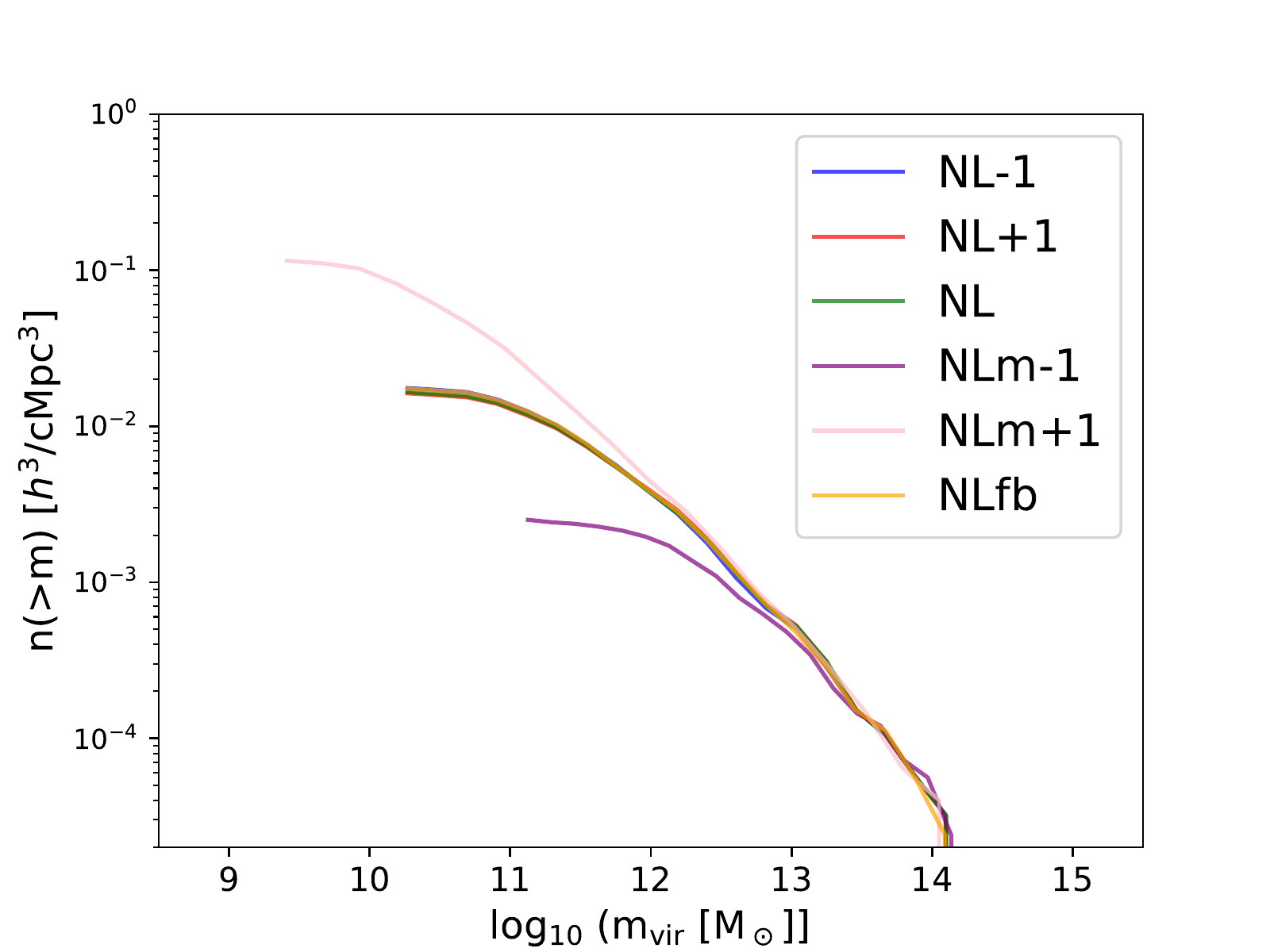}
	}
	\subfloat[$z=-0.59, t \approx t_0+t_\mathrm{H}$ \label{fig:conv-hmf2tH}]{%
		\includegraphics[width=.43\linewidth]{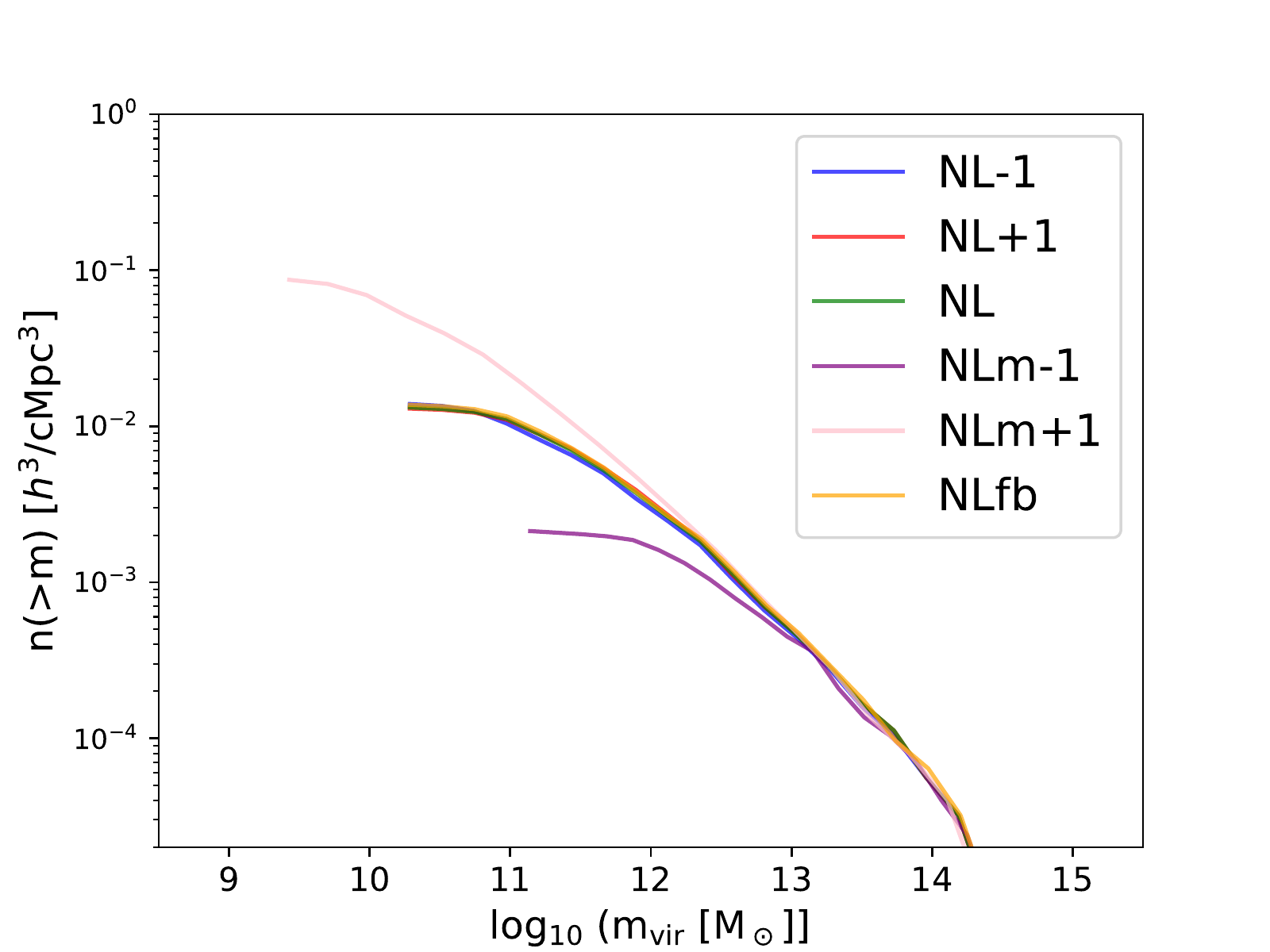}
	}
	\hfill
	\subfloat[$z=-0.82, t \approx t_0+2t_\mathrm{H}$ \label{fig:conv-hmf3tH}]{%
		\includegraphics[width=.43\linewidth]{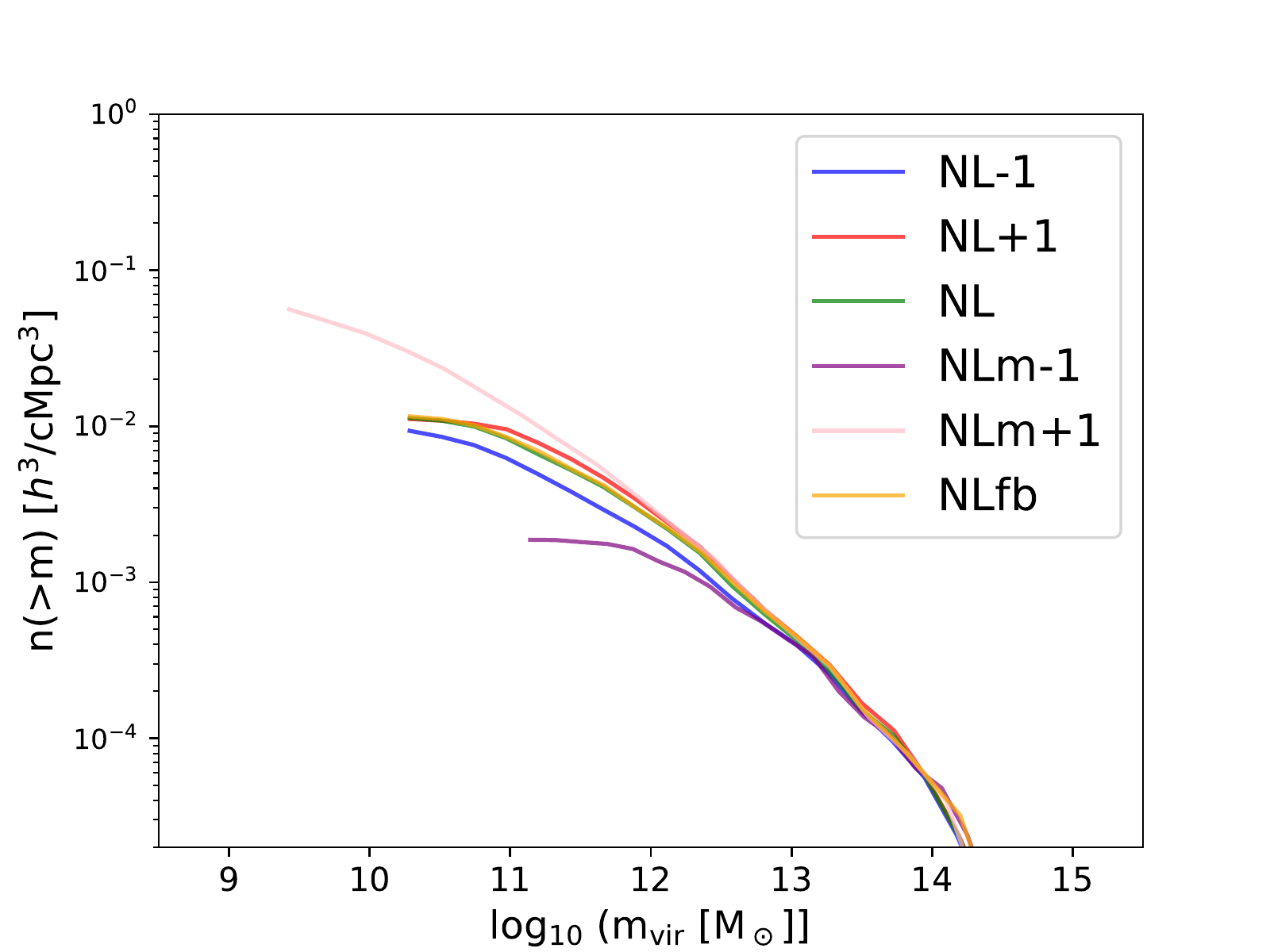}
	}
	\subfloat[$z=-0.92, t \approx t_0+3t_\mathrm{H}$ \label{fig:conv-hmf4tH}]{%
		\includegraphics[width=.43\linewidth]{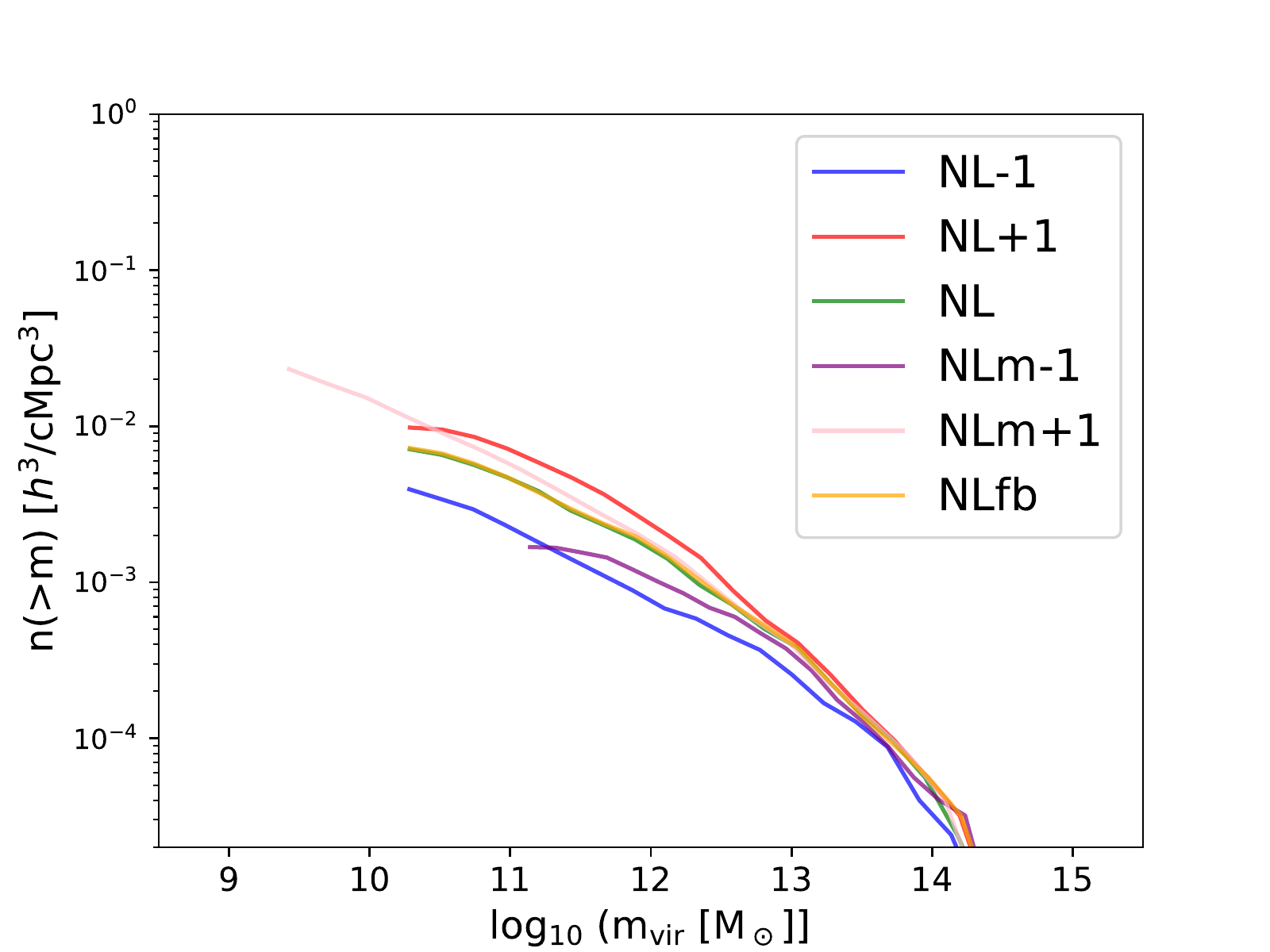}
	}
	\hfill
	\subfloat[$z=-0.98, t \approx t_0+5t_\mathrm{H}$ \label{fig:conv-hmf6tH}]{%
		\includegraphics[width=.43\linewidth]{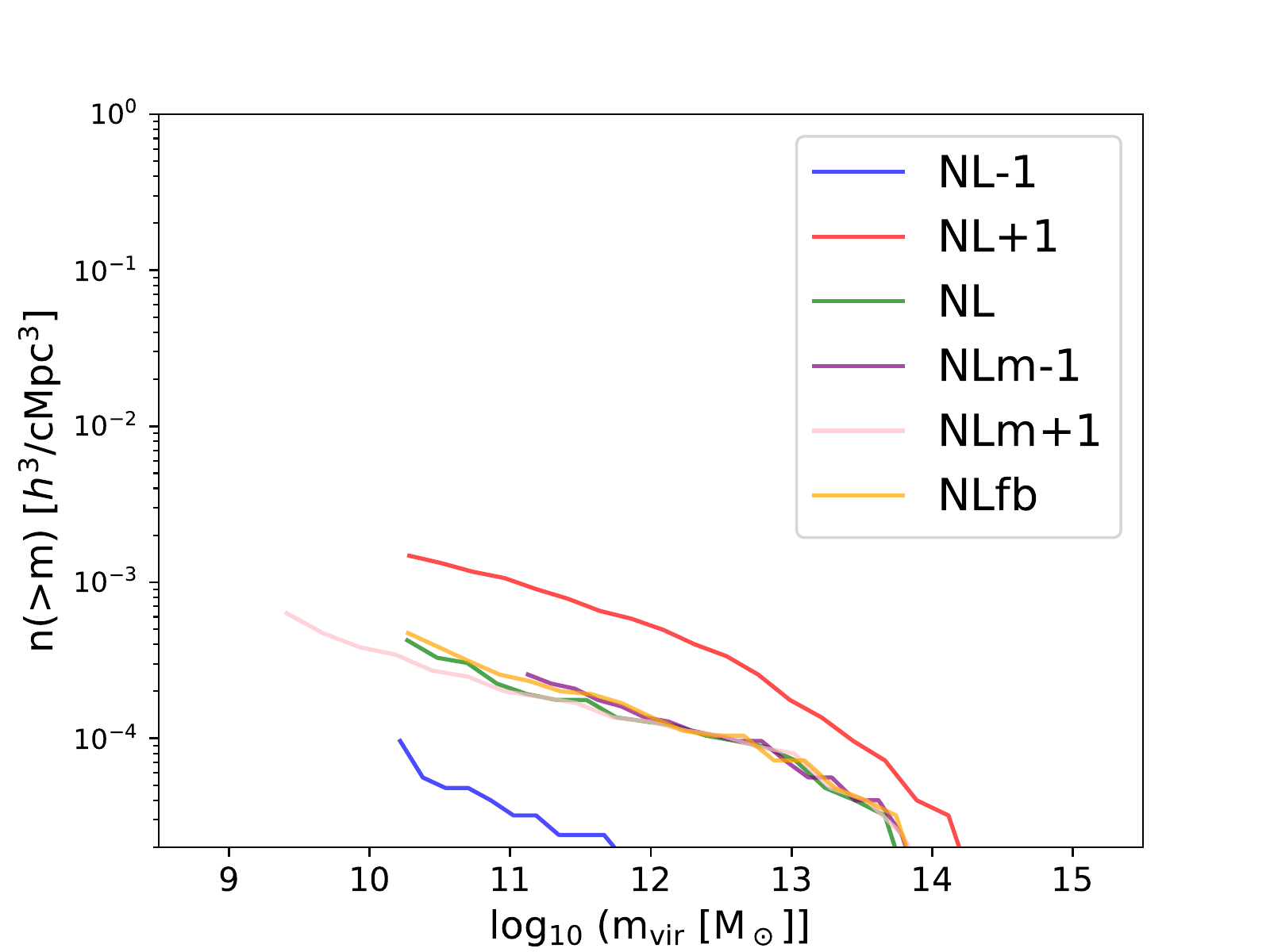}
	}
	\subfloat[$z=-0.99, t \approx t_0+6t_\mathrm{H}$ \label{fig:conv-hmf7tH}]{%
		\includegraphics[width=.43\linewidth]{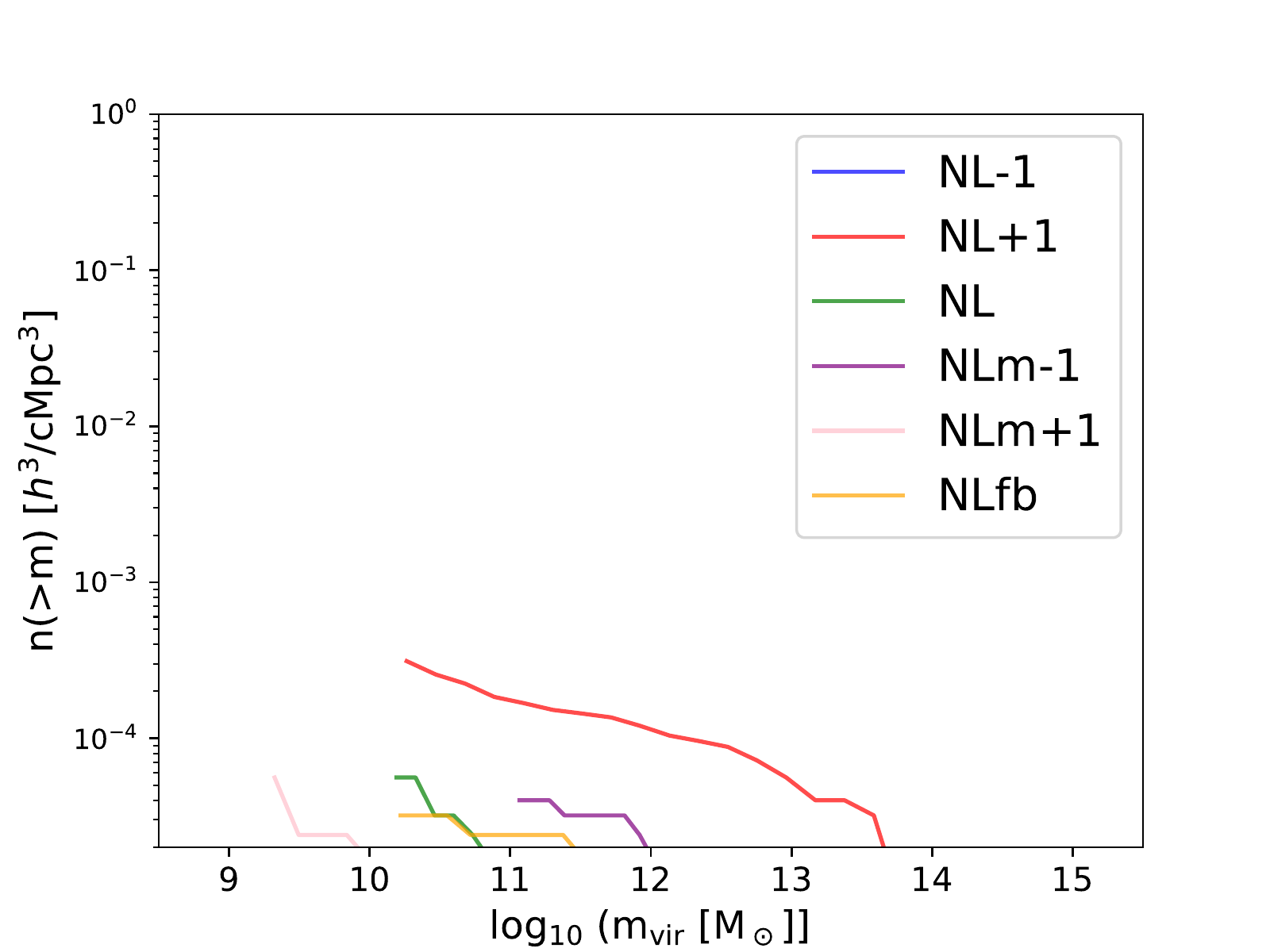}
	}
	\caption{Evolution of the HMFs into the future, considering simulations with varying specifications summarised in Table \ref{tab:naga-setup}. The time interval between each panel is approximately $t_{\rm H}$. The blue, red, green, purple, pink and orange lines correspond to {\sl NL-1}, {\sl NL+1}, {\sl NL}, {\sl NLm-1}, {\sl NLm+1} and {\sl NLfb} respectively. Across time, simulations with different root grid resolutions (NLm+1, NLm-1) have a correspondingly better or worse resolution, affecting their ability to resolve the smaller mass haloes. On the other hand, simulations with different maximum spatial resolutions ({\sl NL-1}, {\sl NL+1}) deviate more significantly further into the future. Lastly, different feedback implementations in simulations ({\sl NLfb}) do not appear to influence the HMF as expected. Refer to Section \ref{sec:hmf-conv} for discussion about the convergence of these results.}
	\label{fig:conv-hmf}
\end{figure*}

In Figure \ref{fig:conv-hmf}, we show that changing the root grid resolution in {\sl NLm-1} (purple) and {\sl NLm+1} (pink) affects the low mass end of the HMFs most significantly. The ratio of the dark matter particle mass between {\sl NLm-1} and {\sl NLm+1} is 64. Since {\tt ROCKSTAR} uses the same minimum number of particles to define a halo, this difference is carried forward to the minimum mass of a resolved halo in both simulations. At the high mass end, there is excellent agreement between all simulations at $z>-0.92$. But this conclusion does not hold for all times. In panels (e) and (f) of Figure \ref{fig:conv-hmf}, the HMFs of {\sl NLm-1}, {\sl NL} and {\sl NLm+1} with identical spatial resolutions are clustered around each other but separated from the rest. Therefore, we do not have convergence at these times, and this separation hints at the sensitivity of the HMFs to spatial resolution. 

The simulation with the worst spatial resolution is {\sl NL-1} (blue). It starts to show signs of deviation earlier than the other simulations, affecting the low mass end of the HMF at $z=-0.82$, and eventually coming to affect the entire mass range (see Section \ref{sec:hmf_evo}). Far enough into the future, the increase in proper grid cell size due to the expansion of the universe will affect even the best spatial resolution simulation ({\sl NL+1}). By $z=-0.98$, the HMFs of the simulations are separated into three distinct bands according to their maximum spatial resolution. In future work, we can thus predict the redshift at which a simulation's HMF will break, marking the reliability limit of the calculation. Lastly, we do not notice any significant disparity between {\sl NLfb} (orange) and {\sl NL} (green) with an identical resolution, confirming that baryonic processes have little influence over their host haloes.

The spatial resolution plays an important role in determining the simulation's ability to form haloes the further it evolves into the future. From the divergence of the HMFs in Figure \ref{fig:conv-hmf} starting from $z=-0.98$, we conclude that the presence of haloes is attributed to the numerical limits of the simulation. To further quantify if this is a purely numerical or physical motivated phenomenon, we have to experiment with simulations of higher spatial resolution and even further into the future. We will discuss the implications of this limitation in the context of star formation in later sections. 

\subsubsection{Properties of the intergalactic medium}\label{sec:conv-igm_evo}
\begin{figure}
	\centering
	\includegraphics[width=\linewidth]{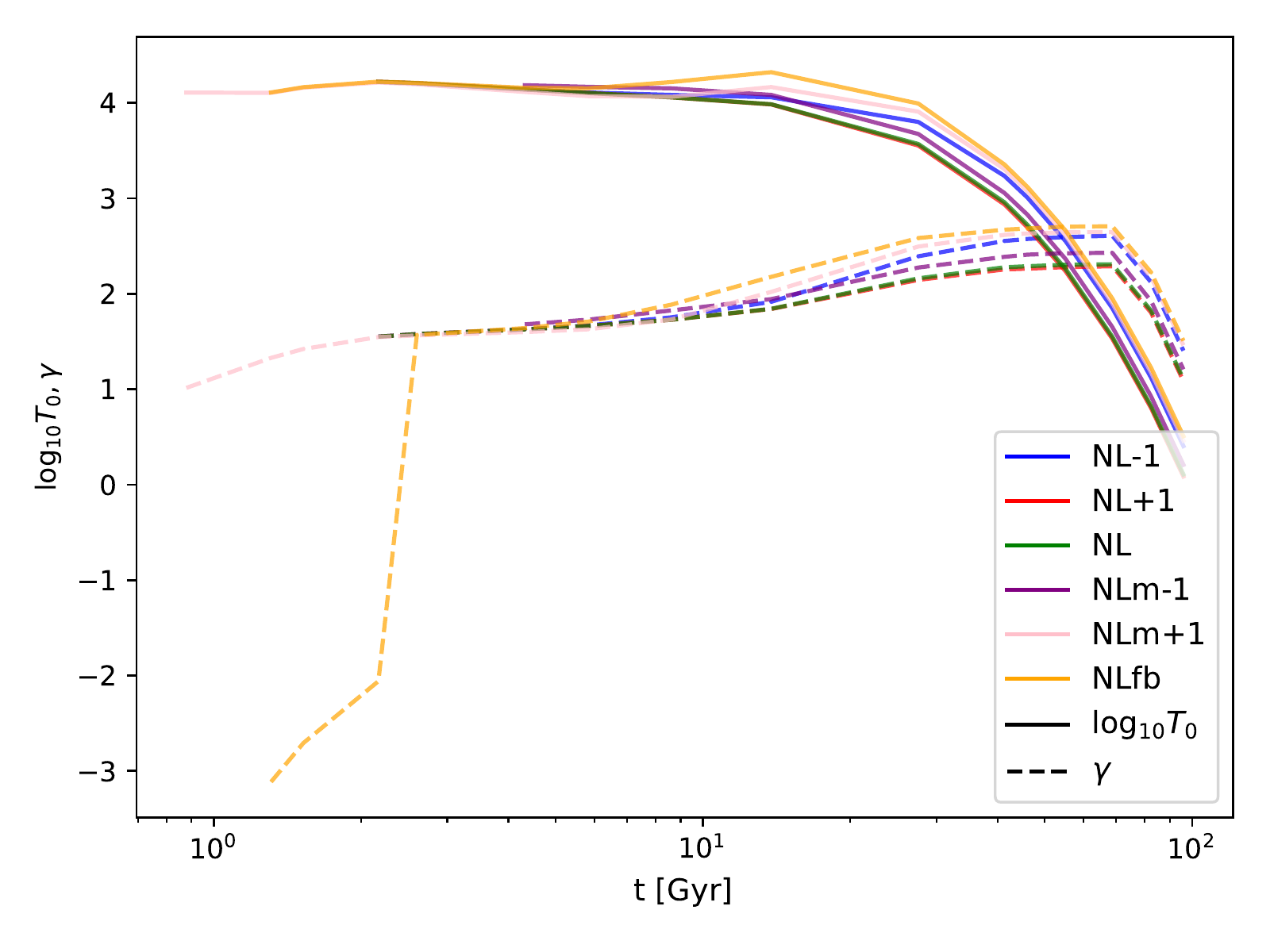}
	\caption{Evolution of the properties of the IGM in time from our suite of simulations summarised in Table \ref{tab:naga-setup}. The lines are coloured in the same way as Figure \ref{fig:conv-hmf}, indicated by the legend. The solid and dashed lines represents the evolution of $\log_{\rm 10}T_0$ and $\gamma$ from Equation \ref{eq:igm_eos} respectively. The main difference between the simulations occurs at early times due to disparities in spatial resolution and star formation criteria specifically for {\sl NLm+1} and {\sl NLfb}. Refer to Section \ref{sec:conv-igm_evo} for detailed discussion.}
	\label{fig:conv-igm_prop_evo}
\end{figure}

In this section, we shift the focus to the IGM, expanding the results in Section \ref{sec:igm_evo} and looking at their convergence. Although we have shown a high degree of variance within the defined density of the IGM in the future, a single equation of state is still one of the most prevalent methods for characterising the IGM. Therefore, we continue to use the power law fit in Equation \ref{eq:igm_eos} in our convergence study. We show the evolution of $\log_{\rm 10}T_0$ (solid lines) and $\gamma$ (dashed lines) from our suite of simulations in Figure \ref{fig:conv-igm_prop_evo}. Generally, we find the evolution to be consistent regardless of resolution and star formation and feedback prescription: the trend closely resembles Figure \ref{fig:igm_prop_evo} (see Section \ref{sec:igm_evo}) with some differences between the simulations.

We notice that the initial redshift when Equation \ref{eq:igm_eos} can be fitted to the simulations differs according to the absence of gas with overdensity of 0.1. It varies significantly between $2 < 1+z < 8$ for {\sl NLm-1}, {\sl NLm+1} and the rest of the simulations except {\sl NLfb}, reflecting the disparity in the root grid resolution. Note that AMR is not in full effect at this early time. With an improved root grid resolution, we can resolve lower mass haloes, which are shallower potential wells that allow gas to reach overdensity of 0.1. 

On the other hand, {\sl NLfb}, which uses a different subgrid prescription, can resolve the IGM at early times, because the difference in the conversion efficiency of gas into stars leaves behind a significant amount of gas. This gas is heavily influenced by the feedback from the stars in the simulation, forming pockets of hot underdense gas, which explains the inverse relation (negative values) observed at very early times for {\sl NLfb} in Figure \ref{fig:conv-igm_prop_evo}.

What about the scatter observed in the late-time evolution of the properties of the IGM in Figure \ref{fig:conv-igm_prop_evo}? We mentioned in Section \ref{sec:igm_evo} that the main drivers for the formation of the power-law fit to the IGM are photo-heating from the UV background, adiabatic cooling due to the expansion of the Universe, and shocks \citep{2016MNRAS.456...47M}. Since the first two factors are consistent across the simulations, the last is the most likely cause for the scatter. We showed that the HMFs are sensitive to the mass and spatial resolution of the simulations (see Section \ref{sec:hmf_evo}), which in turn affects the amount of shocked gas. In {\sl NLfb}, using Setup 1 with a lower conversion efficiency leaves behind a larger gas reservoir as compared to Setup 2. Also, the gas is pushed out to different radii because of the different extent of feedback energy injection, affecting its ability to fall back onto the halo. These factors affect the fraction of shocked gas, leading to slight deviations between simulations.

\subsubsection{SFR} \label{sec:conv-sfrd}

\begin{figure}
	\centering
	\includegraphics[width=\linewidth]{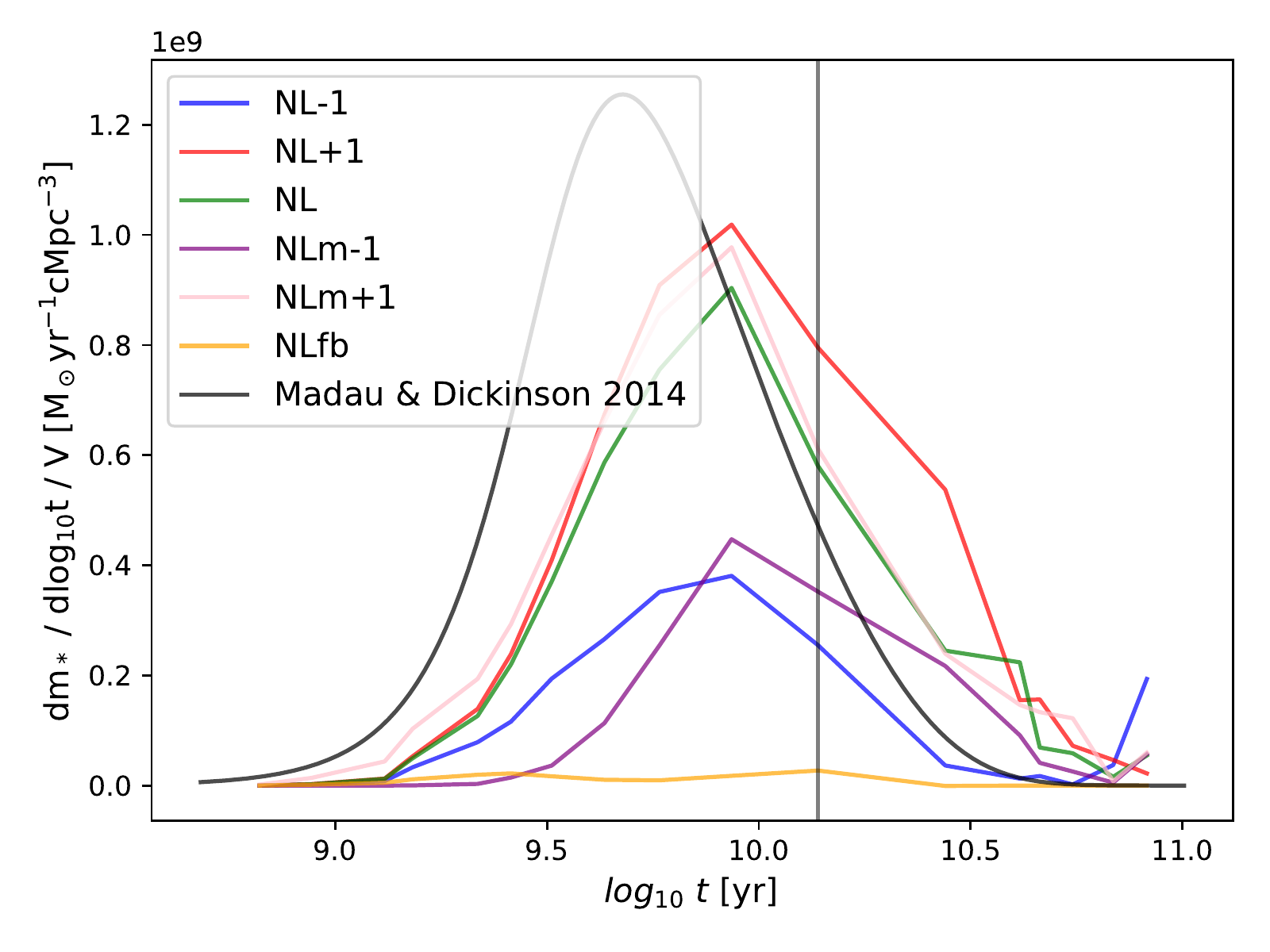}
	\caption{Evolution of SFRD across cosmic time in our suite of simulations. The blue, red, green, purple, pink and orange lines correspond to {\sl NL-1}, {\sl NL+1}, {\sl NL}, {\sl NLm-1}, {\sl NLm+1} and {\sl NLfb} respectively. The black curve represents a modified Equation 15 from \citet{2014ARA&A..52..415M} to match the axis labels. We have also added a vertical grey line to indicate the point of time where $z=0$. Across all simulations, we obtain a similar peak in SFRD of different peak values, albeit at a later time than the black line. Also, we identify a turnaround in the SFRD depending on the resolution of the simulations. The impact of varying resolutions and the turnaround of the SFRD will be discussed in Section \ref{sec:conv-sfrd}.}
	\label{fig:conv-sfr}
\end{figure}

Having investigated the convergence of our results, we have an improved understanding of our ability to follow the haloes that host star formation and the evolution of the IGM that provides the fuel for future star formation. We will estimate how the SFRD behaves as $t\to\infty$, comparing this evolution with the observational fit from \citet{2014ARA&A..52..415M} and its extrapolation. We start by showing the evolution of the SFRD in Figure \ref{fig:conv-sfr} with different coloured lines corresponding to the simulations and observations as indicated in the legend. If we compare all of our simulations as a whole with the analytic fit to observations, we immediately recognise that the peak of the SFRDs from the simulation is lower and occurs at a later time. This difference in part reflects the limited resolution of our simulations, which causes structure formation to be delayed in the simulation, leading to a later onset of star formation. The peak in SFRD is also lower because we do not have the resolution to capture all star-forming haloes.

The peaks in the SFRD are very sensitive to the star formation and feedback prescription, evident from the significant difference between {\sl NLfb} and the rest of the simulations. {\sl NLm-1} and {\sl NL-1} both have a lower SFR peak because of their poorer resolution. Specifically, {\sl NLm-1} has the worst root grid resolution, which means that the mass and spatial resolution is the lowest before AMR kicks in. The deficiency in the number of low mass haloes in {\sl NLm-1} results in delayed structure formation, explaining why it has a lower peak in SFRD. {\sl NL-1} suffers from a different problem that escalates with time. Since {\sl NL-1} has the worst force resolution, gas is prevented from reaching high densities, thus limiting star formation. The other simulations, {\sl NL}, {\sl NLm+1}, {\sl NL+1} exhibit a similar and consistent evolution in SFRD to each other, even if not in complete agreement with the observational fit.

\begin{figure}
	\centering
	\includegraphics[width=\linewidth]{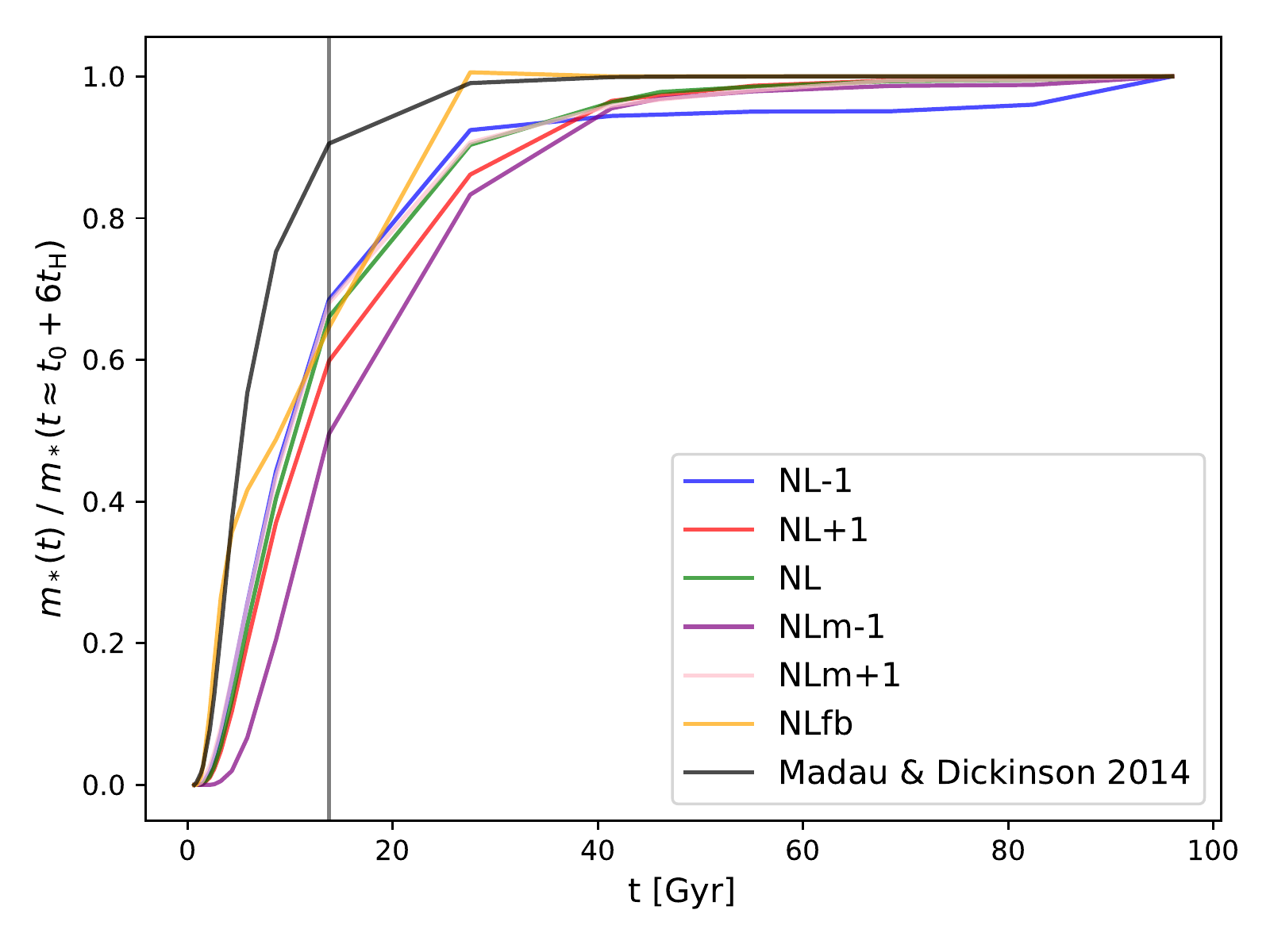}
	\caption{Evolution of the ratio of stellar mass formed by time $t$ to that formed by $t \approx t_0 + 6t_{\rm H}$. The blue, red, green, purple, pink and orange lines correspond to {\sl NL-1}, {\sl NL+1}, {\sl NL}, {\sl NLm-1}, {\sl NLm+1} and {\sl NLfb} respectively. The black curve represents the same ratio obtained from Equation 15 of \citet{2014ARA&A..52..415M}. Across all simulations, at $z=0$ (vertical black line), the percentage of stellar mass formed in simulations is 40\% -- 60\% as compared to 90\% predicted by the black line. Refer to Section \ref{sec:conv-sfrd} for further discussion.}
	\label{fig:conv-mstarfrac}
\end{figure}

We then look at the amount of stellar mass formed by a given time, $t$, as a fraction of the asymptotic total stellar mass formed by $t \approx t_0 + 6\,t_{\rm H}$ in Figure \ref{fig:conv-mstarfrac}. At $z=0$, the observed stellar mass given by Equation 15 from \citet{2014ARA&A..52..415M} is 90\% of the asymptotic stellar mass while the simulations have only formed 40\% -- 60\% of their respective asymptotic stellar masses by $z=0$. This difference reflects the delay in the star formation peak as discussed earlier. As a result, the empirically extrapolated stellar mass converges at an earlier time in contrast to the simulations except for {\sl NLfb}. Although {\sl NLfb} exhibits a comparable peak time to the analytic fit, we know from Table \ref{tab:stellarmass} that the total stellar mass formed is much less than predicted. For the other simulations except for {\sl NL+1}, the mass fraction stagnates at a ratio less than unity before increasing again. {\sl NL-1} illustrates this trend clearly with a plateau at roughly 90\%. This evolution is due to a late time turnaround in the SFRD that will be discussed extensively in the following sections. These discrepancies between the simulated SFRD and the extrapolation of the observational fit from \citet{2014ARA&A..52..415M} are plausibly outcomes of the finite resolution in the simulations.

Although the exact evolution of the SFRD differs between the simulations, we are interested in finding out if the total amount of stellar mass formed is the same. Therefore, we integrate the SFRDs in Figure \ref{fig:conv-sfr} between the specified redshifts and summarise the results in Table \ref{tab:stellarmass}. Other than {\sl NLfb}, the total stellar mass densities within each of the remaining simulations agree with each other and, more importantly, with the value obtained from integrating Equation 15 of \citet{2014ARA&A..52..415M}. The values are of the same order of magnitude with {\sl NL+1} matching within 1\% of the total predicted stellar mass density. In contrast, {\sl NLfb} differs by nearly two orders of magnitude, proving again that the SFRD is highly sensitive to feedback and that Setup 1 is not appropriate for our purposes.

\begin{table}
	\caption{List of simulations with their corresponding total stellar mass density formed between $z=8$ and $z=-0.993$. Refer to Table \ref{tab:naga-setup} for specifications of individual simulations. Except {\sl NLfb}, we obtain a good agreement within the same order of magnitude with the total predicted stellar mass based on the fit to observations provided by \citet{2014ARA&A..52..415M}.}
	\label{tab:stellarmass}
	\centering
	\begin{tabular}{|p{.4\columnwidth}|p{.5\columnwidth}|}
		\hline
		Simulation & Total stellar mass $[M_\odot]$\\
		\hline
		{\sl NL} & $2.57\e{14}$\\
		\hline
		{\sl NL-1} & $1.09\e{14}$\\
		\hline
		{\sl NL+1} & $3.28\e{14}$\\
		\hline
		{\sl NLm-1} & $1.13\e{14}$\\
		\hline
		{\sl NLm+1} & $2.87\e{14}$\\
		\hline
		{\sl NLfb} & $7.74\e{12}$\\
		\hline
		Madau \& Dickinson (2014) & $3.33\e{14}$\\
		\hline
	\end{tabular}
\end{table} 

\subsection{An apparent turnaround of the future SFRD}

Let us now look at the far future, focusing on $t\approx50\,{\rm Gyr}$ in Figure \ref{fig:conv-sfr}. There is an apparent turnaround in the cosmic SFRD, deviating from the extrapolated SFRD of the analytic fit to observations. This onset of a reversal in SFRD occurs at two different times, depending on the spatial resolution of the simulations. The SFRD begin to increase earlier for {\sl NL-1}, followed by {\sl NL}, {\sl NLm-1}, {\sl NLm+1} and {\sl NLfb}. Within the timespan of Figure \ref{fig:conv-sfr}, we do not yet see a SFRD turnaround for {\sl NL+1}. This trend suggests that the result has a numerical origin: a lower spatial resolution in the simulations results in an earlier turnaround. 

\begin{figure}
	\centering
	\includegraphics[width=0.85\linewidth]{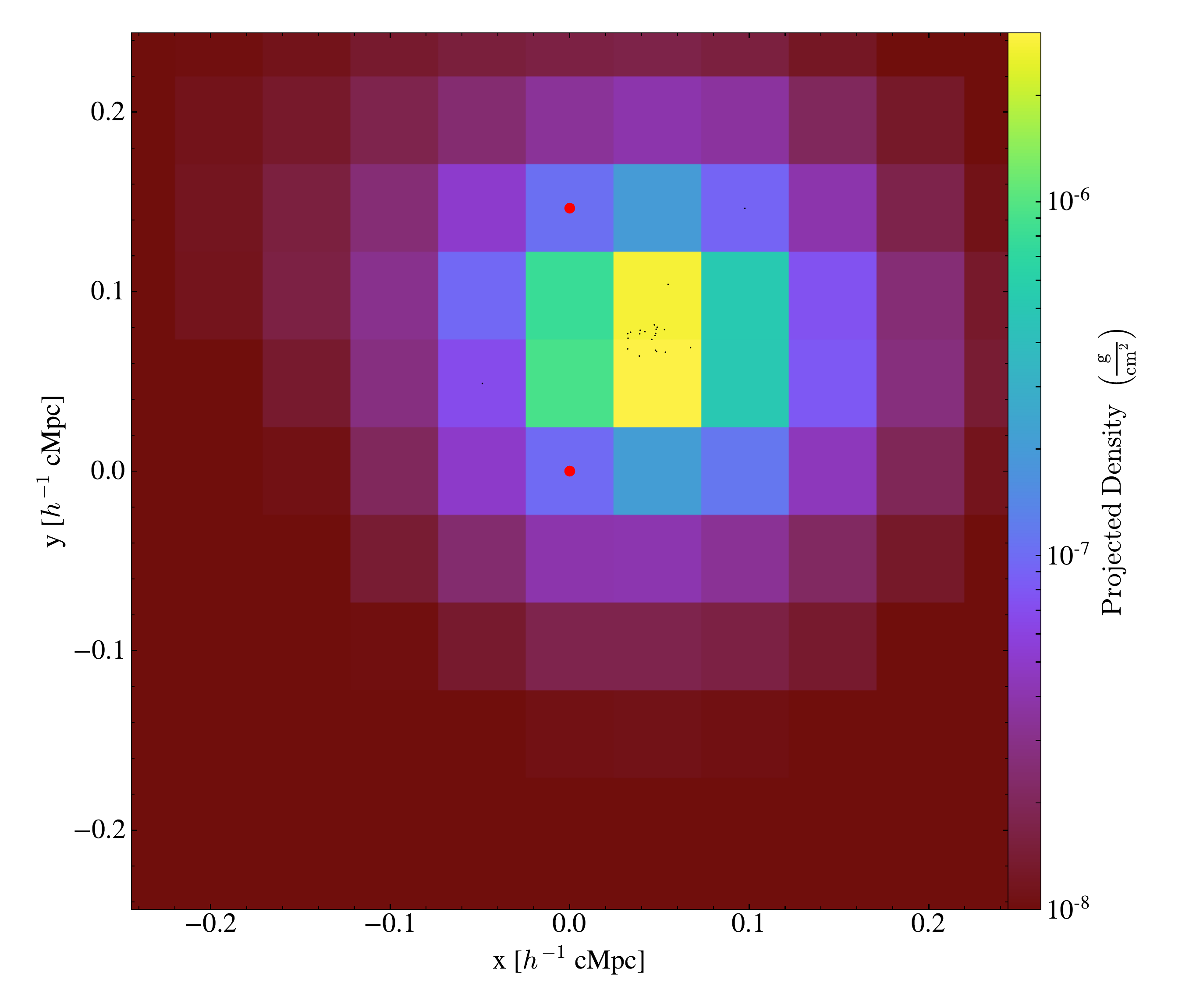}
	\caption{Density projection plot of {\sl NL-1} at $z=-0.99, t=t_0+6t_{\rm H}$ centred around one of the star particles formed within 500 Myr (young stars). Young stars and previously formed stars are indicated by red and black dots respectively. Despite having features similar to a halo (radially decreasing density from a dense centre), it is not classified as one by {\tt ROCKSTAR}. The site of active star formation is extended outwards for the reasons discussed in Section \ref{sec:conv-sfrd}.}
	\label{fig:future_sf}
\end{figure}

In {\tt Enzo}, star formation occurs above a fixed overdensity threshold. Since this value is constant with time, the decrease of the mean matter density into the future translates to a lowered physical density requirement. We illustrate in Figure \ref{fig:future_sf} that star formation can occur at increasing distances from the centre of a halo-like object, characterized by the radially decreasing density from a dense centre. We distinguish stars formed within and beyond $500\,{\rm Myr}$ in the projection plot as young (red dots) and old stars (black dots) respectively. The young stars are further away from the centre of the halo. It is not a halo by the definition of {\tt ROCKSTAR} because at $z=-0.99$, the HMF of {\sl NL-1} (blue line) in Figure \ref{fig:conv-hmf7tH} is deficient. However, the features of Figure \ref{fig:future_sf} resemble that of a halo, 
indicating that {\tt Enzo} is at its operational limits with the given force resolution, which adversely affects {\tt ROCKSTAR}'s ability to locate haloes. This migration of the site of star formation might also be coupled with the presence of hot gas in the central region of the halo-like object (see Figure \ref{fig:sim_phaseplot}).

\begin{figure}
	\centering
	\includegraphics[width=\linewidth]{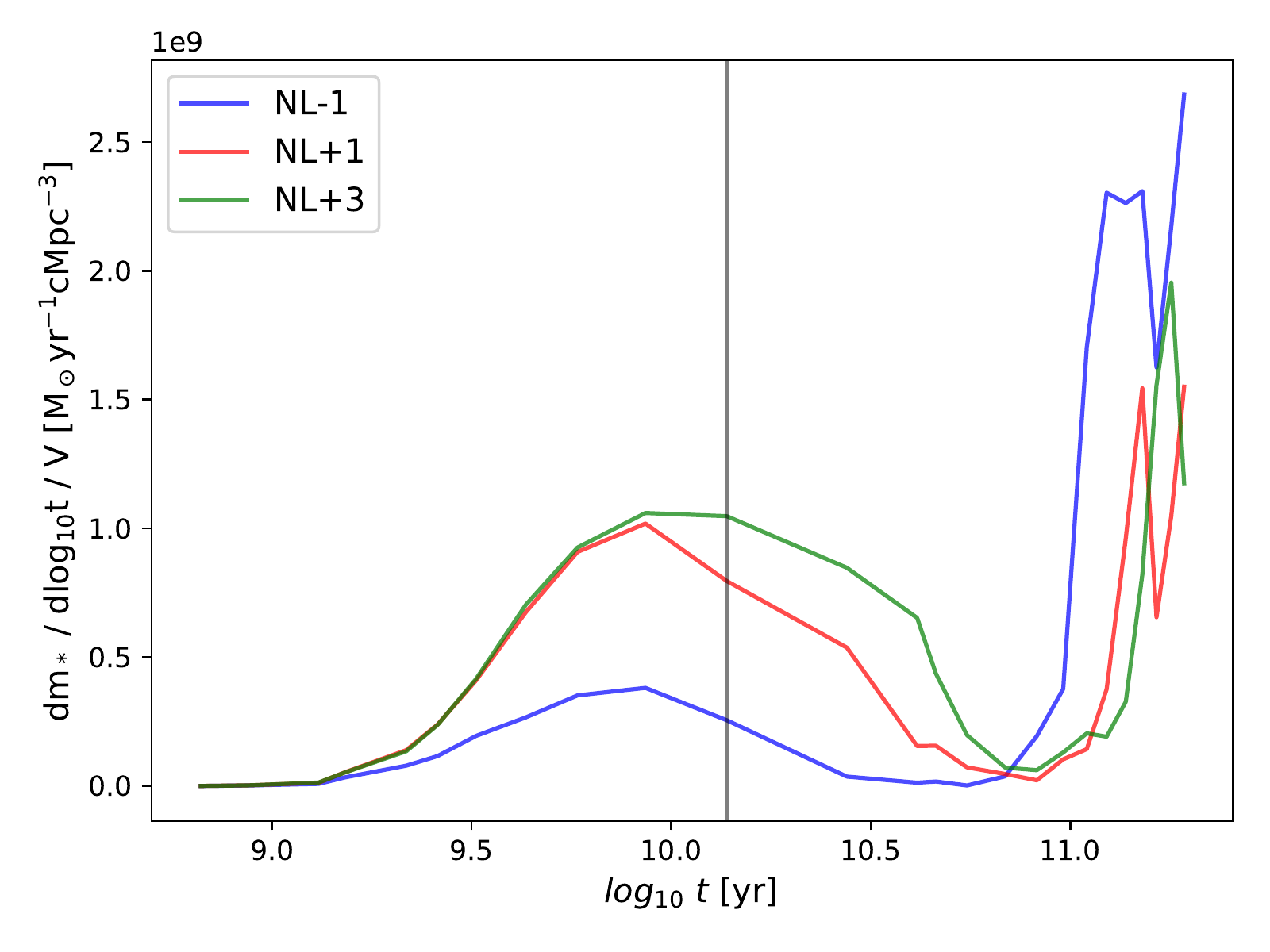}
	\caption{Evolution of the SFRD beyond the timescale in Figure \ref{fig:conv-sfr}. The blue, red and green lines correspond to {\sl NL-1}, {\sl NL+1} and {\sl NL+3} respectively. The vertical black line indicate the present day. As predicted, the SFRD for {\sl NL+1} turns around at a later time. However, the turnaround time for {\sl NL+1} coincides with {\sl NL+3}; we will provide some possible explanations for this in Section \ref{sec:conv-sfrd}.}
	\label{fig:conv-exsfr}
\end{figure}

The turnaround in Figure \ref{fig:future_sf} is delayed when the maximum level of refinement is increased. As a result, we expect it to happen in {\sl NL+1} at a time later than shown. Figure \ref{fig:conv-exsfr} validates this claim by simulating {\sl NL+1} (red line) further into the future. We have also added a simulation named {\sl NL+3} (green line) with a higher maximum spatial resolution of $4.39\,{\rm ckpc}$ to further support the dependence of the turnaround on spatial resolution. In other words, the turnaround in SFRD should occur in the order {\sl NL-1}, {\sl NL+1} then {\sl NL+3}. But contrary to these expectations, it happens at the same time for both {\sl NL+1} and {\sl NL+3}. 

\subsubsection*{Origin of the turnaround}
\begin{figure*}
	\centering
	\subfloat[{\sl NL+1} \label{fig:nl+1}]{%
		\includegraphics[width=.49\linewidth]{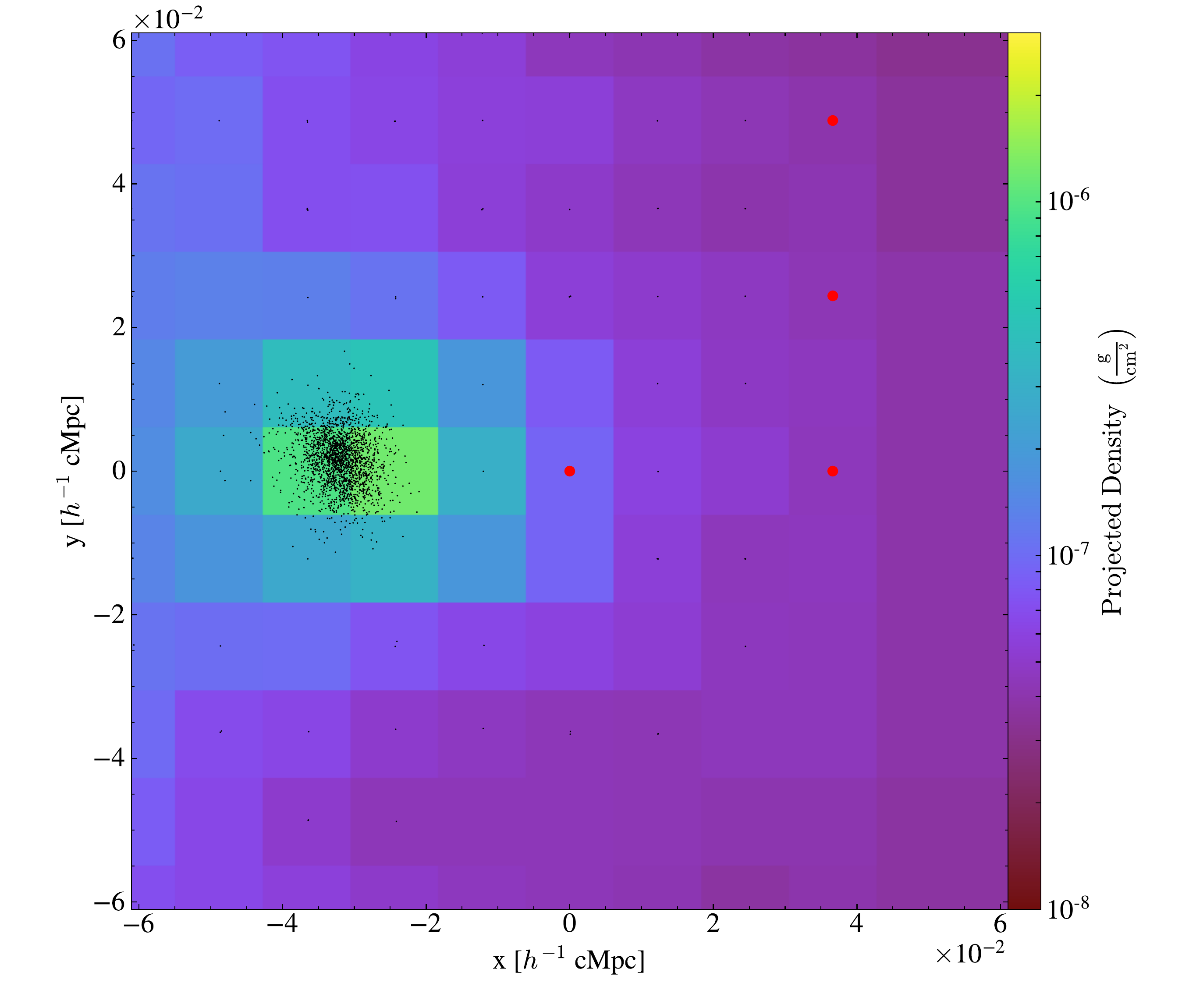}
	}
	\hfill
	\subfloat[{\sl NL+3} \label{fig:nl+3}]{%
		\includegraphics[width=.49\linewidth]{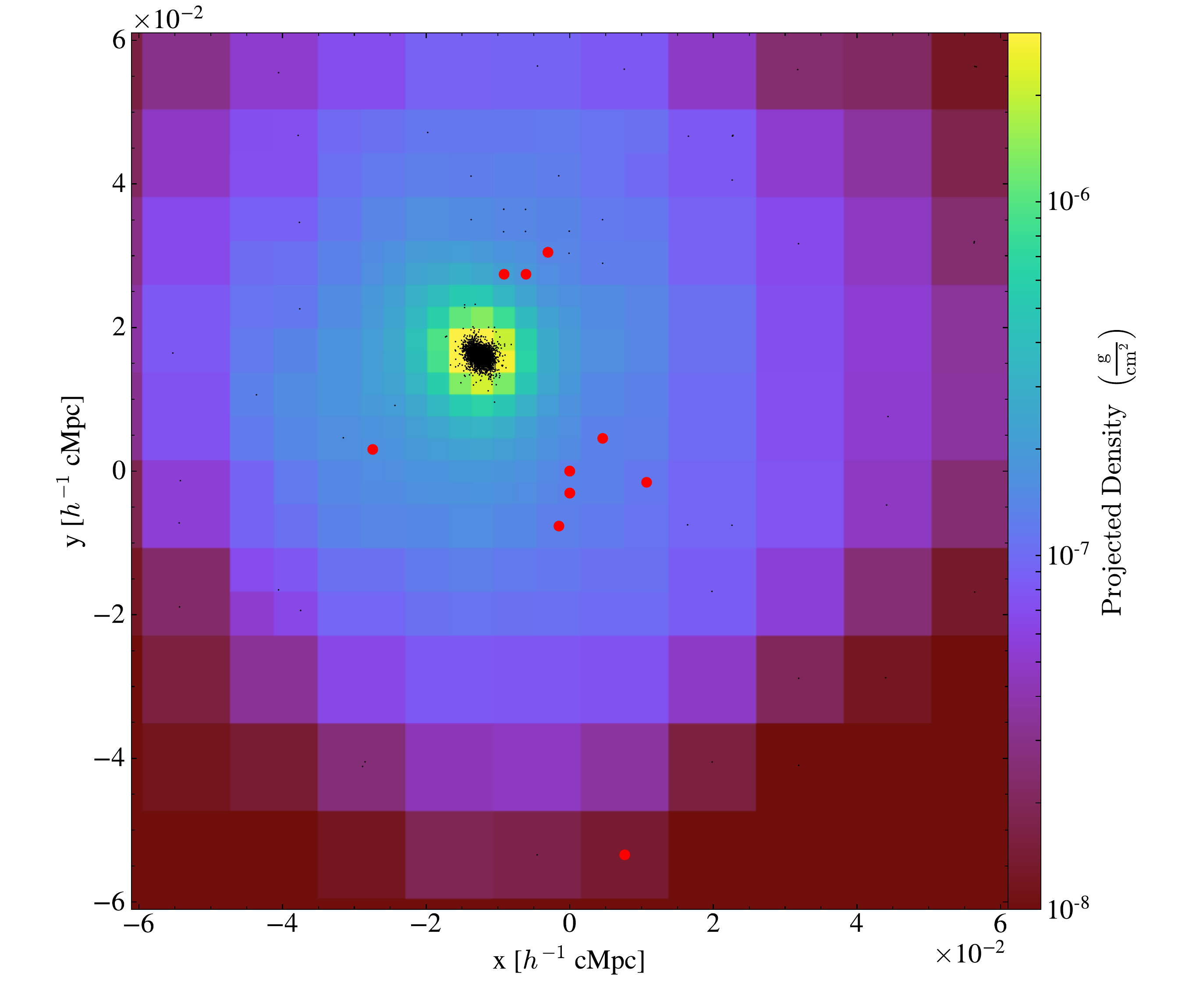}
	}
	\caption{Comparison of density projection plot between {\sl NL+1} and {\sl NL+3} at $z=-0.997$. The dots have identical meanings to those in Figure \ref{fig:future_sf}. Star formation happens further away from the density peak and in cells not at the maximum level of refinement in the simulation in {\sl NL+3} in comparison to {\sl NL+1}. Refer to Section \ref{sec:conv-sfrd} for discussion.}
	\label{fig:NL+1vs+3}
\end{figure*}

It appears that the turnaround in the SFRD does not continue to be delayed according to spatial resolution. We first investigate the temporal resolution applied to obtain Figure \ref{fig:conv-exsfr}. With a time interval of $t_{\rm H} = 13.7\,{\rm Gyr}$ between each point, the exact time where the turnaround occurs may be slightly different between {\sl NL+1} and {\sl NL+3}, but nonetheless breaks with the expected trend. We therefore look at the environments where the young stars form in both simulations, shown in Figure \ref{fig:NL+1vs+3}. It appears that star formation can occur in cells with lower density and at a refinement level that is two levels lower than the maximum level in {\sl NL+3} (bottom region of Figure \ref{fig:nl+3}), i.e., equivalent to the maximum level in {\sl NL+1}. In the {\sl NL+3} simulation, cells that satisfy these requirements are at a spatial resolution identical to the maximum resolution of {\sl NL+1}, concluding that the refinement levels beyond {\sl NL+1} is unnecessarily high for star formation. Therefore, the star formation in {\sl NL+3} will be comparable to {\sl NL+1}, resulting in the similarity of turnaround time. 

Looking at Figure \ref{fig:NL+1vs+3}, the young stars form in an environment with a density that is approximately two orders of magnitude lower than the old stars at the density peak at $z=-0.997$. This difference arises because of the constant star formation overdensity threshold discussed earlier. We should investigate how to modify this parameter in future work for a realistic star formation scenario in the universe at even greater ages.. Similarly to our results, \citet{2018MNRAS.477.3744S} also found a turnaround in star formation rate in their simulations -- but this happens much earlier and the origin is also very different. These authors attributed their turnaround to switching off AGN feedback in their simulations. Since our simulation do not contain this form of feedback, it appears that our stellar feedback alone is sufficient to produce the effects attributed to AGN in other work, suppressing a turnaround in the SFRD unaided.

\subsection{Zoom vs cosmological box simulations}\label{sec:zoom_vs_box}

\begin{figure}
	\centering
	\includegraphics[width=\linewidth]{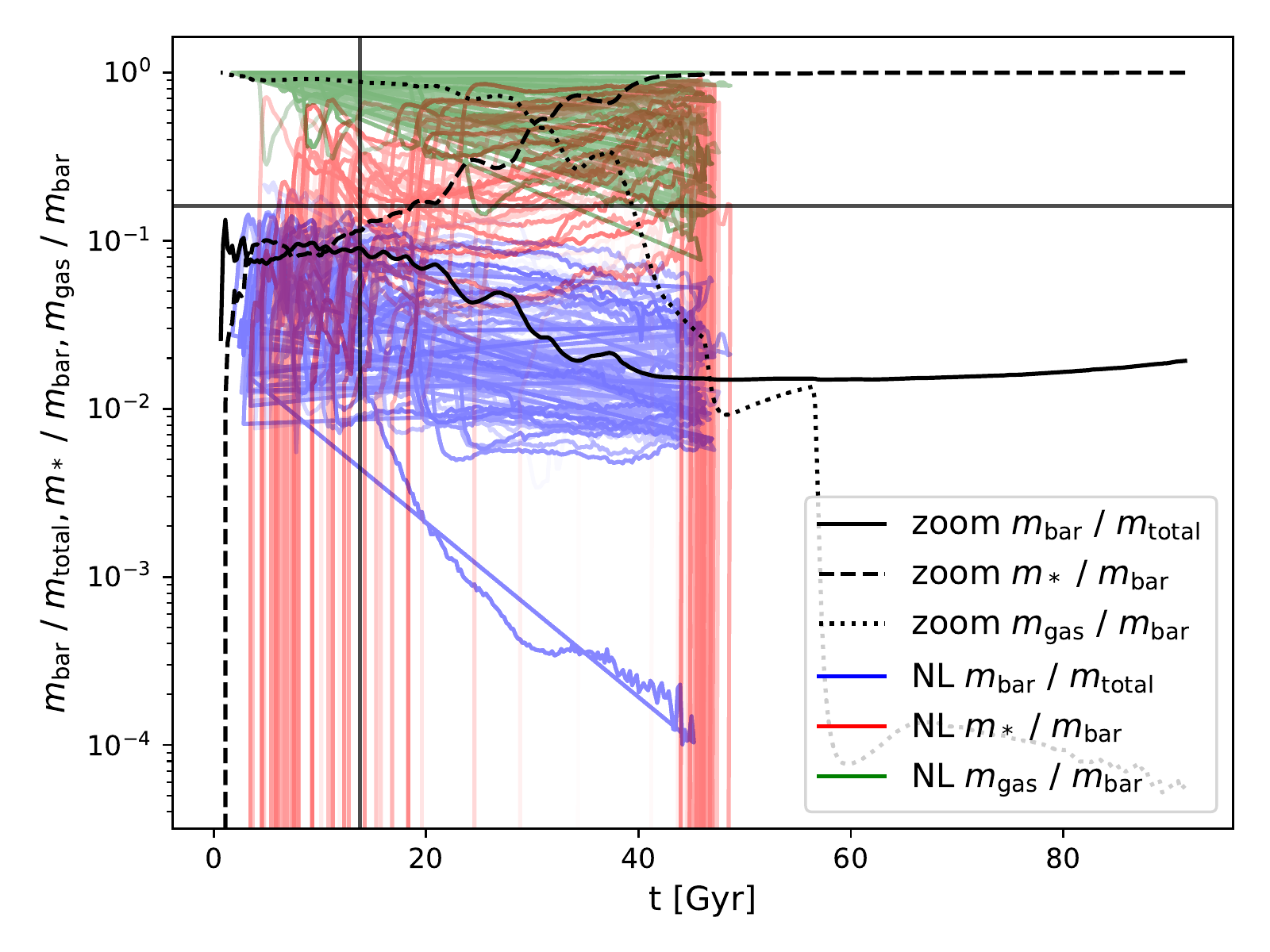}
	\caption{Evolution of various ratios of masses in haloes with time. The black and coloured lines in this figure are from {\sl zoom} and {\sl NL} respectively. We can group several pairs of lines for comparison: solid black and blue lines, dashed black and red lines, and dotted blacked and green lines. They correspond to the baryon fraction ($m_{\rm bar} / m_{\rm total}$), $m_* / m_{\rm bar}$ and $m_{\rm gas} / m_{\rm bar}$ of the haloes respectively. Also, we included a horizontal and vertical line which correspond to the universal baryon fraction in the simulation (0.1618) and $z=0$ respectively. Other than resolution affecting the starting and ending time when the halo can be traced, the MW-sized halo in the {\sl zoom} exhibits a typical evolution as other haloes within the specified mass range in {\sl NL}. Refer to Section \ref{sec:zoom_vs_box} for a detailed discussion.}
	\label{fig:mfracevo}
\end{figure}

Finally, we extend the zoom simulation discussed in \citet{BK20} to $z=-0.995$, labelled {\sl zoom} here. Since this simulation delivered a much higher resolution around a MW-sized halo, we can perform a detailed study of the baryon content within such a halo. Using the virial mass of the MW-sized halo in {\sl zoom} at $z=0$, we identify (sub-)haloes within 10\% of this mass in {\sl NL} at $z=0$. We then trace and present the evolution of several properties of these haloes in Figure \ref{fig:mfracevo}. The superior resolution of {\sl zoom} allows the halo to be tracked over a longer time; but within the period that we are able to track the halo in {\sl NL}, the general evolution of the various properties show that the MW halo in {\sl zoom} is typical of haloes in this mass range in {\sl NL}.

If we focus on at the baryon fraction of the haloes from both {\sl zoom} (black solid line) and {\sl NL} (blue line), it is always below the universal baryon fraction (grey horizontal line). This difference implies that haloes of other masses or the IGM will have a baryon fraction that is above average. An equality between $m_* / m_{\rm bar}$ and $m_{\rm gas} / m_{\rm bar}$ happens for both {\sl zoom} and {\sl NL} at $t\approx30\,{\rm Gyr}$, coinciding roughly with the period of freeze out. Since freeze out reduces the supply of gas from the large scale environment and feedback from stars drives the gas out of halo, it causes $m_{\rm gas}$ in the haloes to decrease. On the other hand, $m_*$ remains relatively constant because of the decreasing SFRD. These factors lead to an increase in $m_* / m_{\rm bar}$ and a corresponding decrease in $m_{\rm gas} / m_{\rm bar}$.

As mentioned, we are able to track the baryon content in a MW-sized halo for a significantly longer period in {\sl zoom}. Although not reaching the value set by the simulation, the baryon fraction remains relatively constant after $40\,{\rm Gyr}$. The same can be said for $m_* / m_{\rm bar}$, staying at a value near unity, implying that all the baryons in the halo are in the form of stars. On the other hand, $m_{\rm gas} / m_{\rm bar}$ decreases into the future. Combining all of these observations, in the far future, baryons are all locked up in stars in the MW-sized halo in {\sl zoom}. In doing so and coupled with feedback, only a small amount of gas will remain in the halo, suggesting a low probability of a turnaround in the star formation rates.

\section{Summary and conclusions} \label{sec:naga-summary}

Our study presents the first suite of simulations using {\tt Enzo} to simulate the evolution of the universe and its star formation into the future. We modify certain aspects of the cooling and chemistry library, {\tt Grackle} and the extrapolation of the UV background. We start with a simulation consistent with \citet{2004NewA....9..573N}, and then vary the resolution in order to check for convergence in the evolution of a range of properties including stellar masses, SFRDs, HMFs, tight power law fit of the IGM, and gas phase distribution. We survey these properties spanning from $z=99$ to $z=-0.995$, which translates to a final age of more than 7 $t_{\rm H}$. We summarise our conclusions as follows:

\begin{itemize}
	\item In order to make a fair comparison between the {\tt GADGET} simulation by \citet{2004NewA....9..573N} and our {\tt Enzo} simulation, we adjusted the root grid resolution and the maximum allowed levels of AMR: a $128^3$ root grid {\tt Enzo} simulation, equivalent to the $64^3$ used in the {\tt GADGET} simulation and four additional levels of AMR to eliminate any discrepancies of the results due to resolution \citep{2005ApJS..160....1O}. We also decided on a final redshift of $-0.995$, producing a universe of a similar age to that of \citet{2004NewA....9..573N}. See Section \ref{sec:resolution} for more details.
	
	\item We changed the method of extrapolation of the UV background in {\tt Enzo} beyond $z=0$ from linear to logarithmic to prevent negative photoheating rates from HI, HeI and HeII, such that the UV background will begin to cool the gas in the IGM. We also modified the {\tt CLOUDY} table and {\tt Grackle} to extend its capability to handle the low densities present in the far future. We also verified the capability of {\tt ROCKSTAR} to locate and track haloes into the far future via a catalogue of idealised NFW haloes.
	
	\item The results from {\sl NL} showed an excellent agreement with \citet{2004NewA....9..573N} concerning the evolution of the distribution of gas on the temperature -- gas overdensity plane, as presented in Section \ref{sec:naga-compare}. The equation of state of the IGM also evolves similarly for both, with a consistent drop beyond the present time, suggesting a further cosmological coincidence problem. However, the agreement is not perfect due to disparities in subgrid physics implemented in both simulations.
	
	\item We changed the mass resolution and maximum allowed spatial resolution in our suite of simulations to test for convergence. The lack of evolution at the high mass end of the HMF in Figure \ref{fig:hmf_evo} is consistent with the prediction of freeze out happening at $z\approx-0.6$. We found that the maximum spatial resolution of the simulation has a drastic impact on the HMF at late times. Since the force resolution is twice the spatial resolution in {\tt Enzo}, the expansion of the universe translates to a deteriorating proper force resolution, leading to the loss of low mass haloes from the HMF. Eventually, the entire HMF is affected, making spatial resolution a key factor for our simulations into the future.
	
	\item Nevertheless, we showed in Figure \ref{fig:conv-igm_prop_evo} that resolution adjustments do not affect the future evolution of the IGM significantly. The balance of the photoheating rate from the UV background and the adiabatic cooling due to the expansion of the universe creates a tight relation between temperature and density of gas in the IGM. This balance is disrupted in the future because the photoheating rate decreases due to the falling star formation rate while the adiabatic cooling rate increases because of the accelerated expansion of the universe.
	
	\item In Figure \ref{fig:conv-sfr}, we showed that our simulations reproduce a peak in the SFRD. This peak is however lowered and delayed somewhat compared to the observational fit by \citet{2014ARA&A..52..415M}, because the limited resolution in the simulations restricts the onset of structure formation. Despite this difference, we obtain a good agreement between the simulated and predicted total asymptotic stellar mass within the same order of magnitude, with the exception of {\sl NLfb}. Differences in star formation and feedback prescription results in a significantly lower star formation rate and total stellar mass formed. The ratio of simulated to predicted total stellar mass reaches as high as 99\% in {\sl NL+1}. We found a turnaround in the SFRD at late times in simulations of poorer resolutions, but its origin is determined to be a numerical artefact.
	
	\item Lastly, we select haloes within 10\% of the virial mass of the MW-halo in {\sl zoom} and compare the evolution of various properties of these haloes from {\sl NL} in Figure \ref{fig:mfracevo}. Although the resolution differs by a large margin, the general evolution of a MW halo in {\sl zoom} is similar haloes of similar mass in {\sl NL}. There is also a crossover of $m_* / m_{\rm bar}$ and $m_{\rm gas} / m_{\rm bar}$ at $t\approx30\,{\rm Gyr}$ which coincides with the period of freeze out in both {\sl zoom} and {\sl NL}.
\end{itemize}

In conclusion, we find a general agreement between many of our results on the long-term evolution of the IGM and those of \citet{2004NewA....9..573N}, despite differences in the methodology of the simulation code, star formation and feedback prescription. The aim of this study of the IGM has been to understand the long-term supply of gas into the haloes, fuelling future star formation. With this purpose in mind, we consider the significance of the presence of cold dense gas in the future of \citet{2004NewA....9..573N}'s simulated universe. Since it is converted into stars in our simulation, this discrepancy can cause possible disparities in the late-time star formation rates, leading to deviations in the asymptotic stellar mass formed.

We have not considered the feedback effects from black holes, which are expected to be subdominant; but it will be interesting to include them to improve the accuracy of the results obtained from the simulations. Considering the HMF, haloes with a mass well above that of the MW exist and will undoubtedly host a supermassive central black hole, whose effects will be important for sufficiently large masses. In a $\Lambda$-dominated universe, freeze out prevents the HMF being dominated by such haloes, but this will not be the case in alternative counter-factual models such as the future evolution of the Einstein--de Sitter universe considered by \citet{2018MNRAS.477.3744S}. Including black hole physics will be desirable in order to allow a fair comparison between the results of that study and an alternative {\tt Enzo} calculation.

As is evident from the present work, the resolution of the simulation strongly influences our ability to predict the future evolution of cosmological entities such as the HMF and SFRD. Future work should therefore direct greater computational resources towards improved resolution. In addition, the criteria for star formation will have to be adjusted in order to prevent an unphysical turnaround in star formation. But we have nevertheless been able to obtain a remarkable convergence of the asymptotic stellar mass in our simulations, which agrees closely with an extrapolation of the observed SFRD. We are therefore encouraged that our framework already provides a useful means of exploring how the cosmological model might influence the asymptotic star formation efficiency, and we plan to explore this in further work.

\section*{Acknowledgements}

BKO and JAP were supported by the European Research Council under grant number 670193 (the COSFORM project). BKO would like to thank Jose O{\~n}orbe and the TMOX group at the Royal Observatory, Edinburgh for many insightful discussions.

\section*{Data Availability}

The data underlying this article will be shared on reasonable request to the corresponding author.




\bibliographystyle{mnras}
\bibliography{myref,myref1,myref2,myref3} 



\appendix

\section{Tests of {\tt ROCKSTAR}'s ability to find haloes}\label{sec:rockstar_test}
As described in Section \ref{sec:postproc}, we conducted a number of experiments to understand the ability of {\tt ROCKSTAR} to locate and identify haloes in the far future. In short, we placed isolated haloes with a truncated NFW density profile down with as little as two particles with a mass resolution of $4.96\times10^9\msunoh$ at three different redshifts ($z=0, -0.5, -0.9$). The haloes are distributed identically in space but with a decreasing comoving virial radius to mimic the affect of the expanding universe.

\begin{figure*}
	\centering
	\subfloat[$z=0$ \label{fig:halo_num_z0}]{%
		\includegraphics[width=.43\linewidth]{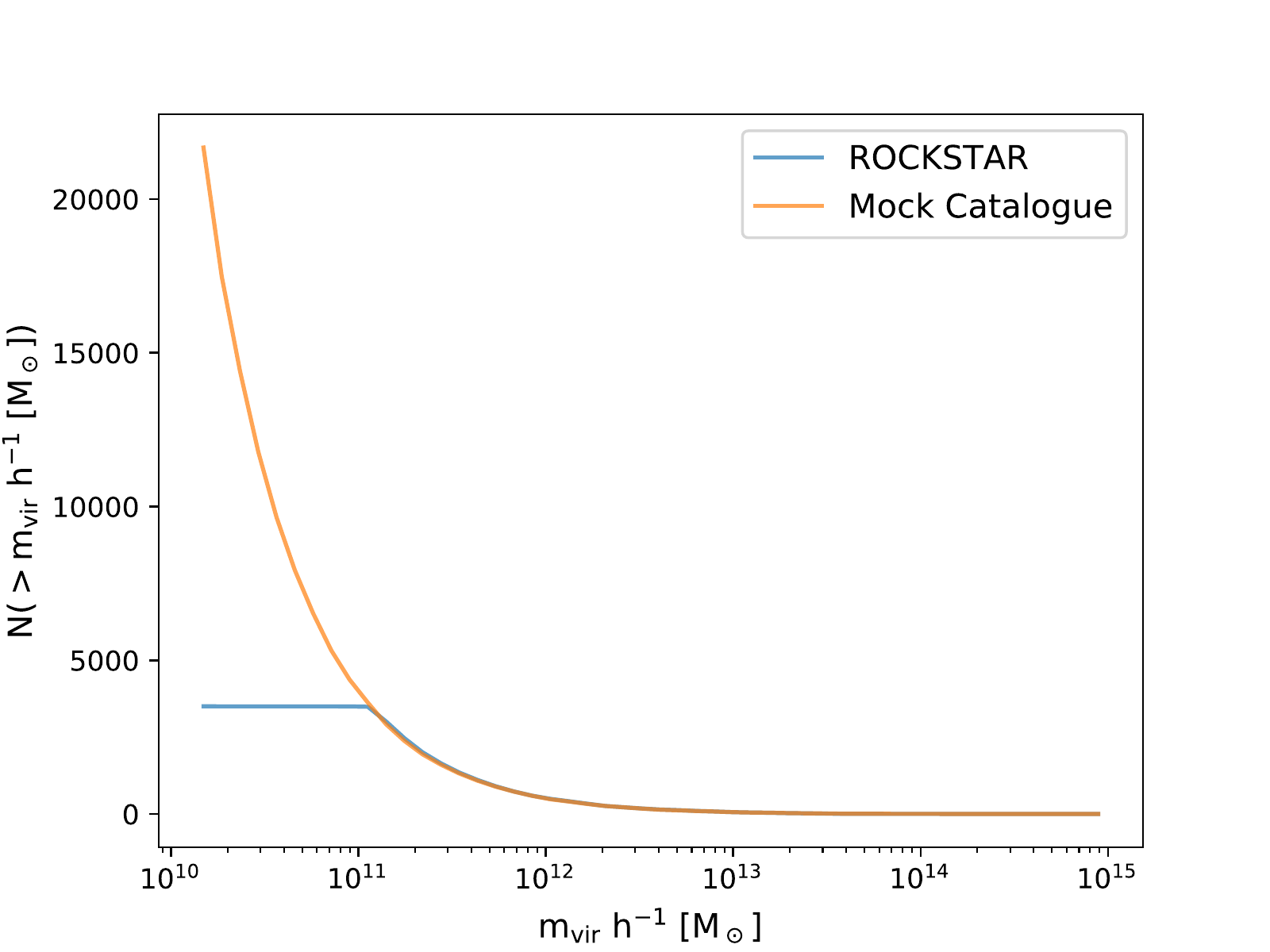}
	}
	\subfloat[$z=-0.5$ \label{fig:halo_num_z05}]{%
		\includegraphics[width=.43\linewidth]{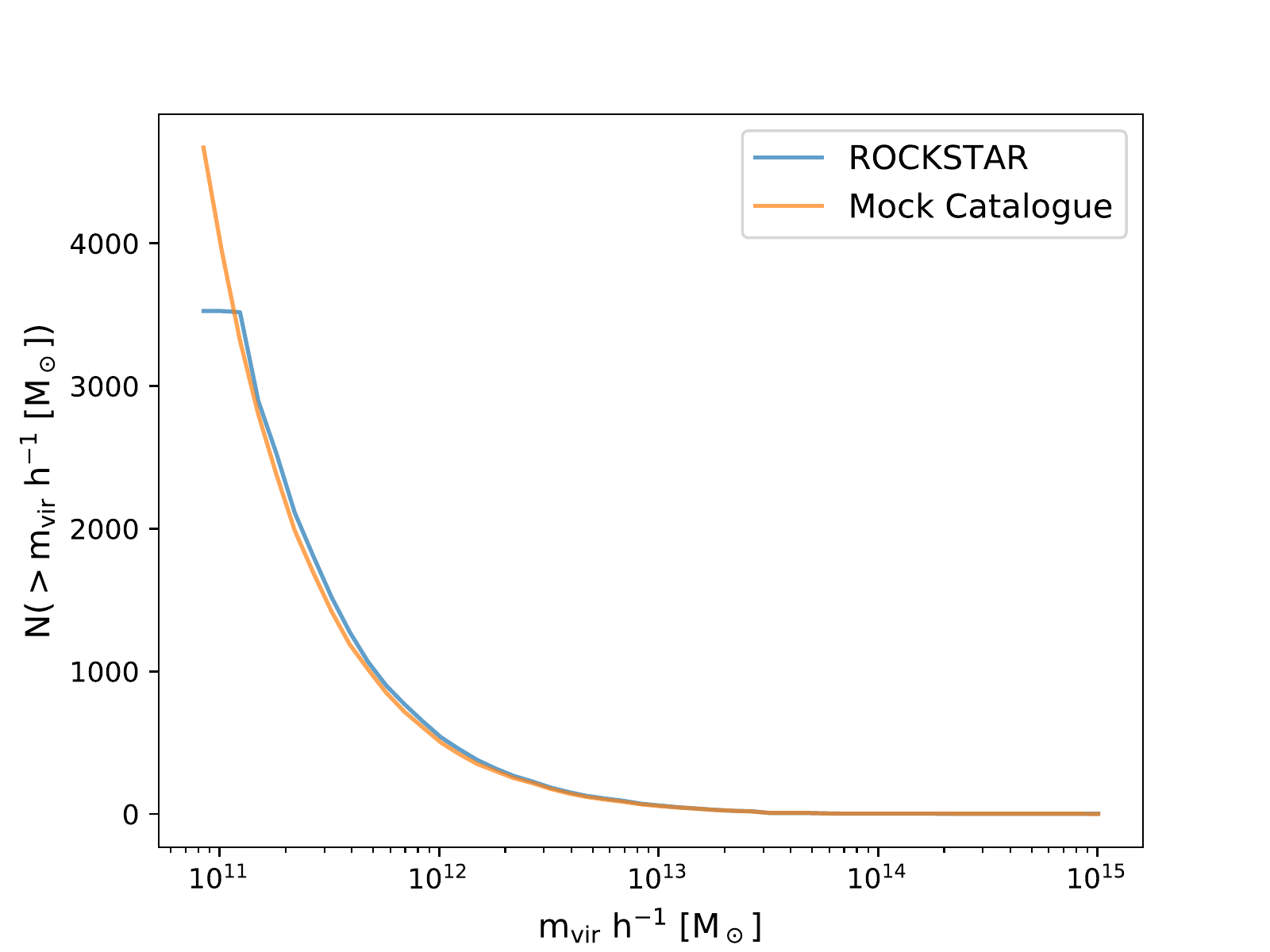}
	}
	\hfill
	\subfloat[$z=-0.9$ \label{fig:halo_num_z09}]{%
		\includegraphics[width=.43\linewidth]{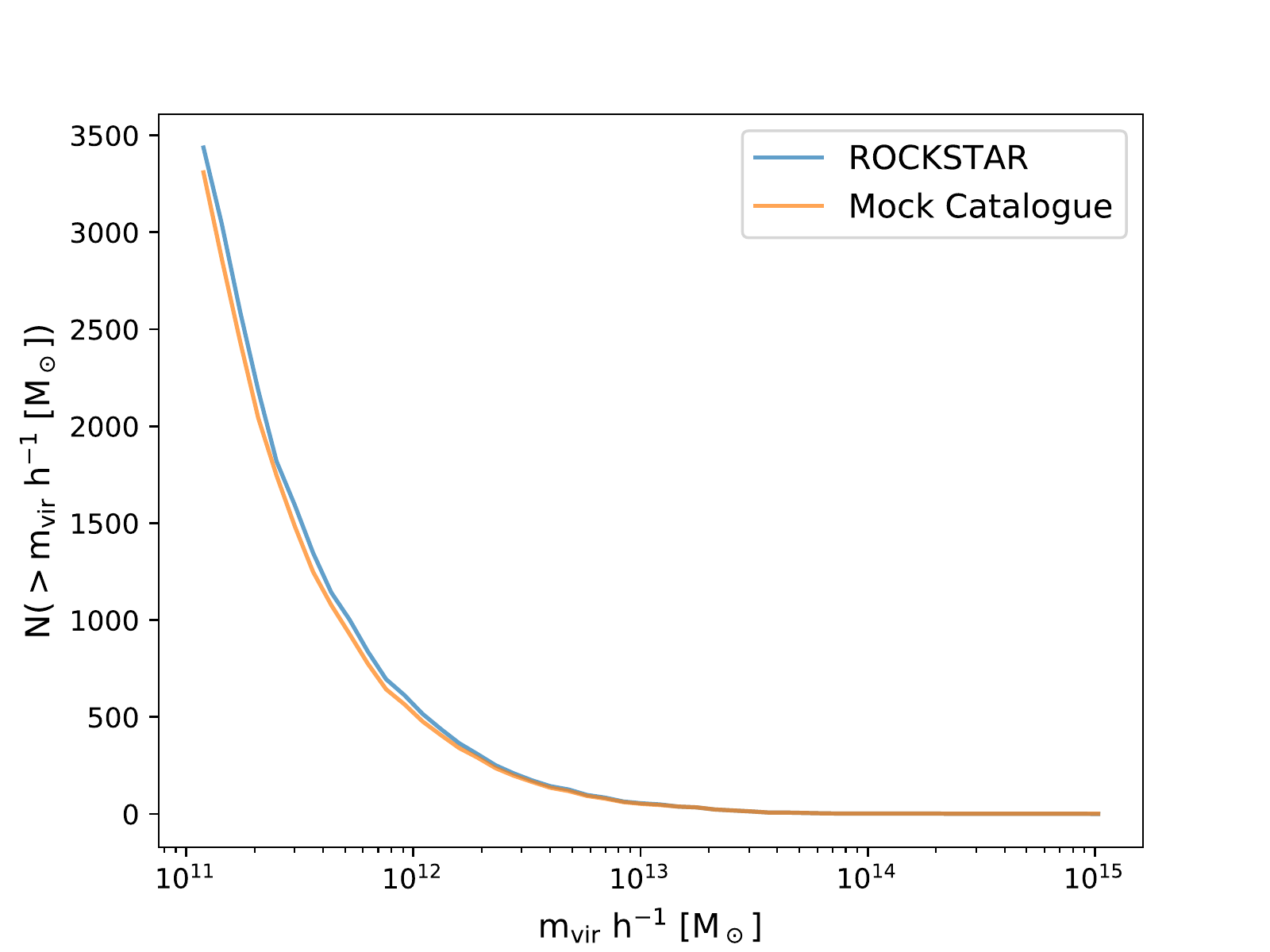}
	}
	\caption{Cumulative number of haloes above a specified virial mass at three different $z$ indicated in the captions. The blue and orange lines represent the numbers obtained from {\tt ROCKSTAR} and the mock catalogue respectively. There is a consistent deviation on the low mass end and agreement on the high mass end across redshifts.}
	\label{fig:rockstar_test}
\end{figure*}

Figure \ref{fig:rockstar_test} compares  the HMFs from the {\tt ROCKSTAR} halo catalogues (blue) and the mock catalogues(orange) at different redshifts. The agreement is remarkable across all redshifts. On the high mass end, we can recover a one to one mapping of the 40 most massive haloes, above a virial mass of $1.4\times10^{13}\msunoh$ at $z=-0.9$ with similar numbers at other redshifts. However, in the intermediate mass range ($1.5\times10^{11}\msunoh$ < $m_{\rm vir}$ < $10^{12}\msunoh$), {\tt ROCKSTAR} slightly overestimates the number of haloes, most apparent in the future. This result is surprising considering the haloes are isolated, eliminating the possibility of the cluttering of the haloes causing {\tt ROCKSTAR} to combine two or more low mass haloes close to each other into one more massive halo. Despite this discrepancy, {\tt ROCKSTAR} has been shown to be capable of accurately identifying haloes in the far future.


\bsp	
\label{lastpage}
\end{document}